\documentclass[a4paper,11pt]{amsart}

\usepackage[margin=1in]{geometry}
\usepackage{graphicx} 
\usepackage[svgnames]{xcolor}
\usepackage[dvipsnames]{xcolor}
\usepackage{url}
\usepackage{algorithm}
\usepackage[font=small,labelfont=bf, width=\textwidth]{caption}
\usepackage{algpseudocode}
\usepackage{mathtools}
\usepackage{tikz-cd}
\usepackage{hyperref}
\usepackage{amsaddr}
\usepackage{lineno}
\usepackage{xr-hyper}
\usepackage[
backend=biber,
style=numeric,
 sortcites,
 url = false,
 clearlang=true,
 maxnames=99
]{biblatex}

\usepackage{soul}

\hypersetup{
  colorlinks   = true, 
  urlcolor     = DarkBlue, 
  linkcolor    = DarkBlue, 
  citecolor   = Maroon 
}

\bibliography{main}

\graphicspath{{figures/}}

\newcommand{\argmin}{\operatornamewithlimits{argmin}}

\definecolor{WildStrawberry}{RGB}{242, 105, 142}

\title[Topological decoding of grid cell activity via path lifting to covering spaces]{Topological decoding of grid cell activity \\ via path lifting to covering spaces}

\author[Yao, Yoon]{Yuxing Jared Yao\textsuperscript{1,2}, Iris H.R. Yoon\textsuperscript{1,3}*}

\address{\small\normalfont\textsuperscript{1}Department of Mathematics and Computer Science, Wesleyan University \\ \textsuperscript{2}Program in Neuroscience and Behavior, Wesleyan University \\ \textsuperscript{3}Department of Mathematics and Statistics, Swarthmore College}

\begin{document}

\begin{abstract}
High-dimensional neural activity often resides in a low-dimensional subspace, referred to as neural manifolds. Grid cells in the medial entorhinal cortex provide a periodic spatial code that is organized near a toroidal manifold, independent of the spatial environment. Due to the periodic nature of this code, it is unclear how the brain utilizes the toroidal manifold to understand its state in a spatial environment. We introduce a novel framework that decodes spatial information from grid cell activity using topology. Our approach uses topological data analysis to extract toroidal coordinates from grid cell population activity and employs path-lifting to reconstruct trajectories in physical space. The reconstructed paths differ from the original by an affine transformation. We validated the method on both continuous attractor network simulations and experimental recordings of grid cells, demonstrating that local trajectories can be reliably reconstructed from a single grid cell module without external position information or training data. These results suggest that co-modular grid cells contain sufficient information for path integration and suggest a potential computational mechanism for spatial navigation.
\end{abstract}

\maketitle


\section{Introduction}

Activity of a population of neurons often resides in a low-dimensional subspace called a neural manifold \cite{hubelReceptiveFieldsFunctional1968, okeefePlaceUnitsHippocampus1976, hafting2005microstructure, villetteInternallyRecurringHippocampal2015, gardner_toroidal_2022, chaudhuriIntrinsicAttractorManifold2019, zhouHyperbolicGeometryOlfactory2018, rybakkenDecodingNeuralData2019} whose structure reflects the information encoded by the neurons. For example, the activity of head direction cells \cite{taube1990head} are organized near a circle \cite{chaudhuriIntrinsicAttractorManifold2019, rybakkenDecodingNeuralData2019}. Grid cells in the medial entorhinal cortex (MEC) exhibit a periodic hexagonal firing pattern that tiles the environment at regular intervals \cite{hafting2005microstructure} and are organized into modules whose cells share scale and orientation but differ by fixed spatial phase offsets \cite{hafting2005microstructure,mcnaughton2006path}. The periodicity of a single-module of grid cell activity implies that the population activity is topologically organized around a torus \footnote{A torus is a space that represents the outside surface of a donut. Here, we represent a torus by identifying the left and right edges and by identifying the top and bottom edges of a parallelogram} \cite{fuhs2006spin, burak2009accurate}.
Such organization has been captured by continuous attractor network (CAN) models of grid cells and has been observed in large-scale recordings \cite{gardner_toroidal_2022}.

Because multiple locations in a spatial environment elicit a similar response among co-modular grid cells, spatial locations are not uniquely encoded in the toroidal neural manifold of grid cells. This insight raises a central question: how much spatial information can be decoded from the activity of a single module of grid cells? Prior efforts to decode position from neural activity often relied on place cell dynamics \cite{solstad2006grid, erdem2012goal, barry2012z, dang2021grid} or on combining phase differences across multiple grid modules \cite{mathis2012optimal, stemmler2015connecting}. Other methods perform cumulative vector integration \cite{bush2015using} or train deep models to map activity to position \cite{livezey_deep_2020, tampuu_efficient_2019, frey_deepinsight_2019, xu_comparison_2019, mitchell_topological_2024}. Theoretical analysis indicates that a single grid module may carry sufficient information to update its internal representation of position \cite{bush2015using,fiete_what_2008,sun_neural-like_1994, masson_decoding_2011}, known as path-integration, though explicit computational demonstrations have been limited, constrained to multiple trials of one-dimensional settings \cite{wen2024one} or relying on access to firing rate maps and training of deep neural networks \cite{peng_grid_2023}.

In this study, we present a novel method for decoding movement trajectories from the activity of a single module of grid cells. The method builds on the insight that the topology of a stimulus space can be recovered directly from neural activity \cite{curto2008cell} and integrates tools from topological data analysis and path lifting in topology: Persistent cohomology reveals the toroidal structure of grid cell activity, circular and toroidal coordinates parametrize grid cell activity on this torus, and path lifting reconstructs the movement path in Euclidean space. To the authors' knowledge, this is the first work integrating path lifting in topology into computational and applied settings. The approach differs from existing decoding work in that  it only uses data from a single module of grid cells and that it doesn't involve any training process. We validate the algorithm in both CAN-simulated and experimental datasets by showing that the reconstructed movement paths differ from the original by an affine transformation. The work highlights the sufficiency of co-modular grid cells for path integration.

\section{Results}

\subsection{An internal representation of space can be constructed from grid cell activity}

\begin{figure}[h!]
    \centering
    \includegraphics[width=\linewidth]{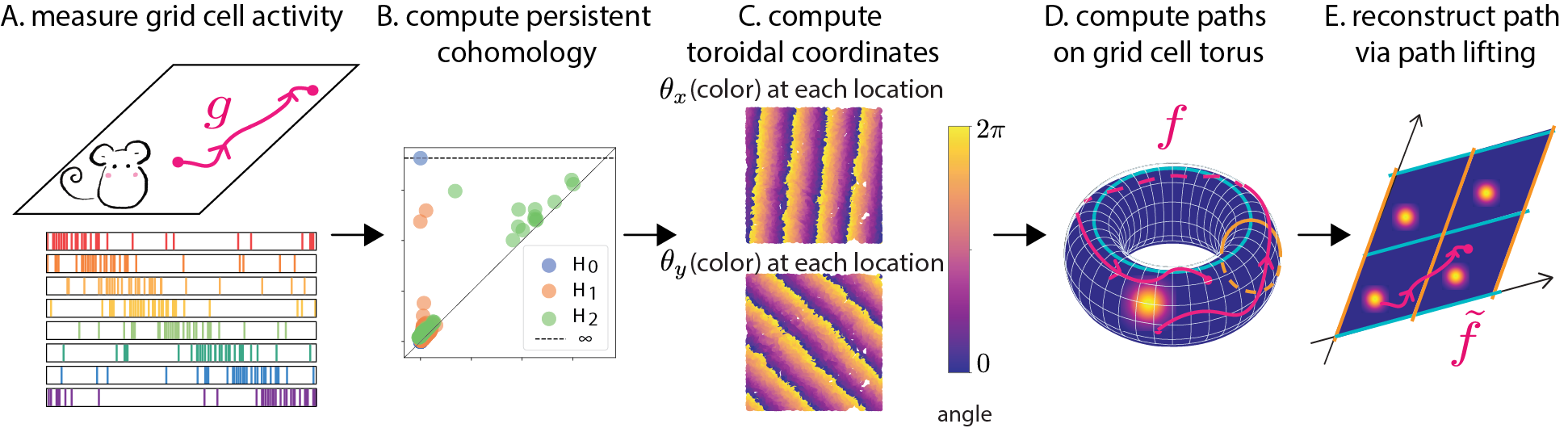}
    \caption{Constructing an internal representation of space from grid cell activity. \textbf{A.} The input data is grid cell activity collected while the mouse moves in an environment. Grid cell population activity is represented as a population vector $P(t)$ evolving over time. \textbf{B.} Persistent cohomology indicates that the population vectors are organized on a torus. \textbf{C.} Each population vector $P(t)$ is assigned toroidal coordinates $(\theta_x^t, \theta^t_y)$. Here, if the mouse is at location $(x,y)$ at time $t$, we show the toroidal coordinates $\theta^t_x$ (top) and $\theta^t_y$ (bottom) by color values over location $(x,y)$. \textbf{D.} The toroidal coordinates form a path $f$ on the grid cell torus. \textbf{E.} We finally lift $f$ to a path $\tilde{f}$ in $\mathbb{R}^2$ that matches the subject's movement up to an affine transformation.}
    \label{fig:pipeline}
\end{figure}

We present a novel algorithm that reconstructs movement trajectories from grid cell activity (Fig.~\ref{fig:pipeline}). The method proceeds in two stages. First, using persistent cohomology and toroidal coordinates, we assign toroidal coordinates to each population vector. This constructs a path on the grid cell torus as the subject moves (Fig.~\ref{fig:pipeline}A-D). Second, we ``lift" this path on the torus to the plane, thereby reconstructing the subject's movement in physical space (Fig.~\ref{fig:pipeline}E).

\subsubsection*{From grid cell activity to path on a torus}

The input is grid cell activity from a subject navigating a spatial environment, represented as a $G \times T$ matrix $A$, where $G$ is the number of grid cells and $T$ is the number of time bins \footnote{In general, $T$ represents the number of time bins. In some cases, as in the one-dimensional experimental data from \cite{wen2024one} analyzed in this paper, grid cell activity is presented over spatial bins, in which case $T$ represents the number of spatial bins.}.
The $(i,j)^{\text{th}}$ entry represents the activity of neuron $i$ at time bin $j$\footnote{The entries of the matrix can be binary, with $A(i,j)=1$ indicating that neuron $i$ fired at time bin $j$ and $A(i,j)=0$ representing that neuron $i$ did not fire at time bin $j$, or non-negative real numbers, with $A(i,j)$ representing the activity level of neuron $i$ at time bin $j$. In this work, the entries represent activity values. }. Each column of $A$ corresponds to the population vector $P(t)$ at time $t$. Although these vectors live in $G$-dimensions, the collection $ \{ P(t)  \}_{t=0}^{T-1}$ resides on a low-dimensional manifold called a torus  \cite{gardner_toroidal_2022}. To confirm this, we construct a Vietoris-Rips filtration, which is a nested sequence of simplicial complexes built by connecting population vectors whose dissimilarity falls below an increasing threshold (see \textit{Materials and Methods}). Persistent (co)homology, computed on this filtration, confirms that this low-dimensional manifold has the homology of a torus: one connected component, two 1-dimensional cycles, and one 2-dimensional void (see Fig.~\ref{fig:pipeline}B). We refer to this manifold as the \emph{grid cell torus}.

Each population vector $P(t)$ is then parametrized by toroidal coordinates \cite{de_silva_persistent_2011, scoccola_toroidal_2023} (\textit{Materials and Methods}), reflecting its position on the grid cell torus.
Formally, we define the map $\Theta: \{0, 1, \dots, T-1 \} \to S^1 \times S^1$ by
\[ \Theta(t) = (\theta^t_x, \theta^t_y),\]
where $\theta^t_x, \theta^t_y \in [0, 2 \pi)$. Here, $S^1$ denotes a circle, and the torus is represented by a product of two circles, $S^1 \times S^1$. Each $\theta^t_x$ and $\theta^t_y$ represents angles on each circle. See Figure~\ref{fig:pipeline}C.

The sequence $\{ \Theta(t) \}_{t = 0}^{T-1}$ forms a (discrete) path on the grid cell torus (Fig.~\ref{fig:pipeline}D).

\subsubsection*{From path on grid cell torus to a path in the plane}

Once the path on the torus is obtained, we finally recover the movement trajectory in $\mathbb{R}^2$. Conceptually, we ``unwrap" the path on the torus into $\mathbb{R}^2$, which we accomplish via path liftings to covering spaces \cite{munkres2017topology}. Given a covering map $p: \mathbb{R}^2 \to S^1 \times S^1$ and a continuous path $f: [0,T-1] \to S^1 \times S^1$ defined on the interval $[0,T-1]$, one can lift the path $f$ to $\tilde{f}:[0,T-1] \to \mathbb{R}^2$ so that $f = p \circ \tilde{f}$, i.e., the following diagram commutes (see SI Section~1.1 for details).
\[
\begin{tikzcd}
\quad & \mathbb{R}^2 \arrow[d, "p"] \\
\text{[} 0,T-1 \text{]} \arrow[ru, dashed, "\tilde{f}"] \arrow[r, "f"] & S^1 \times S^1
\end{tikzcd}
\]

Conceptually, the map $p$ folds $\mathbb{R}^2$ into a square torus by tiling the plane into parallelograms and mapping each parallelogram to one copy of the square torus (see SI Fig. 2). In an idealized setting where grid cell responses at the same physical location are identical, the toroidal coordinates $(\theta_x^*, \theta_y^*)$ will be identical at every time point $t$ at which the subject visits location $(x,y)$. In such idealized settings, the map $p$ is defined as $p(x,y) = (\theta_x^*, \theta_y^*)$.

\begin{figure}[t!]
    \centering
    \includegraphics[width=0.9\linewidth]{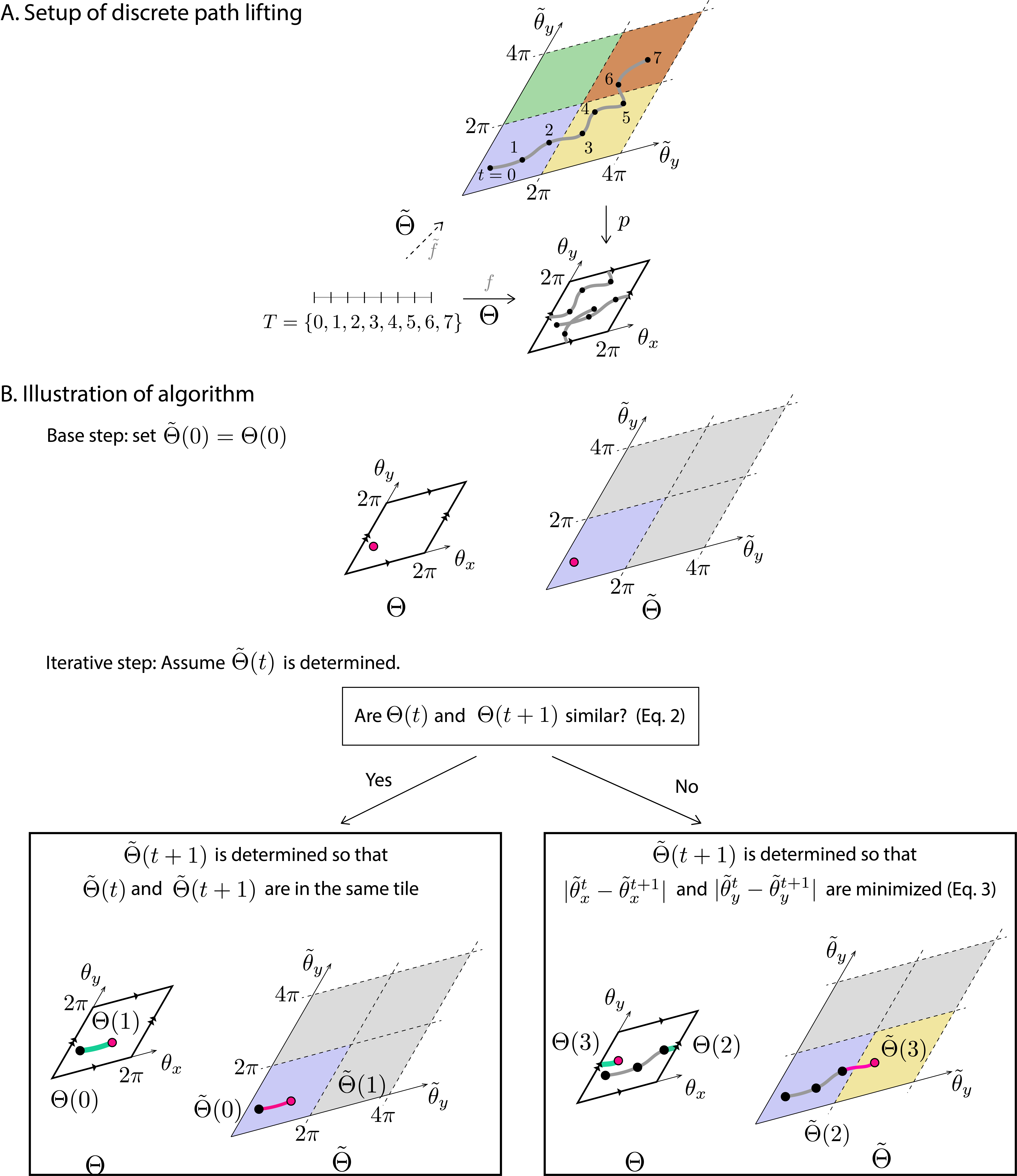}
    \caption{Lifting a discrete path $\Theta$ on the torus to a path $\tilde{\Theta}$ in $\mathbb{R}^2$. \textbf{A.} Setup: given a discrete path $\Theta: \{ 0, 1, \dots, T-1 \} \to S^1 \times S^1$ on the torus, and the goal is to construct a lifted path $\tilde{\Theta}: \{0, 1, \dots, T-1\} \to \mathbb{R}^2$ such that $p \circ \tilde{\Theta} = \Theta$, where $p: \mathbb{R}^2 \to S^1 \times S^1$ is a covering map.
   \textbf{B.} Algorithm. \textit{Base step:} $\tilde{\Theta}(0)$ is placed in the tile closest to the origin (blue). \textit{Iterative step:} Given $\tilde{\Theta}(t)$, the next lift $\tilde{\Theta}(t+1)$ is determined by comparing the consecutive toroidal coordinates $\Theta(t)$ and $\Theta(t+1)$ via Eq.~\ref{eq:similarity}. If they are similar (``Yes'' branch), the underlying path (green) is assumed to not cross a torus edge and $\tilde{\Theta}(t+1)$ is placed in the same tile as $\tilde{\Theta}(t)$. Otherwise (``No'' branch), the underlying path (green) is assumed to cross at least one edge and $\tilde{\Theta}(t+1)$ is placed in an adjacent tile, chosen to minimize $|\tilde{\theta}_x^t - \tilde{\theta}_x^{t+1}|$ and $|\tilde{\theta}_y^t - \tilde{\theta}_y^{t+1}|$ (Eq.~\ref{eq:determine_tile}).}
    \label{fig:algorithm}
\end{figure}

Here, $f$ denotes a continuous path on the grid cell torus traced by the population vectors throughout the experiment. We treat the sequence $\{\Theta(t) \}_{t=0}^{T-1}$ as samples of the path $f$. The goal is to create a discrete path $\tilde{\Theta}: \{0, 1, \dots, T-1\} \to \mathbb{R}^2$ such that the sequence $\{ \tilde{\Theta}(t) \}_{t = 0}^{T-1}$ are samples of $\tilde{f}$. In particular, $p \circ \tilde{\Theta} = \Theta$. See Fig.~\ref{fig:algorithm}A for a visualization of this setup.

We define $\tilde{\Theta}$ by lifting segments of the discrete path $\{\Theta(t) \}_{t=0}^{T-1}$ to various parallelograms of $\mathbb{R}^2$. This will be done by defining $\tilde{\Theta}(t) = (\tilde{\theta}^t_x, \tilde{\theta}^t_y)$ via
\begin{equation}
\label{eq:tile_defining}
\tilde{\theta}_x^{t} = \theta_x^{t} + 2 \pi M^t \quad \text{and} \quad  \tilde{\theta}_y^{t} = \theta_y^{t} + 2 \pi N^t,
\end{equation}
where $M^t$ and $N^t$ are integers specifying the tile in which the lifted point inhabits.

The algorithm inductively determines $M^t$ and $N^t$. Conceptually, we define $\tilde{\Theta}$ so that if two consecutive time points $t$ and $t+1$ have similar toroidal coordinates $\Theta(t)$ and $\Theta(t+1)$, then their lifts $\tilde{\Theta}(t)$ and $\tilde{\Theta}(t+1)$ live in the same tile. Otherwise, the lifted points live in different tiles. We say that the toroidal coordinates are similar if
\begin{equation}
    \label{eq:similarity}
    | \theta_x^{t} - \theta_x^{t+1} | \leq \varepsilon  \quad \text{and} \quad   | \theta_y^{t} - \theta_y^{t+1} | \leq \varepsilon
\end{equation}
for some proximity threshold $\varepsilon$. The proximity threshold $\varepsilon$ controls which consecutive time points are tested for nontrivial lifts: if the coordinate difference exceeds $\varepsilon$, the algorithm evaluates whether the path has crossed a torus edge; otherwise, the two points are lifted to the same tile. Importantly, $\varepsilon$ only flags candidates -- whether a nontrivial lift actually occurs is decided by a distance comparison in Equation~\ref{eq:determine_tile}. We select $\varepsilon$ from the distribution of consecutive coordinate differences, which concentrate near $0$ and $2\pi$, by choosing a value that lies below the cluster near $2\pi$. This ensures that nearly all potential edge crossings are tested. (See \textit{Materials and Methods} and Fig.~\ref{fig:epsilon_selection_simulation} for details). The path reconstruction is largely insensitive to the precise choice of $\varepsilon$ (SI Fig. 9).

We now describe the algorithm that defines $M^t$ and $N^t$ from Equation~\ref{eq:tile_defining} inductively. First, we choose the tile closest to the origin (Fig.~\ref{fig:algorithm}, base step) and define $\tilde{\Theta}(0) = \Theta(0)$ by setting $M^0 = N^0 = 0$.

For each pair of consecutive time points $t$ and $t+1$, we test if the toroidal coordinates $\Theta(t)$ and $\Theta(t+1)$ are similar via Equation~\ref{eq:similarity}. If the two coordinates are similar, then $\Theta(t+1)$ is lifted to the same tile as $\tilde{\Theta}(t)$ by setting $M^{t+1} = M^t$ and $N^{t+1} = N^t$.

If $\Theta(t)$ and $\Theta(t+1)$ fail to satisfy Equation~\ref{eq:similarity}, there are two possibilities for the underlying path $f|_{[t, t+1]}$. The first possibility is that the path $f|_{[t, t+1]}$ did not cross any of the edges of the torus (Fig.~\ref{fig:algorithm}, ``Yes" branch, green path), in which case $\tilde{\Theta}(t+1)$ should remain in the same tile as $\tilde{\Theta}(t)$. We set $M^{t+1} = M^t$ and $N^{t+1} = N^t$.

The second possibility is that $f|_{[t, t+1]}$ crossed at least one edge of the square torus (Fig.~\ref{fig:algorithm}, ``No" branch, green path segment), in which case $\tilde{\Theta}(t+1)$ should lie in a tile adjacent to that of $\tilde{\Theta}(t)$. This amounts to setting $M^{t+1} = M^t \pm 1$ and (or) $N^{t+1} = N^t \pm 1$ \footnote{If two consecutive time points $t$ and $t+1$ are far apart, possibly because the time sampling is too sparse, then the subject could have traveled a large distance between them. Then, the lifted point $\tilde{\Theta}(t+1)$ may need to be placed in a tile that is not adjacent to the tile containing $\tilde{\Theta}(t)$. Here, we assume that temporal sampling is dense enough relative to the subject's movement speed that consecutive population vectors always correspond to the same or adjacent tiles.}.

In practice, the underlying path $f$ is unobserved, so we do not know which case applies. We therefore compare both cases and choose the lift that results in a smaller distance to $\tilde{\Theta}(t) = (\tilde{\theta}^t_x, \tilde{\theta}^t_y)$ for each coordinate. That is,
\begin{equation}
\label{eq:determine_tile}
M^{t+1} = \argmin_{M \in \{ M^t, M^t \pm 1 \}} | \tilde{\theta}^t_x - \big( \theta^{t+1}_x + 2 \pi M \big)|, \quad  \quad N^{t+1} = \argmin_{N \in \{ N^t, N^t \pm 1 \}} | \tilde{\theta}^t_y - \big(\theta^{t+1}_y + 2 \pi N \big) |.
\end{equation}

We then use $M^{t+1}$ and $N^{t+1}$ to define $\tilde{\Theta}(t+1) = (\tilde{\theta}^{t+1}_x, \tilde{\theta}^{t+1}_y)$ via  Equation~\ref{eq:tile_defining}. See Fig.~\ref{fig:algorithm} for an illustration of the algorithm. We repeat this process for all time points to obtain the full lifted sequence $\{ \tilde{\Theta}(t) \}_{t = 0}^{T-1}$. We consider $\{ \tilde{\Theta}(t) \}_{t = 0}^{T-1}$ as the reconstructed movement path in $\mathbb{R}^2$ (Fig.~\ref{fig:algorithm}A). In the following sections, we show that this reconstructed path closely matches the subject's true trajectory up to an affine transformation. We emphasize that the subject's physical location is not used in the path reconstruction process.

\subsubsection*{Measuring the quality of the path reconstruction.}
We evaluate reconstruction quality at two scales. The global reconstruction error measures the fidelity of the entire reconstructed trajectory, computed as the mean Euclidean distance between the original and reconstructed paths after optimal affine alignment, normalized by the size of the environment (see \textit{Materials and Methods}). Local reconstruction error, on the other hand, captures whether the reconstruction preserves local geometry even when the global shape may be distorted. To compute local reconstruction error, the original movement path and the reconstructed path are first split into shorter path segments. For each segment, we align the corresponding movement segment and reconstructed path segment and compute the normalized, mean Euclidean distance as described in \textit{Materials and Methods}. As we will show, the distinction between these two scales is important: a reconstruction can be locally faithful while the global geometry is distorted (see Fig.~\ref{fig:Gardner-comp}), possibly because
local reconstruction can remain accurate even when small lifting errors accumulate and distort the global path (see SI Section 4.3).

\subsection{Grid cell activity accurately reflects the geometry of the environment}

\begin{figure}[h!]
    \centering
    \includegraphics[width=0.75\linewidth]{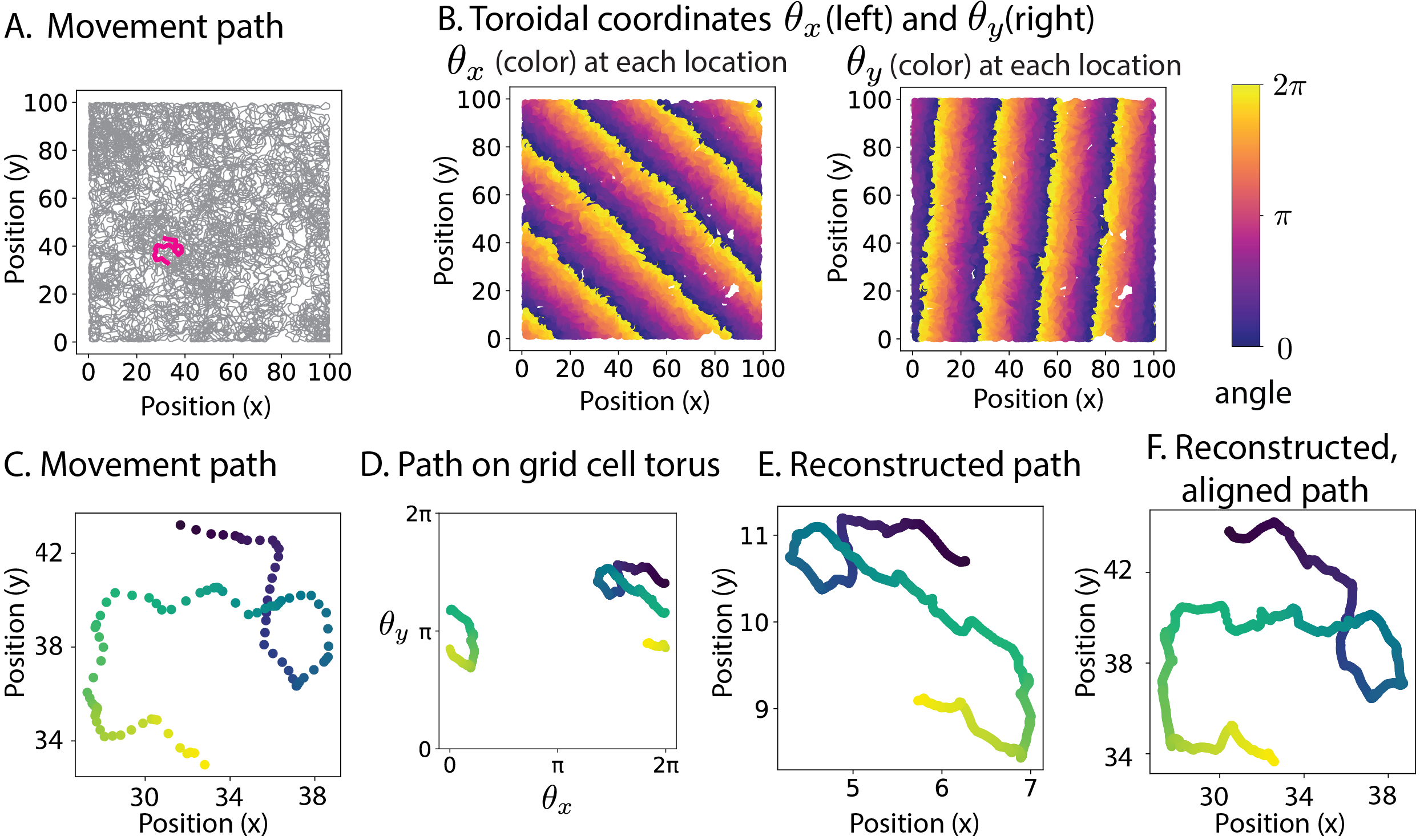}
    \caption{Illustration of path lifting on a simulated CAN data (56 $\times$ 44 grid cell network, $T = 599,999$ time bins.). \textbf{A}. A simulated movement path, with a highlighted segment. \textbf{B}. Toroidal coordinates for each location on the map. The repeated values indicate that the map is large enough to require nontrivial lifting during path reconstruction. \textbf{C}. Enlarged view of the highlighted segment. The color indicates that the simulated mouse moves from dark to light. \textbf{D}. The toroidal coordinates corresponding to the path segment in panel C. \textbf{E}. The output of the reconstruction algorithm resembles the original path in panel C. \textbf{F}. The reconstructed path, post affine transformation, recovers the original movement path in panel C.}
    \label{fig:simulation_segment}
\end{figure}

\begin{figure}[h!]
    \centering
    \includegraphics[width=1\linewidth]{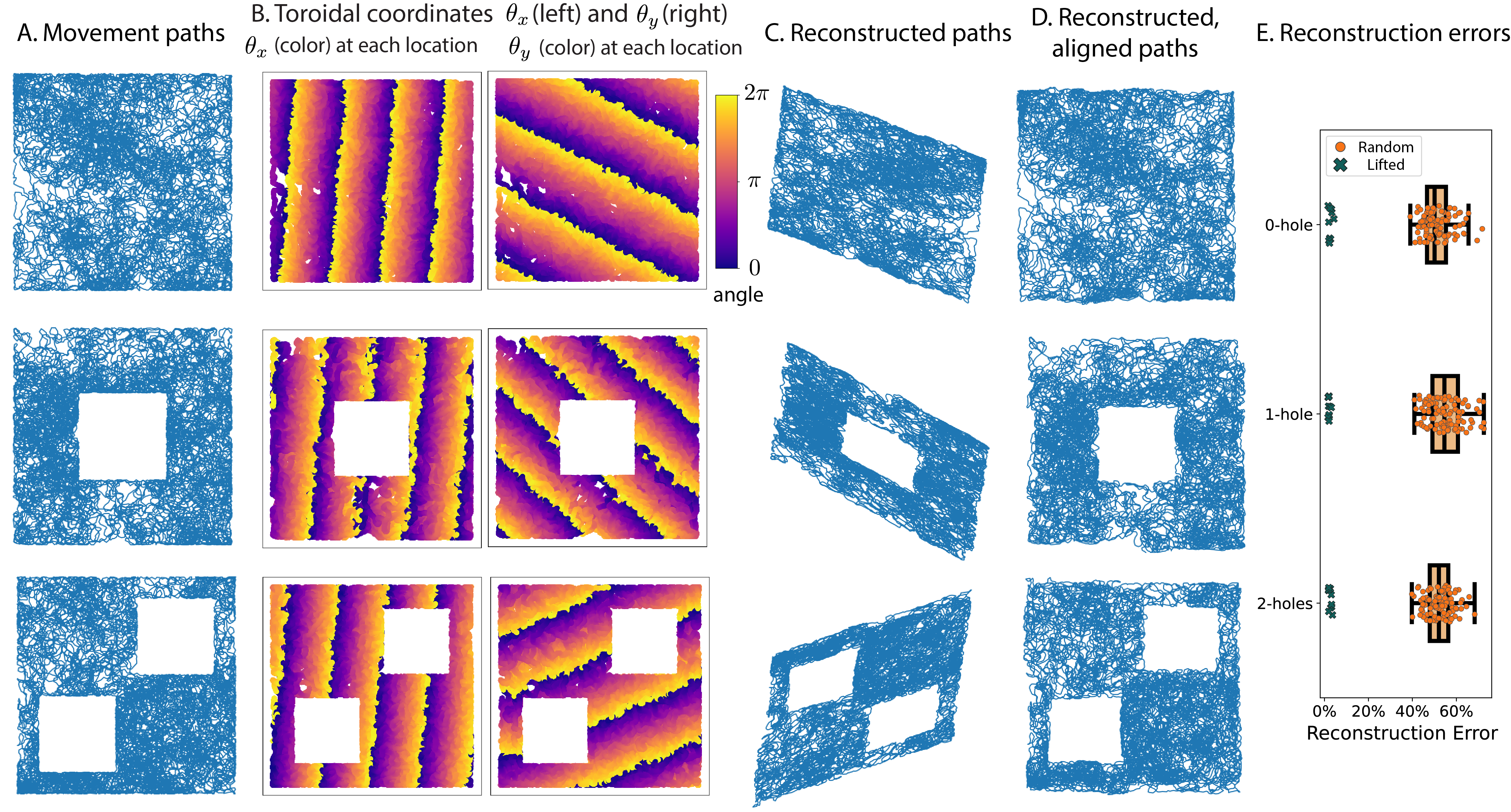}
    \caption{Path lifting on CAN-simulated grid cell activity ($G=2,464$, $T = 599,999$ time bins per simulation.) reconstructs the original movement path. \textbf{A}. Simulated movement trajectories in environments with 0, 1, and 2 holes.  \textbf{B}. Toroidal coordinates for each location on the map. \textbf{C}. Reconstructed paths from the simulation of mouse movement on maps with 0, 1, and 2 holes reflect the topology of the maps. \textbf{D}. After optimal affine alignment, the reconstructed paths resemble the original movements in panel A. \textbf{E}. Reconstruction errors across 10 independent trials. For each environment, the error between simulated movement paths and reconstructed paths (teal) are compared against random baseline (orange), computed as the error between pairs of independently simulated trajectories in the same environment. The reconstruction errors are significantly smaller than the random baselines in all three environments. }
    \label{fig:simulation_result}
\end{figure}

A fundamental question is whether the proposed method can faithfully reconstruct the true movement trajectory from grid cell activity. To test this, we first consider simulated grid cell activity where the ground-truth trajectory is known.

We simulated mouse trajectories of length $25,000$ in three environments with zero, one, and two holes (Fig.~\ref{fig:simulation_result}A, \textit{Materials and Methods}). Grid cell activity was generated using a continuous attractor network (CAN) model \cite{burak2009accurate, gardner_toroidal_2022}, implemented on a $56 \times 44$ grid cell network with a shared spatial resolution. This produced simulated activity of $G=2,464$ grid cells over approximately $T=599,999$ time bins (see \textit{Materials and Methods}). We then applied the proposed pipeline to reconstruct the movement paths (Fig.~\ref{fig:simulation_segment}). A visualization of the toroidal coordinates over the physical environment reveals that multiple locations elicit similar toroidal coordinates, indicating that path reconstruction will require nontrivial lifts (Fig.~\ref{fig:simulation_result}B).

The reconstructed trajectories preserved the global topology of the environment. Visualizations of the reconstructed paths captured the presence of holes in the environment (Fig.~\ref{fig:simulation_result}C). Indeed, the persistence diagrams computed from the reconstructed paths revealed the correct number of connected components and one-dimensional holes in the physical map (SI Fig.~20C). These results demonstrate that activity of a single grid cell module provides sufficient information to recover the topological features of the environment.

We then investigated how well the reconstructed path resembles the geometry of the original movement path. We first aligned the reconstructed path to the ground-truth trajectory via an optimal affine transformation (\emph{Materials and Methods}). In each simulation, the aligned reconstructed path closely matched the original, not only preserving the holes but also reflecting spatial patterns such as frequently visited regions (Fig.~\ref{fig:simulation_result}D). Quantitatively, the mean reconstruction error across 10 simulations (Fig.~\ref{fig:simulation_result}E, teal) was significantly lower than the errors from random reconstructions (Fig.~\ref{fig:simulation_result}E, orange) (z-scores: -7.6, -5.6, -8.4), where random reconstruction error refers to the error between two randomly simulated paths in the same environment (see \textit{Materials and Methods}). The local reconstruction errors were computed using paths of length 10,000 time bins, resulting in 59 local paths. The mean and standard deviation of local reconstruction errors for the 1-hole environment were 1.2 $\pm$ 0.3\%. See SI Fig.~20E for example local paths and their reconstructions.

We next tested robustness to noise in neural activity. To mimic spontaneous firing, we added Gaussian-shaped noise events of peak height $h$ and variance $\sigma^2$ at random time points to the simulated grid cell activity. The number of noise events added was controlled by a proportion parameter $p$ (\textit{Materials and Methods}).

\begin{figure}[h!]
    \centering
    \includegraphics[width=0.55\linewidth]{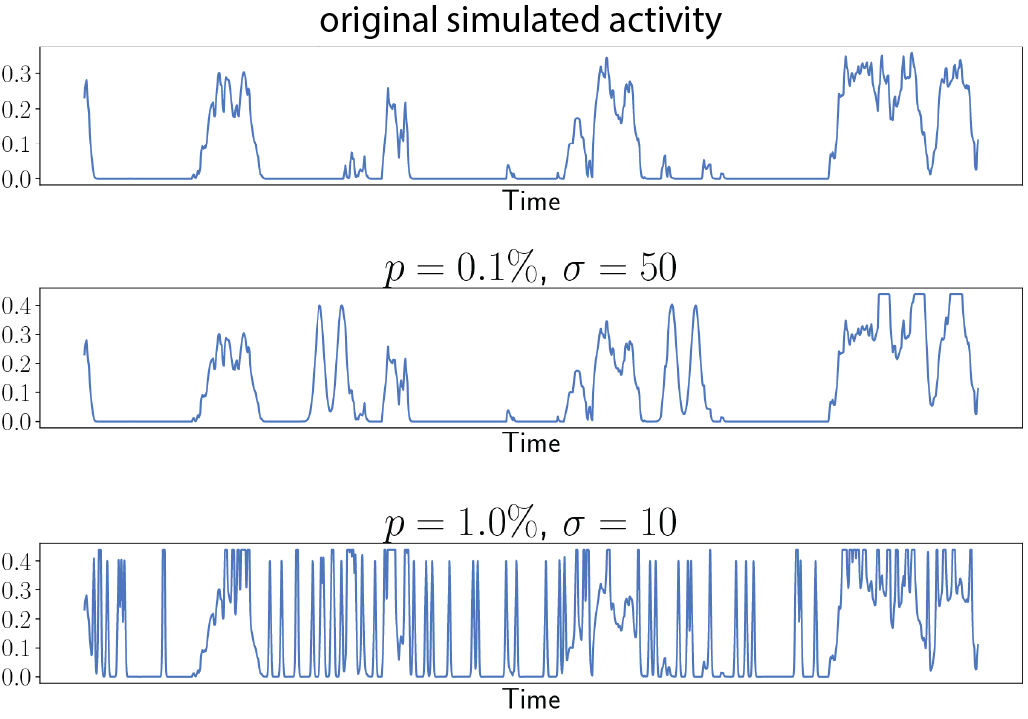}
    \caption{Example simulated grid cell activity with spontaneous firings that lead to low path reconstruction errors. (Top) Example activity trace from a CAN simulation. (Center) Simulated activity with additional spontaneous activity, generated with $h = 0.4$, $p = 0.1 \%$ and $\sigma = 50$. The mean global reconstruction error for such noisy activity is $2.115\%$ (see Table \ref{tab:noisy-table}). (Bottom) Activity trace with additional spontaneous activity, generated with $h = 0.4$, $p = 1 \%$ and $\sigma = 10$. The mean reconstruction error is $4.894\%$ (see Table \ref{tab:noisy-table}). }
    \label{fig:noisy-data}
\end{figure}

When spontaneous activity is introduced to a relatively small proportion of time points, for example, $p < 10 \%$, the reconstructions remained highly accurate even under large noise variance $\sigma^2$, with errors comparable to those from the original data (see Fig.~\ref{fig:noisy-data} and Table~\ref{tab:noisy-table}). Even when the resulting activity traces look visibly very different from the original, the reconstruction errors remained small (see Fig.~\ref{fig:noisy-data} and Table~\ref{tab:noisy-table}). However, at higher $p$, the toroidal structure of the grid cell activity degraded, leading to poor or failed reconstructions (see Table~\ref{tab:noisy-table} and SI Fig. 4). These results show that the reconstruction method is robust to moderate levels of noise but fails when spurious activity overwhelms the toroidal organization of grid cells. See SI Section 3 for a more thorough analysis of the robustness of the method against various kinds of noise in neural activity, including spontaneous firing, neural activity suppression, and temporal shifts.

\renewcommand{\arraystretch}{1.4}
\begin{table}[h!]
    \centering
    \resizebox{1\textwidth}{!}{ 
    \begin{tabular}{|p{0.4cm}|p{1.5cm}||p{3.5cm}|p{3.5cm}|p{3.5cm}|p{3.5cm}|} \hline
     \multicolumn{2}{|c||}{} & \multicolumn{4}{c|}{\textbf{Standard Deviation ($\sigma$)}} \\ \cline{3-6}
      \multicolumn{2}{|c||}{\%} &  \textbf{1} &  \textbf{10} &  \textbf{50} &  \textbf{100} \\ \hline \hline
       & \textbf{0.1 \%} & $1.647 \pm 0.0295$ & $1.723 \pm 0.0413$ & $2.115 \pm 0.1768$ & $49.480 \pm 32.2798$ \\
    \cline{2-6}
      & \textbf{0.5 \%} & $1.683 \pm 0.0327$ & $1.903 \pm 0.0571$ & $42.327 \pm 2.2691$ & $44.736 \pm 3.7464$ \\
    \cline{2-6}
             \smash{\rotatebox[origin=c]{90}{\textbf{Proportion ($p$)}}}
      & \textbf{1 \%}   & $1.695 \pm 0.0492$ & $4.894 \pm 7.3985$ & $41.785 \pm 2.0993$ & $45.528 \pm 5.9802$ \\
    \cline{2-6}
      & \textbf{5 \%}   & $1.819 \pm 0.0922$ & $44.294 \pm 4.6828$ & N/A & N/A \\
    \cline{2-6}
      & \textbf{10 \%}  & $14.744 \pm 15.2993$ & $44.176 \pm 5.4732$ & N/A & N/A \\
         \hline
    \end{tabular}
   }
   \caption{ Average reconstruction errors (\%) between the original trajectory and the reconstructed paths over 10 trials. Here, the maximum height of the spontaneous firing is fixed at $h=0.4$. The rows represent the proportion of times during which a grid cell randomly fired, and the columns represent the variance $\sigma$ of the noise added. An entry of N/A indicates that the method failed to compute toroidal coordinates in all 10 trials. }
    \label{tab:noisy-table}
\end{table}

\subsection{Path reconstructions recovers one-dimensional environment from grid cell activity}
\begin{figure}[h!]
    \centering
    \includegraphics[width=\linewidth]{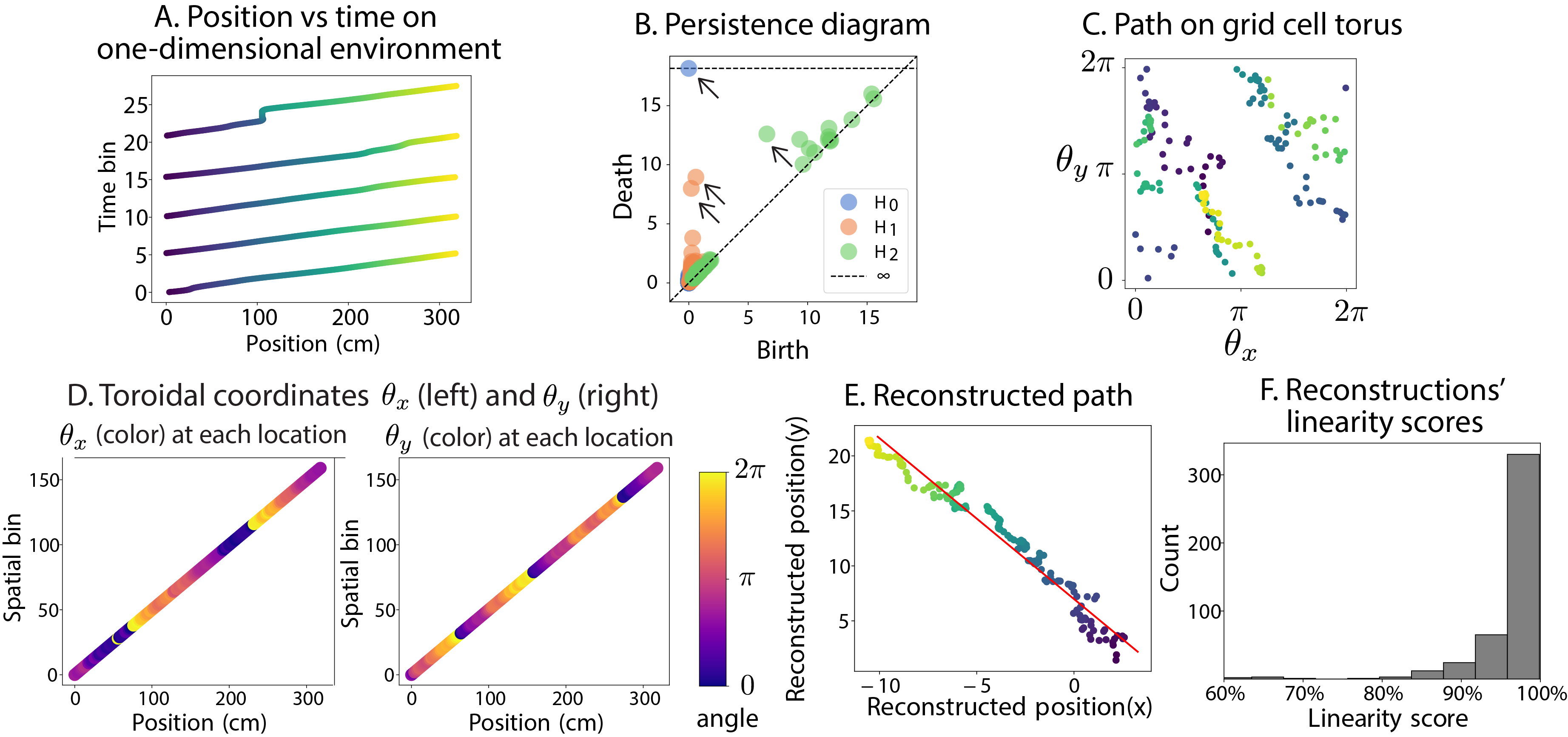}
    \caption{ Path reconstruction recovers one-dimensional environment from grid cell activity. Data are from mouse $N2$ (dataset ``N2 200203 buildup track"; \cite{wen2024one}) navigating a $320$ cm virtual build-up track. 44 co-modular grid cells were identified. Firing rates were provided in $2cm$ spatial bins (160 bins per run, 441 total runs). \textbf{A}. Mouse position over time across 5 runs; each rising segment corresponds to one traversal of the track, after which the mouse is teleported to the start. A single run is highlighted in pink. \textbf{B}. The persistence diagram confirms that grid cells are organized on a torus: one connected component ($H_0$), two one-dimensional cycles ($H_1$), and one two-dimensional void ($H_2$). \textbf{C}. Example path on the grid cell torus corresponding to a single run. For each time point $t$, the corresponding toroidal coordinates $\theta_x$ and $\theta_y$ are plotted. \textbf{D}. Toroidal coordinates from panel C visualized over position. Because the firing rate data is provided in $2cm$ spatial bins, the toroidal coordinates are also computed for each spatial bin. Each point on the plot corresponds to one spatial bin in a fixed run, plotted at its track position ($x$-axis) and spatial bin index ($y$-axis). Color encodes the toroidal coordinates $\theta_x$ (left) and $\theta_y$(right).  \textbf{E}. The reconstructed path lies close to a one-dimensional line. The red line indicates the line spanned by the first principal component (PC1) of PCA.  \textbf{F}. Distribution of linearity scores (variance explained by PC1) across 441 runs; median = $98.8\%$.}
    \label{fig:Wen-comp}
\end{figure}

We applied the method to experimental data, asking whether one-dimensional trajectories can be reconstructed from real grid cell recordings. We analyzed a publicly available dataset of grid cell activity in mice navigating a 320 cm virtual build-up track \cite{wen2024one}. Here, whenever the mouse reached the end of a track, the mouse was teleported to the start without visual discontinuity (Fig.~\ref{fig:Wen-comp}A). We analyzed 441 such runs on the linear track. The co-modular grid cells were identified via clustering on spectrograms (see \cite{wen2024one} for details). The dataset provides firing rates that were preprocessed by the original authors as follows: spikes were binned into 2-cm spatial bins along the track, divided by the time occupancy per bin, and smoothed with a 2-bin Gaussian kernel. Each continuous run on the 320 cm track thus produced grid cell firing rates over 160 spatial bins.

From 44 identified grid cells, the persistence diagrams confirmed a toroidal organization of population activity (Fig.~\ref{fig:Wen-comp}B). While toroidal coordinates were computed from the 44 grid cells over the full duration of the experiment, path reconstructions were performed separately on each of the 441 continuous runs of the track. For each run, the toroidal coordinates $(\theta_x, \theta_y)$ cross the edges of the grid cell torus (Fig.~\ref{fig:Wen-comp}C,D), indicating that the path reconstruction will involve nontrivial liftings. The reconstructed paths recovered the one-dimensional structure of the environment (see Fig.~\ref{fig:Wen-comp}E). The one-dimensional nature of a reconstructed path was quantified by the variance explained by the first principal component, which we call ``linearity score." Across all 441 runs, the median linearity score was 98.8\% (Fig.~\ref{fig:Wen-comp}F). For comparison, the median linearity score of 59 local paths from a simulated trajectory in the 0-hole world (see SI Fig. 20E for examples) was 79.2\%. This analysis demonstrates that one-dimensional spatial structure can be reliably decoded from experimental grid cell activity.

\subsection{Local paths can be reconstructed from grid cell activity in a two-dimensional environment}

\begin{figure}[h!]
    \centering
    \includegraphics[width=\linewidth]{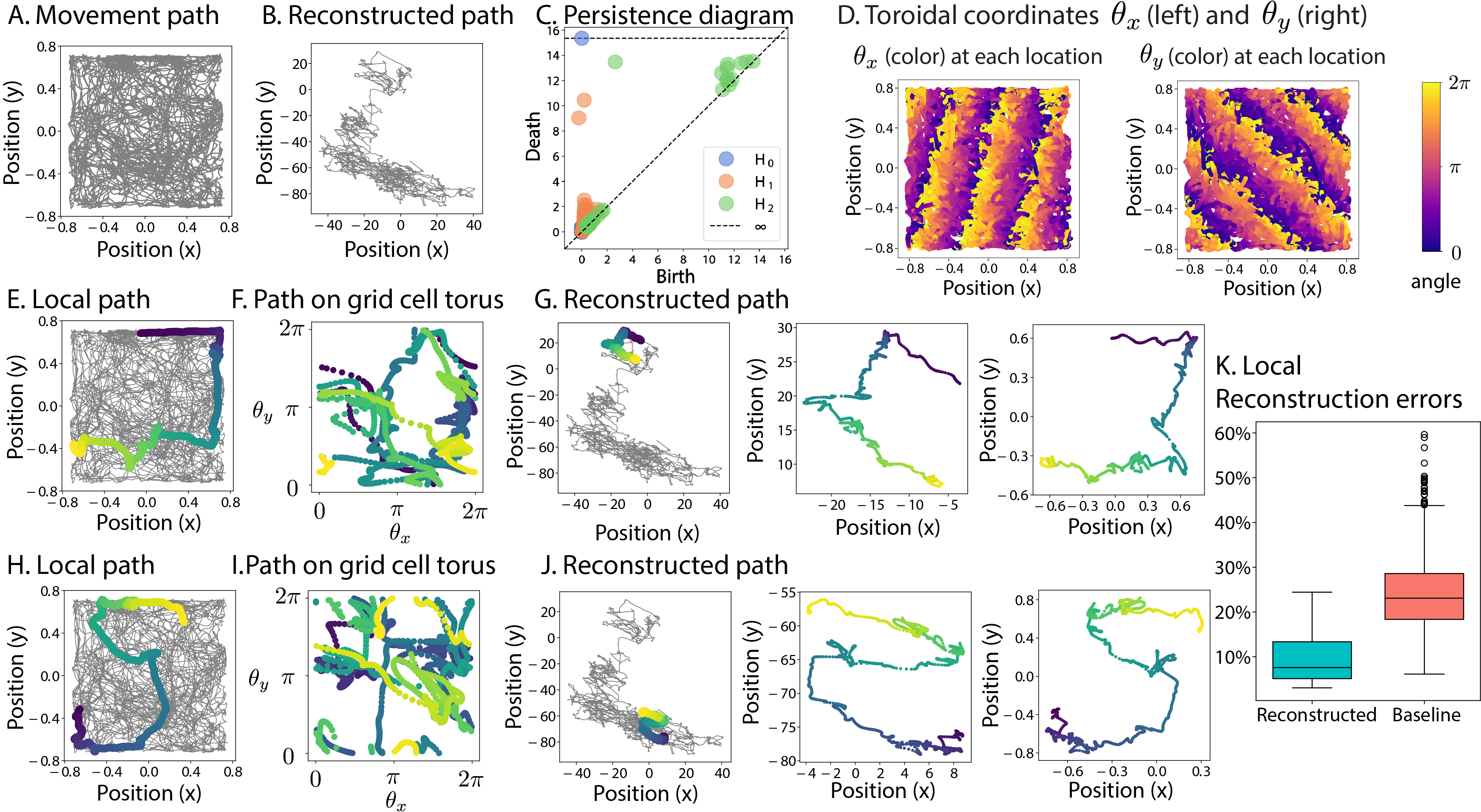}    \caption{Reconstruction of local paths from two-dimensional experimental data \cite{gardner_toroidal_2022} (rat $R$, module $1$, day $2$, open-field session; 111 co-modular grid cells ). \textbf{A}. The original trajectory of a rat exploring a $1.5 m \times 1.5 m $ open-field arena. \textbf{B}.  The reconstructed global trajectory, which differs in overall shape from the original path. \textbf{C}. The persistence diagram indicates that the grid cells are organized on a torus. \textbf{D}. A visualization of the toroidal coordinates for each location. \textbf{E}. An example local path. \textbf{F}. The toroidal coordinates corresponding to panel E involve non-trivial liftings. \textbf{G}. A highlight of the reconstructed segment in panel B (left), the reconstructed path, before affine transformation (center), and after affine transformation (right). \textbf{H - J}. Another example local path and its reconstruction. \textbf{K}. Distribution of local reconstruction errors: pairs of original local paths and reconstructed paths (left) show significantly smaller errors than baseline consisting of mismatched local paths (right) ($t(2014) = -14.6, p < 0.0001$). }
    \label{fig:Gardner-comp}
\end{figure}

Finally, we tested the method on two-dimensional experimental data that is publicly shared in \cite{gardner_toroidal_2022}. Extracellular spikes from grid cells in layers II and III of MEC-parasubiculum were recorded while rats explored a square open-field arena of size $1.5 m \times 1.5 m$ (Fig.~\ref{fig:Gardner-comp}A). Six grid modules were identified by clustering. Here, we report the analysis on one of the datasets (rat $R$, module $1$, day $2$, OF), which consisted of 111 co-modular grid cells recorded over $21.1$ minutes ($T = 126,728$ time bins, 1 time bin = 10ms) of open-field foraging.

Globally, the reconstructed path differed in overall shape from the true trajectory (Fig.~\ref{fig:Gardner-comp}B). However, when the analysis was restricted to shorter local paths corresponding to 20-second intervals, the reconstructed paths were highly consistent with the original movement paths (Fig.~\ref{fig:Gardner-comp}E-J). For example, the reconstructed paths in Fig.~\ref{fig:Gardner-comp}G and J resemble the geometry of the original local paths in Fig.~\ref{fig:Gardner-comp}E and H, despite requiring many non-trivial lifts across torus edges (Fig.~\ref{fig:Gardner-comp}F, I).

To quantify the quality of local path reconstructions, we compared the local reconstruction errors between the original and lifted path segments against a baseline distribution of mismatched segment pairs. Local reconstructions had significantly lower errors (mean 9.5\%) than the null baseline (mean 23.7\%, s.d. 7.6\%, Fig.~\ref{fig:Gardner-comp}K), confirmed by an independent t-test ($t(2014) = -14.6, p < 0.0001$). These results establish that while global reconstructions may deviate from the true path, local trajectories can be faithfully recovered.

\begin{figure}[h!]
    \centering
    \includegraphics[width=\linewidth]{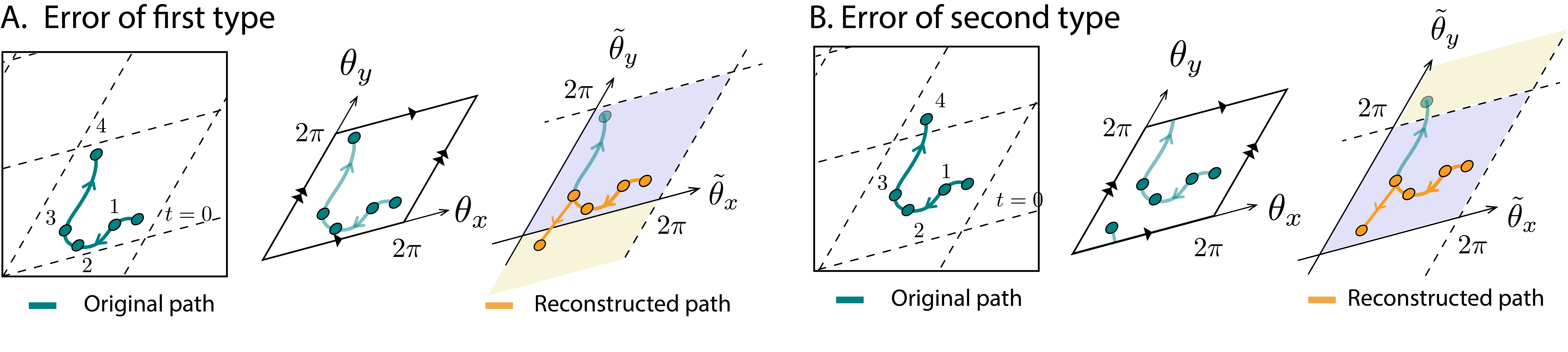}
    \caption{Two possible errors in path reconstruction arising from sparsity of time points. \textbf{A.}
The first type of error occurs when two consecutive toroidal coordinates are lifted to two distinct tiles when they should be lifted to a single tile. (Left) Original movement path. Circles indicate the location at select time points. (Center) The corresponding toroidal coordinates. (Right) The algorithm lifts the toroidal coordinates $\Theta(0), \dots, \Theta(3)$ to the blue tile. Because $\theta^3_y$ and $\theta^4_y$ are dissimilar, $\tilde{\Theta}(4)$ is in a different tile, shown in yellow. The resulting reconstructed path (orange) deviates from the original path (green).
\textbf{B.} The second type of error occurs when two consecutive toroidal coordinates are lifted to the same tile when they should be lifted to different tiles. Here, the toroidal coordinates $\theta^3_y$ and $\theta^4_y$ have a small enough difference so the algorithm lifts $\Theta(3)$ and $\Theta(4)$ to the same tile. Again, the reconstructed path (orange) deviates from the original (green).  }
    \label{fig:lifting_issues}
\end{figure}

Several factors can contribute to the discrepancy between local and global path reconstructions. First, noise in the toroidal coordinates can distort the lifted path  (compare Fig.~\ref{fig:Gardner-comp}D to Fig.~\ref{fig:simulation_segment}B). Second, when the sampled time points are too sparse, two types of lifting errors can occur: toroidal coordinates $\Theta(t)$ and $\Theta(t+1)$ that should be lifted to the same tile may be lifted to different tiles (see Fig.~\ref{fig:lifting_issues}A), or coordinates $\Theta(t)$ and $\Theta(t+1)$ that should be lifted to distinct tiles may end up lifting to the same tile (see Fig.~\ref{fig:lifting_issues}B). Accumulated errors of this type can cause large-scale distortions in the global path reconstruction (see SI Section 4.3).

\section{Discussion}
In this study, we introduced and validated a topological framework for decoding spatial trajectories from grid cell population activity. By identifying toroidal coordinates through persistent cohomology and lifting paths from the torus to the plane, we effectively reconstruct movement trajectories up to an affine transformation. This approach decodes spatial trajectories without access to external positional cues or knowledge of grid phases, offering a new computational perspective on spatial representation of the grid cell system. Our results show that the toroidal organization of grid cell activity is not merely a descriptive feature but can be functionally leveraged to recover spatial information.

The proposed method builds on the work of Gardner et al \cite{gardner_toroidal_2022}, who used persistent cohomology to reveal the toroidal structure of grid cell activity and assigned toroidal coordinates to population vectors using the circular coordinates construction of de Silva et al \cite{de_silva_persistent_2011}. While prior work characterizes the structure of the neural manifold, present work shows that these toroidal coordinates can be used to decode spatial trajectories from it. The key methodological contribution is path lifting -- unwrapping paths on the grid cell torus into paths in the plane via covering maps.

The proposed method complements previous decoding approaches in that it only requires the activity of a single grid module (as opposed to multiple grid modules) \cite{mathis2012optimal, stemmler2015connecting} and that it doesn't involve any training process. Furthermore, the method doesn't require any phase information of the grid modules, enhancing the applicability of the method.

Recent work has shown that individual grid cell spike trains are often too irregular to convey the lattice structure of the firing fields, and that extracting lattice structure from spike trains requires parameter regimes in which the animal's trajectory passes through neighboring grid fields in sequence without omissions \cite{dabaghian2023grid}. In contrast, our method operates on population vectors -- the simultaneous activity of co-modular grid cells at each time step. The collection of these population vectors forms a torus, which is then used for path reconstruction. Because we analyze the population-level activity, we do not require any individual cell to exhibit regular firing activity.

In order for the method to reconstruct paths accurately, several assumptions must be met. The presented method requires that the grid cells are recorded for a sufficiently long time so that the toroidal structure of the population vectors becomes clear. Furthermore, accurate path reconstruction depends on sufficiently dense temporal sampling. As discussed in Fig.~\ref{fig:lifting_issues}, having insufficient time points can lead to poor lifting. On the other hand, including too many time points can lead to slow computation, especially when computing the toroidal coordinates. In particular, the analysis of two-dimensional experimental data involves a preprocessing that selects the relevant time points. Such preprocessing must be done in a manner that preserves the toroidal structure of grid cells.

The affine transformation used to align the reconstructed path to the ground-truth trajectory serves as an evaluation tool to quantify reconstruction accuracy; it is not part of the decoding algorithm itself. It is unclear whether the brain needs to resolve this affine ambiguity to perform path integration. The path reconstruction in this work can be interpreted as performing path integration over basis vectors of $\mathbb{R}^2$ that are not necessarily orthogonal. If the brain's internal coordinate system similarly operates in a non-orthogonal basis, then no affine correction would be required to track relative position.

If, however, alignment with the geometry of the physical environment is needed, the affine ambiguity could in principle be resolved in at least two ways. First, combining information from multiple grid modules with different spacings and orientations is sufficient to recover absolute position, because the ambiguity inherent in any single module's periodic code is eliminated when multiple modules are combined \cite{stemmler2015connecting}. Second, other spatially tuned cell populations, including place cells and boundary cells, as well as visual landmarks, could provide an external anchor for mapping from the brain's toroidal coordinate system to the physical environment. These possibilities are consistent with the broader view that spatial navigation relies on the coordinated activity of multiple cell types.

There are several directions for future research. Incorporating interpolation and probabilistic inference during path lifting could improve robustness to sparse or noisy data. Combining grid modules of different phases may provide a more comprehensive and precise decoding framework. Future research should also examine how the toroidal organization of grid cells integrates with other spatially tuned cell populations such as place cells to support path integration and spatial navigation. Because place cells provide a non-periodic spatial code, they could help disambiguate the choice of lift during path reconstruction from grid cell activity. For example, a place cell that fires at two similar time points would indicate that the lifted coordinates at the corresponding times should occupy a similar region in $\mathbb{R}^2$, providing a constraint that guides the path reconstruction algorithm. Topological methods have been used to reconstruct stimulus spaces from place cell activity\cite{curto2008cell}, and integrating such information with the present framework is a natural direction for future work.

Beyond neuroscience, the method offers new perspectives on artificial navigation systems. Recent work has shown that grid-like representations emerge in deep networks trained to perform path integration, enabling vector-based navigation in artificial agents \cite{banino2018vector}. Grid cell-inspired path integration and state estimation have also been explored for mobile robots \cite{simkuns2025deep, zou2021neurobiologically}. Our framework suggests a complementary approach: if an artificial agent's internal representation exhibits toroidal structure, whether learned or engineered, path lifting could recover spatial trajectories and aid in position estimation without external cues.

\section{Materials and Methods}
\subsection{Simulation of grid cells}
\subsubsection{Simulated Mice Trajectory}\
\label{sec:simulated_mice_trajectory}

A two-dimensional random walk simulation was developed to model the exploratory behavior of mice within a bounded environment ($100\times 100$). The environment was defined by spatial boundaries and obstacle parameters, including the sizes and positions of holes. A mouse agent was initially placed at a randomly selected valid location within these boundaries. At each time step, candidate positions were computed within an angular window of  $\pm75^\circ$ relative to the current heading. A step size is randomly drawn up to a predefined maximum, and candidate positions are generated by adding a scaled directional vector to the current position. Only positions that remained within the environment and avoided designated holes were accepted. If no valid candidate position was available, the agent remained stationary for that time step and subsequently adopted a new random heading from the full 360° range. Repeated iterations of this process produced a random trajectory that models the exploratory behavior of mice. All simulated trajectories had length $25,000$.

\subsubsection{Grid cell simulation with CAN model}\

The simulated mice trajectories were then used to simulate grid cell activity using a noiseless CAN model with purely inhibitory recurrent connectivity, following \cite{burak2009accurate, couey2013recurrent}. We use the same model and parameters that were used in \cite{gardner_toroidal_2022}. There, the CAN model was used to simulate grid cell activity in response to position data. Here, we use it to simulate grid cell activity in response to simulated mice trajectory.

The network consists of a $56 \times 44$ neuronal sheet with periodic (toroidal) boundary conditions. The simulated random walk trajectory was then provided as input to the model, which computed the speed $v(t)$ and head direction $\theta(t)$ at each time step. The activity of neuron $i$ at time $t$ is determined by

\[s_i (t+1) = s_i (t) + \frac{dt}{\tau}\left(-s_i(t) + \left[I + \sum_j W_{ij}\, s_j (t) + \alpha\, v(t) \cos\!\bigl(\theta(t) - \tilde{\theta}_i\bigr)\right]_+\right),\]
where $[\, x \, ]_+ = \max( x, 0)$ is the threshold-linear function, and $\tilde{\theta}_i$ is the preferred direction of neuron $i$. The parameters followed the implementation in \cite{gardner_toroidal_2022}: $I=1$ (constant external input), $\alpha=0.15$ (velocity modulation), $dt = 1$ (integration time step)
, and $\tau = 10$ (neuronal time constant).
$W_{ij}$ denotes the strength of connection from neuron $j$ to neuron $i$, and it was computed as described in \cite{couey2013recurrent} with $W_0=-0.02$ and connectivity radius $R = 15$.

The activity patterns were initialized by setting $90\%$ of neurons to $s_i = 1$ (the rest were set to 0) and performing $2,000$ updates to allow the hexagonal bump pattern to stabilize. For computational efficiency, the activity was set to 0 whenever $s_i < 0.0001$.

For each simulation, the result was simulated activity of $G = 56 \times 44 = 2,464$ grid cells over $T = 599,999$ time bins. The activity values ranged from 0 to approximately 0.45 per time bin, with a mean peak activity of approximately 0.44 per time bin. These amplitudes and time bins are inherent to the model dynamics and should not be interpreted as firing rates in physiological units such as Hz. The grid fields had an average diameter of 14.8 units in an environment of size $100$ by $100$. See SI Fig. 20D for example firing fields.

We refer the reader to \cite{couey2013recurrent} for the original model and \cite{gardner_toroidal_2022} for details of the implementation we adapted.

\subsubsection{Simulation of grid cells with additional spontaneous activity.}\
To emulate the random firing of grid cells, we incorporated one-dimensional Gaussian noise into the simulated grid cell activity. Given a simulated activity $r(t)$ with $t \in [0, T]$, we modify $r(t)$ by adding one-dimensional Gaussian functions $g_{h, \sigma}(t)$ centered at some random value in $[0,T]$, with peak height $h$ and variance $\sigma^2$. Letting $r_{\max} = \max_t r(t)$, the maximum value in the simulated data, we construct a noisy activity trace by
\[ r^*(t) = \min \Bigl\{ r(t) + \sum_i g_{h, \sigma}(t), \, r_{\max} \Bigr\}.\]
We clip the result so that $r^*(t)$ doesn't exceed the maximum value $r_{\max}$ from the original simulation. Here, the number of Gaussian functions added can vary, and the number is determined as some proportion $p$ of $T$.

For computational efficiency, a fast approximation routine precomputes a truncated Gaussian curve by identifying the index at which the noise amplitude falls below a specified threshold of 1e-4, thereby limiting the range over which noise is applied.

\subsection{Topological features of grid cell activity}

\subsubsection{Persistent cohomology}
Persistent (co)homology is a tool in topological data analysis that can be used to identify structural features in neural manifolds. Here, it is used firstly to verify that the grid cells' population activity is organized in a torus and secondly to compute the toroidal coordinates of each population vector.

Given an activity matrix $A$ with $G$ grid cells and $T$ time bins as input to persistent cohomology computation, the general persistent cohomology computation on the population activity would first create a symmetric $T \times T$ pairwise dissimilarity matrix. However, due to the large size of $T$, persistent cohomology is usually computed on a smaller number of time points $T^*$ ($T^* = 250$ in simulated data;  $T^* = 1,200$ in both experimental data, where the selection of $T^*$ time points is described in the following section). Let $\{\delta_i\}_{i=1}^m$ be a collection of nonnegative real numbers satisfying $0\leq \delta_{0} <\delta_1 < \dots < \delta_m$.  Given a parameter $\delta_i$, we construct a Vietoris-Rips complex $VR_{\delta_i}$ that consists of $T^*$ vertices and has an $n$-simplex $[v_0, \dots, v_{n}]$ precisely when all pairwise dissimilarity among the listed elements is at most $\delta_i$. We then obtain a filtration of Vietoris-Rips complexes
\[VR_{\delta_0} \subseteq VR_{\delta_1} \subseteq \cdots \subseteq VR_{\delta_m}.\]
Computing (co)homology in dimensions 0, 1, and 2 with a field coefficient, we obtain a sequence of vector spaces summarizing the connected components, circular features, and voids in each Vietoris-Rips complex. The birth and death of these topological features are summarized in a persistence diagram. A topological feature born at parameter $b$ that dies at parameter $d$ is represented by a point in the plane with coordinates $(b, d)$ (see Fig.~\ref{fig:pipeline}B). Ripser \cite{bauer2021ripser, tralie2018ripser} was used for persistent cohomology computations. For a general introduction to persistent cohomology, we refer the reader to SI Section 1.2 and \cite{ghrist2008barcodes, edelsbrunner2008persistent, carlsson2009topologydata, edelsbrunner2002topological}.

For the simulated data, the dissimilarity between two population vectors were computed using Euclidean distance. Persistent cohomology was computed using DREiMac \cite{DREIMAC_Perea2023}, which selects $T^* = 250$ landmarks for its computation. For the experimental data, where more noise is present, we compute persistent cohomology on $T^* = 1,200$ time bins with cosine dissimilarity. The selection of $T^*$ time bins is described below.

\subsubsection{Preprocessing and persistent cohomology computation of experimental grid cell activity}
\label{sec:preprocessing}

When analyzing the one-dimensional and two-dimensional experimental data, we followed the preprocessing pipeline and publicly available code of \cite{gardner_toroidal_2022} to construct the grid activity matrix, as described below.

The first step is to obtain a firing rate estimate for each grid cell. For the one-dimensional experimental data \cite{wen2024one}, the dataset provides firing rates that were spatially binned (2-cm bins) and smoothed by the original authors (see \textit{Data Availability}). For the two-dimensional experimental data, the spike trains were converted to firing rate estimates by convolving with a Gaussian kernel and then sampling, following the preprocessing steps described in \cite{gardner_toroidal_2022}.

Given $G$ co-modular grid cells observed over $T$ bins (spatial bins for the one-dimensional experimental data; time bins for the two-dimensional experimental data), let $A$ be a $G \times T$ activity matrix, where row $i$ of $A$ is the activity of neuron $i$. Computing persistent cohomology on large point clouds is computationally expensive and sensitive to outliers, so the population vectors were downsampled and dimension-reduced prior to the computation, following the steps described in \cite{gardner_toroidal_2022}. First, 15,000 population vectors with the highest total activity were selected. These were z-scored and projected onto their first six principal components via PCA. The resulting six-dimensional point cloud was further subsampled to $T^* = 1,200$ points using a neighborhood-based selection procedure adapted from UMAP, which iteratively selects points with the strongest local neighborhoods. A $1,200 \times 1,200$ symmetric dissimilarity matrix was then computed from these points in the fuzzy topological representation described in \cite{gardner_toroidal_2022}. The dissimilarity was computed using cosine dissimilarity. The performance of the path lifting algorithm was largely insensitive to the choice of metric at this stage (see SI Section 4.5). This matrix served as input to the persistent cohomology computation. We refer the reader to the \textit{Methods} section and \textit{Supplementary Methods} of \cite{gardner_toroidal_2022} for details.

\subsubsection{Circular and toroidal coordinates}\

Given a point cloud $P$ that is arranged in a circular fashion, \emph{circular coordinates} parametrize $P$ using a circle-valued map $f: P \to S^1$, where $S^1$ denotes a circle. In practice, the range of the circular coordinates will be the angles $[0, 2\pi)$, considered as a circle after identifying $0$ and $2\pi$. Originally presented in \cite{de_silva_persistent_2011}, the construction is motivated by the fact that there is a bijection between the equivalence classes of continuous maps from a CW complex $K$ to the circle $S^1$ and the cohomology of $K$ with integer coefficients
\[ [K, S^1] = H^1(K; \mathbb{Z}). \]

Given a point cloud $P$, persistent cohomology is used to fix a Vietoris-Rips complex $VR_{\delta}$ such that there is a nontrivial circular structure. Then a generator $[\eta] \in H^1(VR_{\delta}; \mathbb{Z})$ is used to assign circular coordinates to each point of $P$. See \cite{de_silva_persistent_2011} for details.

Since the torus $S^1 \times S^1$ is a product of two circles, any point on the torus can be parametrized using two circular coordinates as $(\theta_x, \theta_y)$, where $\theta_x, \theta_y \in[0, 2 \pi )$.

On the simulated data, we compute the toroidal coordinates using DREiMac \cite{DREIMAC_Perea2023}, which implements the toroidal coordinates algorithm of \cite{scoccola_toroidal_2023}. This algorithm extends the circular coordinates construction by decorrelating the two circle-valued maps, ensuring that the resulting toroidal coordinates are geometrically independent.

On the experimental data, we assign toroidal coordinates to each population vector using the cohomological decoding framework described in \cite{gardner_toroidal_2022}, which adapts the original circular coordinates algorithm \cite{de_silva_persistent_2011}. Persistent cohomology and toroidal coordinates were computed for the $T^* = 1,200$ selected time points and interpolated to the full set of population vectors. See \cite{gardner_toroidal_2022} for details.

We use different methods for computing toroidal coordinates on the simulated and experimental data for the following reason. DREiMac requires fewer preprocessing of neural activity matrix than the decoding framework of \cite{gardner_toroidal_2022}, making it well-suited for simulated data, where the data involved has less noise. Moreover, it allows us to isolate the effect of individual factors — such as neuron count, temporal resolution, or noise level — on reconstruction quality, without confounding these factors with choices made during the preprocessing of experimental data (see SI Section 4). For the experimental data, however, applying DREiMac directly to the raw population vectors often failed to produce toroidal coordinates (resulting in N/A), due to the higher noise levels and variability inherent in the recordings. We therefore adopted the decoding framework of \cite{gardner_toroidal_2022} for the experimental analysis. We note that the two methods yield comparable reconstruction errors when applied to simulated data (SI Fig. 17).

\subsection{Path lifting}

\subsubsection{Aligning paths via affine transformation}\

An affine transformation is a linear mapping that preserves points, straight lines, and planes under rotation, translation, and scaling.
Affine transformations of $2$-dimensional vectors is defined by a $2 \times 2$ matrix $B$ and a vector $\vec{b} \in \mathbb{R}^2$. The image of a vector $\vec{x} \in \mathbb{R}^2$ under this transformation is $B \vec{x} + \vec{b}$.

Given two ordered collections $P = \{ \vec{p}_1, \dots, \vec{p}_k \}$ and $Q = \{ \vec{q}_1, \dots, \vec{q}_k \}$ of $2$-dimensional vectors, the goal of optimal affine transformation is to find the matrix $B$ and $\vec{b}$ that minimizes the total distance between the transformed source points and their target counterparts. In this work, $P$ refers to the (discrete) reconstructed path and $Q$ refers to the original movement path. We computed the optimal affine transformation using the \texttt{estimateAffine2D} function from the OpenCV package \cite{opencv_library}, which uses RANSAC (Random Sample Consensus)\cite{fischler1981random} to iteratively fit candidate transformations to random subsets of three point correspondences and select the transformation that best agrees with the majority of the data.
The final transformation is refined using the Levenberg-Marquardt optimization algorithm. We refer to the result of applying the optimal affine transformation to the reconstructed path as the ``reconstructed, aligned path."

\subsubsection{Reconstruction error}
\label{sec:reconstruction_error}

Let $\Psi: \{0,1, \dots T-1\} \to \mathbb{R}^2$ be the discrete movement path in physical space, where $T$ is the number of bins of the grid cell activity (time bins for the simulated and two-dimensional experimental data; spatial bins for the one-dimensional experimental data). Given a reconstructed path $\tilde{\Theta}(t): \{0,1, \dots T-1\} \to \mathbb{R}^2 $, we quantify the dissimilarity between $\Psi$ and $\tilde{\Theta}$ as follows.
We first compute the optimal affine transformation aligning $\tilde{\Theta}$ to $\Psi$ as described above, and let $\tilde{\Theta}^*$ denote the aligned path. The reconstruction error is defined as the mean Euclidean distance between points of $\Psi$ and $\tilde{\Theta}^*$, normalized by the environment size:
\[ \frac{1}{S} \cdot \frac{1}{T}\sum_{t=0}^{T-1}\|\Psi(t) -\tilde{\Theta}^*(t)\| . \]

Here, $S$ is a normalization factor that allows the error to be interpreted as a fraction of the environment size. When computing the global reconstruction error, we use $S = 100$ (the length and width of the map) for the simulation study and $S=1.5$ (the length and width of the physical environment in m) for the two-dimensional experimental study. When computing local reconstruction errors of shorter path segments, we use $S$ to be the larger of the length and width of the path segment. A lower reconstruction error indicates closer agreement between the original and the reconstructed path.

Throughout this paper, we compare the global and local reconstruction errors against baseline errors computed between pairs of independent movement trajectories in the same environment (Fig.~\ref{fig:simulation_result}E, Fig.~\ref{fig:Gardner-comp}K). Given two discrete paths $\Psi: \{0,1, \dots T-1\} \to \mathbb{R}^2$ and $\Omega: \{0,1, \dots T-1\} \to \mathbb{R}^2$ of equal length, the error between two paths is computed analogously:
\[ \frac{1}{S} \cdot \frac{1}{T}\sum_{t=0}^{T-1}\|\Psi(t) - \Omega^*(t)\|,\]
where $\Omega^*$ is the transformed path of $\Omega$ after aligning it to $\Psi$. When computing the baseline global reconstruction errors from independent simulations (Fig.~\ref{fig:simulation_result}E), we use $S = 100$. When computing the baseline local errors between two random local paths (Fig.~\ref{fig:Gardner-comp}K), we set $S$ to be the mean size of two paths, where the size of a path is the maximum of the length and width.

\subsubsection{Proximity parameter selection}
The parameter $\varepsilon$ determines which consecutive toroidal coordinates are tested for nontrivial lifts. Specifically, if two toroidal coordinates $\Theta(t)$ and $\Theta(t+1)$ satisfy  Equation~\ref{eq:similarity}, then, they are lifted to the same tile; otherwise, the algorithm evaluates whether the toroidal coordinates should be lifted to adjacent tiles using Equation~\ref{eq:determine_tile}.

A large $\varepsilon$ causes more consecutive coordinates to be classified as similar, reducing the number of time points tested for nontrivial lifts. A small $\varepsilon$ increases the number of time points tested. However, because the parameter only identifies candidates for nontrivial lifts -- whether a lift actually occurs depends on the distance comparison in Equation~\ref{eq:determine_tile} -- a smaller $\varepsilon$ generally improves reconstruction accuracy without significant drawbacks. An analysis of the impact of $\varepsilon$ on the reconstruction error shows that the performance of the path reconstruction is stable across a wide range of $\varepsilon$ parameters (SI Fig. 9).

\begin{figure}[h!]
    \centering
    \includegraphics[width=1\linewidth]{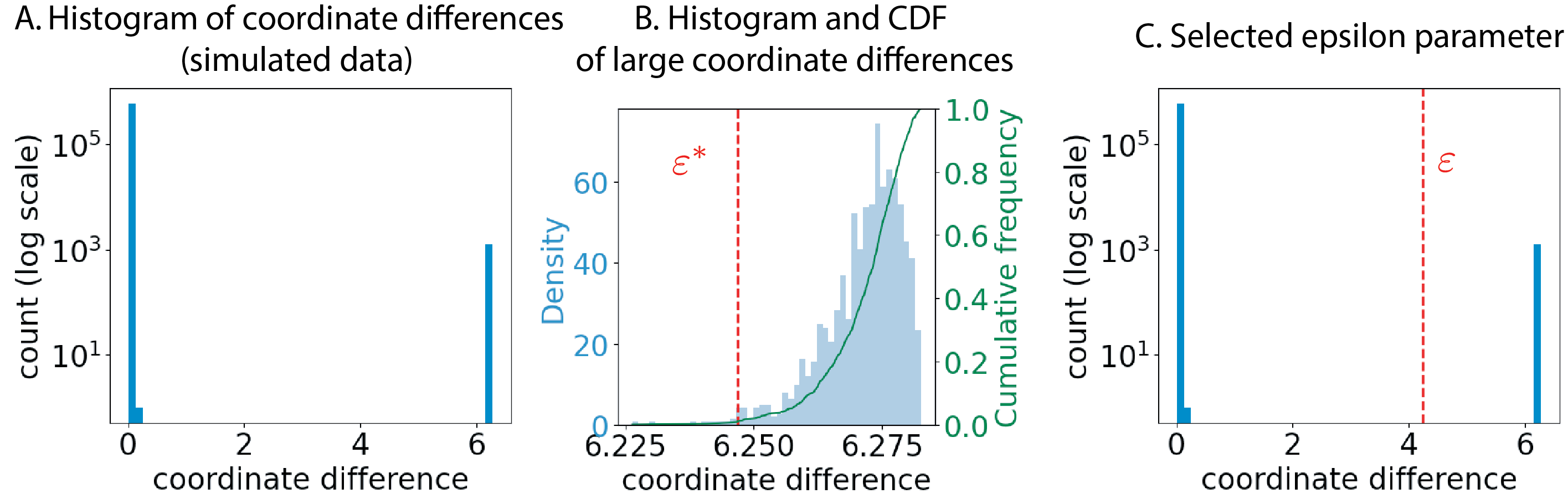}
    \caption{Selection of the proximity parameter $\varepsilon$ for CAN-simulated data. \textbf{A.} Histogram of maximal coordinate differences $\max\{| \theta^t_x - \theta^{t+1}_x|, \, |\theta^t_y - \theta^{t+1}_y |\}$ across all consecutive time points, shown on a log scale. The differences concentrate near $0$ and $2\pi$. \textbf{B.} Histogram and empirical cumulative distribution function (green curve) of the coordinate differences restricted to the interval $[2, 2\pi)$. The red dashed line indicates the threshold $\varepsilon^*$ at which 99\% of the maximal coordinate differences exceed $\varepsilon^*$. \textbf{C.} The final parameter is set to $\varepsilon = \varepsilon^* - 2$.}
    \label{fig:epsilon_selection_simulation}
\end{figure}

To select $\varepsilon$, we examine the distribution of maximal coordinate differences $\max\{| \theta^t_x - \theta^{t+1}_x|, \, |\theta^t_y - \theta^{t+1}_y |\}$ across all consecutive time points. In both simulated (Fig.~\ref{fig:epsilon_selection_simulation}A) and experimental data (SI Fig. 9B, left), these differences concentrate near $0$ (reflecting small local movements) and $2 \pi$ (reflecting crossing a torus edge), with the vast majority near $0$. Differences near $2 \pi$ correspond to time points at which the path crosses a torus edge and requires a nontrivial lift. We therefore choose $\varepsilon$ to be smaller than the cluster of values near $2 \pi$.

Concretely, we restrict the maximal coordinate difference to $[2, 2 \pi)$ (Fig.~\ref{fig:epsilon_selection_simulation}B, SI Fig. 9B, center) and compute their empirical cumulative distribution function. We select $\varepsilon^*$ to be the value at which $P(X > \varepsilon^*) = 0.99$,  ensuring that $99\%$ of the larger coordinate differences lie above $\varepsilon^*$. We then set the final parameter to $\varepsilon = \varepsilon^* - 2$, which provides additional margin for identifying potential lifts (Fig.~\ref{fig:epsilon_selection_simulation}C, SI Fig. 9B, right). The time points whose toroidal coordinate differences are greater than this $\varepsilon$ are potential time points for nontrivial lifts. Whether the nontrivial lifts occur, and if so, in what manner, is determined by Equation~\ref{eq:determine_tile}.

\subsubsection{Null baselines for assessing reconstruction quality}
To assess whether the reconstructed paths are meaningfully similar to the original trajectories, we compare the reconstruction errors against null baselines that quantify the expected error when no true correspondence exists between two paths. We use different null constructions for the simulated and experimental settings.

\noindent\textit{Simulated data.} For each environment (0, 1, and 2 holes), we independently simulated pairs of trajectories using the same random walk model and environment parameters described in \textit{Materials and Methods} Section~\ref{sec:simulated_mice_trajectory}. For each pair, we computed the optimal affine transformation aligning one trajectory to the other and computed the error as described in \textit{Materials and Methods} Section~\ref{sec:reconstruction_error}.
This process was repeated 100 times to obtain a distribution of null errors. Because the two trajectories in each pair are generated independently, any alignment between them is purely coincidental, and the resulting error distribution represents the expected reconstruction error in the absence of a meaningful relationship between the reconstructed and original paths (Fig.~\ref{fig:simulation_result}E).

\noindent\textit{Experimental data (two-dimensional).} To assess the quality of the reconstruction on local paths, we constructed a null baseline from mismatched segment pairs. Specifically, given the set of all 20-second local path segments, for each pair of path segments, we aligned one trajectory to the other and computed the error as described in \textit{Materials and Methods} Section~\ref{sec:reconstruction_error}. There were 63 local path segments, so we obtained a total of 1,953 baseline errors. The distribution of these errors serves as a baseline against which the true (matched) reconstruction errors are compared (Fig.~\ref{fig:Gardner-comp}K). A reconstruction error significantly lower than this baseline indicates that the reconstructed path preserves the geometry of the specific original segment from which it was derived, rather than reflecting generic properties shared across all segments.

\noindent\textit{Null models for path lifting}. To verify that the inductive lifting procedure is necessary for accurate path reconstruction, we compared the proposed method against two null models: a no-lifting model, in which the toroidal coordinates are used directly as the reconstructed path without any tile transitions, and a random-lifting model, in which the coordinates at each time point are lifted to a tile chosen randomly among the preceding lift's tile and its eight neighbors. Both null models were applied to simulated trajectories in the one-hole environment. See SI Section 2 for details.

\subsection{Code Availability} Code can be found at the following Github repository \newline\href{https://github.com/irishryoon/GridPathLifting}{https://github.com/irishryoon/GridPathLifting}. For simulations, we modified the code shared in \href{https://github.com/erikher/GridCellTorus}{https://github.com/erikher/GridCellTorus}.

All computations were performed on a single core of an Intel Xeon Gold 6326 (2.90 GHz) node with 512 GB RAM. For a single simulated dataset (2,464 neurons, 600,000 time bins), the full pipeline was completed in approximately 12 minutes (3 minutes for grid cell activity simulation, 9 minutes for the toroidal coordinates computation, 0.5 seconds for path lifting). For the two-dimensional experimental data (111 neurons, 126,728 time bins), the full pipeline was completed in less than a minute. For reference, the same pipeline required approximately 80 minutes for the simulated data and less than a minute for the experimental data on a standard laptop (MacBook Pro, Apple M2, 16 GB RAM, single core).

\subsection{Data Availability} All experimental data in this manuscript are from publicly available datasets.

The one-dimensional movement dataset \cite{wen2024one} can be found at \newline\href{https://data.mendeley.com/datasets/rgtk6jygjc/1}{https://data.mendeley.com/datasets/rgtk6jygjc/1}. We analyzed the dataset ``N2\_200203\_buildup\_track.mat," which contains recordings from mouse N2 navigating a 320 cm virtual build-up track. The dataset provides spatially binned firing rates (2-cm bins, $T=160$ bins per lap) for $G =44$ co-modular grid cells identified via clustering on power spectra. We analyzed 441 continuous runs on the track.

The two-dimensional open field movement dataset \cite{gardner_toroidal_2022} can be found at \newline\href{https://figshare.com/articles/dataset/Toroidal_topology_of_population_activity_in_grid_cells/16764508}{https://figshare.com/articles/dataset/Toroidal\_topology\_of\_population\_activity\_in\_grid\_cells/16764508}. We analyzed data from rat R, module 1, day 2, open-field session, in which the rat explored a 1.5 m $\times$ 1.5 m square arena for 21.1 minutes. Spikes from grid cells in layers II and III of MEC-parasubiculum were recorded. Six grid modules were identified by clustering; we analyzed $G =111$ co-modular grid cells. Spike trains were preprocessed into continuous firing rate estimates as described in Section~\ref{sec:preprocessing}. The resulting firing rates were reported over $T =126,728$ time bins, where 1 time bin corresponds to 10 ms.

Both datasets have been made public by the authors of the corresponding publications.

\newpage
\printbibliography

@article{zou2021neurobiologically,
  title={A neurobiologically inspired mapping and navigating framework for mobile robots},
  author={Zou, Qiang and Cong, Ming and Liu, Dong and Du, Yu},
  journal={Neurocomputing},
  volume={460},
  pages={181--194},
  year={2021},
  month={October},
  doi={10.1016/j.neucom.2021.07.025},
  publisher={Elsevier}
}

@article{banino2018vector,
  author  = {Banino, Andrea and Barry, Caswell and Uria, Benigno and Blundell, Charles and Lillicrap, Timothy and Mirowski, Piotr and Pritzel, Alexander and Chadwick, Martin J. and Degris, Thomas and Modayil, Joseph and Wayne, Greg and Soyer, Hubert and Viola, Fabio and Zhang, Brian and Goroshin, Ross and Rabinowitz, Neil and Pascanu, Razvan and Beattie, Charlie and Petersen, Stig and Sadik, Amir and Danihelka, Ivo and Riedmiller, Martin and Fidjeland, Andreas and Grewe, Dominik and Hassabis, Demis and Kumaran, Dharshan},
  title   = {Vector-based navigation using grid-like representations in artificial agents},
  journal = {Nature},
  volume  = {557},
  number  = {7705},
  pages   = {429--433},
  year    = {2018},
  doi     = {10.1038/s41586-018-0102-6}
}

@article{simkuns2025deep,
  author  = {Simkuns, Arturs and Saltanovs, Rodions and Ivanovs, Maksims and Kadikis, Roberts},
  title   = {Deep Learning-Emerged Grid Cells-Based Bio-Inspired Navigation in Robotics},
  journal = {Sensors},
  volume  = {25},
  number  = {5},
  pages   = {1576},
  year    = {2025},
  doi     = {10.3390/s25051576}
}

@article{dabaghian2023grid,
  author  = {Dabaghian, Yuri},
  title   = {Grid Cell Percolation},
  journal = {Neural Computation},
  volume  = {35},
  number  = {10},
  pages   = {1609--1626},
  year    = {2023},
}

@article{curto2008cell,
  author    = {Carina Curto and Vladimir Itskov},
  title     = {Cell Groups Reveal Structure of Stimulus Space},
  journal   = {PLoS Computational Biology},
  volume    = {4},
  number    = {10},
  pages     = {e1000205},
  year      = {2008},
  doi       = {10.1371/journal.pcbi.1000205}
}

@article{rybakkenDecodingNeuralData2019,
	title = {Decoding of {Neural} {Data} {Using} {Cohomological} {Feature} {Extraction}},
	volume = {31},
	issn = {0899-7667},
	number = {1},
	urldate = {2024-02-22},
	journal = {Neural Computation},
	author = {Rybakken, Erik and Baas, Nils and Dunn, Benjamin},
	month = jan,
	year = {2019},
	pages = {68--93},
}

@article{zhouHyperbolicGeometryOlfactory2018,
	title = {Hyperbolic geometry of the olfactory space},
	volume = {4},
	url = {https://www.science.org/doi/10.1126/sciadv.aaq1458},
	doi = {10.1126/sciadv.aaq1458},
	number = {8},
	urldate = {2023-12-25},
	journal = {Science Advances},
	author = {Zhou, Yuansheng and Smith, Brian H. and Sharpee, Tatyana O.},
	month = aug,
	year = {2018},
	note = {Publisher: American Association for the Advancement of Science},
	pages = {eaaq1458},
}

@article{chaudhuriIntrinsicAttractorManifold2019,
	title = {The intrinsic attractor manifold and population dynamics of a canonical cognitive circuit across waking and sleep},
	volume = {22},
	copyright = {2019 The Author(s), under exclusive licence to Springer Nature America, Inc.},
	issn = {1546-1726},
	url = {https://www.nature.com/articles/s41593-019-0460-x},
	doi = {10.1038/s41593-019-0460-x},
	language = {en},
	number = {9},
	urldate = {2023-12-25},
	journal = {Nature Neuroscience},
	author = {Chaudhuri, Rishidev and Gerçek, Berk and Pandey, Biraj and Peyrache, Adrien and Fiete, Ila},
	month = sep,
	year = {2019},
	note = {Number: 9
Publisher: Nature Publishing Group},
	pages = {1512--1520},
}

@article{villetteInternallyRecurringHippocampal2015,
	title = {Internally {Recurring} {Hippocampal} {Sequences} as a {Population} {Template} of {Spatiotemporal} {Information}},
	volume = {88},
	issn = {1097-4199},
	doi = {10.1016/j.neuron.2015.09.052},
	language = {eng},
	number = {2},
	journal = {Neuron},
	author = {Villette, Vincent and Malvache, Arnaud and Tressard, Thomas and Dupuy, Nathalie and Cossart, Rosa},
	month = oct,
	year = {2015},
	pmid = {26494280},
	pmcid = {PMC4622933},
	pages = {357--366},
}

@article{okeefePlaceUnitsHippocampus1976,
	title = {Place units in the hippocampus of the freely moving rat},
	volume = {51},
	issn = {0014-4886},
	url = {https://www.sciencedirect.com/science/article/pii/0014488676900558},
	doi = {10.1016/0014-4886(76)90055-8},
	number = {1},
	urldate = {2024-03-03},
	journal = {Experimental Neurology},
	author = {O'Keefe, John},
	month = jan,
	year = {1976},
	pages = {78--109},
}

@article{hubelReceptiveFieldsFunctional1968,
	title = {Receptive fields and functional architecture of monkey striate cortex},
	volume = {195},
	issn = {0022-3751},
	doi = {10.1113/jphysiol.1968.sp008455},
	language = {eng},
	number = {1},
	journal = {The Journal of Physiology},
	author = {Hubel, D. H. and Wiesel, T. N.},
	month = mar,
	year = {1968},
	pmid = {4966457},
	pmcid = {PMC1557912},
	pages = {215--243},
}

@misc{peng_grid_2023,
	title = {Grid cells perform path integration in multiple reference frames during self-motion-based navigation},
	copyright = {© 2023, Posted by Cold Spring Harbor Laboratory. This pre-print is available under a Creative Commons License (Attribution-NonCommercial-NoDerivs 4.0 International), CC BY-NC-ND 4.0, as described at http://creativecommons.org/licenses/by-nc-nd/4.0/},
	url = {https://www.biorxiv.org/content/10.1101/2023.12.21.572857v1},
	doi = {10.1101/2023.12.21.572857},
	language = {en},
	urldate = {2025-07-22},
	publisher = {bioRxiv},
	author = {Peng, Jing-Jie and Throm, Beate and Jazi, Maryam Najafian and Yen, Ting-Yun and Monyer, Hannah and Allen, Kevin},
	month = dec,
	year = {2023},
	note = {Pages: 2023.12.21.572857
Section: New Results},
}

@inproceedings{masson_decoding_2011,
	address = {Dordrecht},
	title = {Decoding the {Grid} {Cells} for {Metric} {Navigation} {Using} the {Residue} {Numeral} {System}},
	isbn = {978-90-481-9695-1},
	booktitle = {Advances in {Cognitive} {Neurodynamics} ({II})},
	publisher = {Springer Netherlands},
	author = {Masson, Cécile and Girard, Benoît},
	editor = {Wang, Rubin and Gu, Fanji},
	year = {2011},
	pages = {459--464},
}

@article{fiete_what_2008,
	title = {What {Grid} {Cells} {Convey} about {Rat} {Location}},
	volume = {28},
	copyright = {Copyright © 2008 Society for Neuroscience 0270-6474/08/286858-14\$15.00/0},
	issn = {0270-6474, 1529-2401},
	url = {https://www.jneurosci.org/content/28/27/6858},
	doi = {10.1523/JNEUROSCI.5684-07.2008},
	language = {en},
	number = {27},
	urldate = {2025-07-22},
	journal = {Journal of Neuroscience},
	author = {Fiete, Ila R. and Burak, Yoram and Brookings, Ted},
	month = jul,
	year = {2008},
	pmid = {18596161},
	note = {Publisher: Society for Neuroscience
Section: Articles},
	pages = {6858--6871},
}

@inproceedings{sun_neural-like_1994,
	title = {A neural-like network approach to residue-to-decimal conversion},
	volume = {6},
	url = {https://ieeexplore.ieee.org/document/374831},
	doi = {10.1109/ICNN.1994.374831},
	urldate = {2025-07-22},
	booktitle = {Proceedings of 1994 {IEEE} {International} {Conference} on {Neural} {Networks} ({ICNN}'94)},
	author = {Sun, Hong and Yao, Tian-Ren},
	month = jun,
	year = {1994},
	pages = {3883--3887 vol.6},
}

@article{livezey_deep_2020,
	title = {Deep learning approaches for neural decoding across architectures and recording modalities},
	volume = {22},
	issn = {1477-4054},
	url = {https://doi.org/10.1093/bib/bbaa355},
	doi = {10.1093/bib/bbaa355},
	number = {2},
	journal = {Briefings in Bioinformatics},
	author = {Livezey, Jesse A and Glaser, Joshua I},
	month = dec,
	year = {2020},
	note = {\_eprint: https://academic.oup.com/bib/article-pdf/22/2/1577/36654842/bbaa355.pdf},
	pages = {1577--1591},
}

@article{tampuu_efficient_2019,
	title = {Efficient neural decoding of self-location with a deep recurrent network},
	volume = {15},
	issn = {1553-7358},
	url = {https://journals.plos.org/ploscompbiol/article?id=10.1371/journal.pcbi.1006822},
	doi = {10.1371/journal.pcbi.1006822},
	language = {en},
	number = {2},
	urldate = {2025-07-22},
	journal = {PLOS Computational Biology},
	author = {Tampuu, Ardi and Matiisen, Tambet and Ólafsdóttir, H. Freyja and Barry, Caswell and Vicente, Raul},
	month = feb,
	year = {2019},
	note = {Publisher: Public Library of Science},
	pages = {e1006822},
}

@misc{frey_deepinsight_2019,
	title = {Deepinsight: a general framework for interpreting wide-band neural activity},
	copyright = {© 2019, Posted by Cold Spring Harbor Laboratory. This pre-print is available under a Creative Commons License (Attribution-NonCommercial-NoDerivs 4.0 International), CC BY-NC-ND 4.0, as described at http://creativecommons.org/licenses/by-nc-nd/4.0/},
	shorttitle = {Deepinsight},
	url = {https://www.biorxiv.org/content/10.1101/871848v1},
	doi = {10.1101/871848},
	language = {en},
	urldate = {2025-07-22},
	publisher = {bioRxiv},
	author = {Frey, Markus and Tanni, Sander and Perrodin, Catherine and O’Leary, Alice and Nau, Matthias and Kelly, Jack and Banino, Andrea and Doeller, Christian F. and Barry, Caswell},
	month = dec,
	year = {2019},
	note = {Pages: 871848
Section: New Results},
}

@article{xu_comparison_2019,
	title = {A {Comparison} of {Neural} {Decoding} {Methods} and {Population} {Coding} {Across} {Thalamo}-{Cortical} {Head} {Direction} {Cells}},
	volume = {13},
	issn = {1662-5110},
	url = {https://www.frontiersin.org/journals/neural-circuits/articles/10.3389/fncir.2019.00075/full},
	doi = {10.3389/fncir.2019.00075},
	language = {English},
	urldate = {2025-07-22},
	journal = {Frontiers in Neural Circuits},
	author = {Xu, Zishen and Wu, Wei and Winter, Shawn S. and Mehlman, Max L. and Butler, William N. and Simmons, Christine M. and Harvey, Ryan E. and Berkowitz, Laura E. and Chen, Yang and Taube, Jeffrey S. and Wilber, Aaron A. and Clark, Benjamin J.},
	month = dec,
	year = {2019},
	note = {Publisher: Frontiers},
}

@article{mitchell_topological_2024,
	title = {A topological deep learning framework for neural spike decoding},
	volume = {123},
	issn = {0006-3495, 1542-0086},
	url = {https://www.cell.com/biophysj/abstract/S0006-3495(24)00041-9},
	doi = {10.1016/j.bpj.2024.01.025},
	language = {English},
	number = {17},
	urldate = {2025-07-22},
	journal = {Biophysical Journal},
	author = {Mitchell, Edward C. and Story, Brittany and Boothe, David and Franaszczuk, Piotr J. and Maroulas, Vasileios},
	month = sep,
	year = {2024},
	pmid = {38402607},
	note = {Publisher: Elsevier},
	pages = {2781--2789},
}

@article{DREIMAC_Perea2023,
  doi = {10.21105/joss.05791},
  url = {https://doi.org/10.21105/joss.05791},
  year = {2023},
  publisher = {The Open Journal},
  volume = {8},
  number = {91},
  pages = {5791},
  author = {Jose A. Perea and Luis Scoccola and Christopher J. Tralie},
  title = {DREiMac: Dimensionality Reduction with Eilenberg-MacLane Coordinates},
  journal = {Journal of Open Source Software}
}

@article{stemmler2015connecting,
  title={Connecting multiple spatial scales to decode the population activity of grid cells},
  author={Stemmler, Martin and Mathis, Alexander and Herz, Andreas VM},
  journal={Science Advances},
  volume={1},
  number={11},
  pages={e1500816},
  year={2015},
  publisher={American Association for the Advancement of Science}
}

@article{mathis2012optimal,
  title={Optimal population codes for space: grid cells outperform place cells},
  author={Mathis, Alexander and Herz, Andreas VM and Stemmler, Martin},
  journal={Neural computation},
  volume={24},
  number={9},
  pages={2280--2317},
  year={2012},
  publisher={MIT Press One Rogers Street, Cambridge, MA 02142-1209, USA journals-info~…}
}

@article{de_silva_persistent_2011,
	title = {Persistent {Cohomology} and {Circular} {Coordinates}},
	volume = {45},
	issn = {1432-0444},
	url = {https://doi.org/10.1007/s00454-011-9344-x},
	doi = {10.1007/s00454-011-9344-x},
	language = {en},
	number = {4},
	urldate = {2025-03-04},
	journal = {Discrete \& Computational Geometry},
	author = {de Silva, Vin and Morozov, Dmitriy and Vejdemo-Johansson, Mikael},
	month = jun,
	year = {2011},
	pages = {737--759},
}

@article{gardner_toroidal_2022,
	title = {Toroidal topology of population activity in grid cells},
	volume = {602},
	rights = {2022 The Author(s)},
	issn = {1476-4687},
	url = {https://www.nature.com/articles/s41586-021-04268-7},
	doi = {10.1038/s41586-021-04268-7},
	pages = {123--128},
	number = {7895},
    year = {2022},
	journal = {Nature},
	author = {Gardner, Richard J. and Hermansen, Erik and Pachitariu, Marius and Burak, Yoram and Baas, Nils A. and Dunn, Benjamin A. and Moser, May-Britt and Moser, Edvard I.},
	urldate = {2024-09-26},
	date = {2022-02},
	langid = {english},
	note = {Publisher: Nature Publishing Group}
}

@article{scoccola_toroidal_2023,
	title = {Toroidal Coordinates: Decorrelating Circular Coordinates with Lattice Reduction},
	volume = {258},
	rights = {Creative Commons Attribution 4.0 International license, info:eu-repo/semantics/{openAccess}},
	issn = {1868-8969},
	url = {https://drops.dagstuhl.de/entities/document/10.4230/LIPIcs.SoCG.2023.57},
	doi = {10.4230/LIPICS.SOCG.2023.57},
	shorttitle = {Toroidal Coordinates},
	pages = {57:1--57:20},
	journal = {{LIPIcs}, Volume 258, {SoCG} 2023},
    year = {2023},
	author = {Scoccola, Luis and Gakhar, Hitesh and Bush, Johnathan and Schonsheck, Nikolas and Rask, Tatum and Zhou, Ling and Perea, Jose A.},
	editora = {Chambers, Erin W. and Gudmundsson, Joachim},
	editoratype = {collaborator},
	urldate = {2024-09-26},
	date = {2023},
	langid = {english},
	note = {Artwork Size: 20 pages, 6894838 bytes
{ISBN}: 9783959772730
Medium: application/pdf
Publisher: Schloss Dagstuhl – Leibniz-Zentrum für Informatik},
}

@article{hafting2005microstructure,
  title={Microstructure of a spatial map in the entorhinal cortex},
  author={Hafting, Torkel and Fyhn, Marianne and Molden, Sturla and Moser, May-Britt and Moser, Edvard I},
  journal={Nature},
  volume={436},
  number={7052},
  pages={801--806},
  year={2005},
  publisher={Nature Publishing Group UK London}
}

@article{mcnaughton2006path,
  title={Path integration and the neural basis of the'cognitive map'},
  author={McNaughton, Bruce L and Battaglia, Francesco P and Jensen, Ole and Moser, Edvard I and Moser, May-Britt},
  journal={Nature Reviews Neuroscience},
  volume={7},
  number={8},
  pages={663--678},
  year={2006},
  publisher={Nature Publishing Group UK London}
}

@article{burak2009accurate,
  title={Accurate path integration in continuous attractor network models of grid cells},
  author={Burak, Yoram and Fiete, Ila R},
  journal={PLoS computational biology},
  volume={5},
  number={2},
  pages={e1000291},
  year={2009},
  publisher={Public Library of Science San Francisco, USA}
}

@article{couey2013recurrent,
  title={Recurrent inhibitory circuitry as a mechanism for grid formation},
  author={Couey, Jonathan J and Witoelar, Aree and Zhang, Sheng-Jia and Zheng, Kang and Ye, Jing and Dunn, Benjamin and Czajkowski, Rafal and Moser, May-Britt and Moser, Edvard I and Roudi, Yasser and others},
  journal={Nature neuroscience},
  volume={16},
  number={3},
  pages={318--324},
  year={2013},
  publisher={Nature Publishing Group US New York}
}

@article{tralie2018ripser,
  title={Ripser. py: A lean persistent homology library for python},
  author={Tralie, Christopher and Saul, Nathaniel and Bar-On, Rann},
  journal={Journal of Open Source Software},
  volume={3},
  number={29},
  pages={925},
  year={2018}
}

@article{bauer2021ripser,
  title={Ripser: efficient computation of Vietoris--Rips persistence barcodes},
  author={Bauer, Ulrich},
  journal={Journal of Applied and Computational Topology},
  volume={5},
  number={3},
  pages={391--423},
  year={2021},
  publisher={Springer}
}

@article{opencv_library,
    author = {Bradski, G.},
    journal = {Dr. Dobb's Journal of Software Tools},
    title = {{The OpenCV Library}},
    year = {2000}
}

@article{solstad2006grid,
  title={From grid cells to place cells: a mathematical model},
  author={Solstad, Trygve and Moser, Edvard I and Einevoll, Gaute T},
  journal={Hippocampus},
  volume={16},
  number={12},
  pages={1026--1031},
  year={2006},
  publisher={Wiley Online Library}
}

@article{erdem2012goal,
  title={A goal-directed spatial navigation model using forward trajectory planning based on grid cells},
  author={Erdem, U{\u{g}}ur M and Hasselmo, Michael},
  journal={European Journal of Neuroscience},
  volume={35},
  number={6},
  pages={916--931},
  year={2012},
  publisher={Wiley Online Library}
}

@article{barry2012z,
  author  = {Barry, Caswell and Bush, Daniel},
  title   = {From {A} to {Z}: a potential role for grid cells in spatial navigation},
  journal = {Neural Systems \& Circuits},
  year    = {2012},
  volume  = {2},
  number  = {1},
  pages   = {6},
  month   = may,
  doi     = {10.1186/2042-1001-2-6},
  pmid    = {22647296},
  pmcid   = {PMC3423065},
  url     = {https://pmc.ncbi.nlm.nih.gov/articles/PMC3423065/}
}

@article{bush2015using,
  title={Using grid cells for navigation},
  author={Bush, Daniel and Barry, Caswell and Manson, Daniel and Burgess, Neil},
  journal={Neuron},
  volume={87},
  number={3},
  pages={507--520},
  year={2015},
  publisher={Elsevier}
}

@article{fuhs2006spin,
  title={A spin glass model of path integration in rat medial entorhinal cortex},
  author={Fuhs, Mark C and Touretzky, David S},
  journal={Journal of Neuroscience},
  volume={26},
  number={16},
  pages={4266--4276},
  year={2006},
  publisher={Society for Neuroscience}
}

@article{dang2021grid,
  title={Why grid cells function as a metric for space},
  author={Dang, Suogui and Wu, Yining and Yan, Rui and Tang, Huajin},
  journal={Neural Networks},
  volume={142},
  pages={128--137},
  year={2021},
  publisher={Elsevier}
}

@article{ghrist2008barcodes,
  title={Barcodes: the persistent topology of data},
  author={Ghrist, Robert},
  journal={Bulletin of the American Mathematical Society},
  volume={45},
  number={1},
  pages={61--75},
  year={2008}
}

@article{edelsbrunner2008persistent,
  title={Persistent homology-a survey},
  author={Edelsbrunner, Herbert and Harer, John and others},
  journal={Contemporary mathematics},
  volume={453},
  number={26},
  pages={257--282},
  year={2008},
  publisher={Citeseer}
}

@article{carlsson2009topologydata,
author = {Carlsson, Gunnar},
year = {2009},
month = {04},
pages = {255-308},
title = {Topology and Data},
volume = {46},
journal = {Bulletin of The American Mathematical Society - BULL AMER MATH SOC},
doi = {10.1090/S0273-0979-09-01249-X}
}

@article{edelsbrunner2002topological,
  title={Topological persistence and simplification},
  author={Edelsbrunner and Letscher and Zomorodian},
  journal={Discrete \& computational geometry},
  volume={28},
  pages={511--533},
  year={2002},
  publisher={Springer}
}

@book{munkres2017topology,
  title={Topology},
  author={Munkres, J.},
  isbn={9780134689517},
  lccn={2016055057},
  series={Pearson Modern Classics for Advanced Mathematics Series},
  url={https://books.google.com/books?id=51n8MAAACAAJ},
  year={2017},
  publisher={Pearson}
}

@article{taube1990head,
  title={Head-direction cells recorded from the postsubiculum in freely moving rats. I. Description and quantitative analysis},
  author={Taube, Jeffrey S and Muller, Robert U and Ranck, James B},
  journal={Journal of Neuroscience},
  volume={10},
  number={2},
  pages={420--435},
  year={1990},
  publisher={Society for Neuroscience}
}

@article{wen2024one,
  title={One-shot entorhinal maps enable flexible navigation in novel environments},
  author={Wen, John H and Sorscher, Ben and Aery Jones, Emily A and Ganguli, Surya and Giocomo, Lisa M},
  journal={Nature},
  volume={635},
  number={8040},
  pages={943--950},
  year={2024},
  publisher={Nature Publishing Group UK London}
}

@article{fischler1981random,
  author    = {Martin A. Fischler and Robert C. Bolles},
  title     = {Random Sample Consensus: {A} Paradigm for Model Fitting with Applications to Image Analysis and Automated Cartography},
  journal   = {Communications of the ACM},
  volume    = {24},
  number    = {6},
  pages     = {381--395},
  year      = {1981},
  doi       = {10.1145/358669.358692}
}

\end{document}


\linenumbers

\maketitle

\tableofcontents

\section{Mathematical Preliminaries}
\subsection{Lifting paths on torus to paths in $\mathbb{R}^2$}

Given a topological space $X$, a path in $X$ is a continuous map $f: I \to X$, where $I = [0,1]$ denotes the unit interval. If one considers $t \in [0,1]$ as representing time, then $f(t)$ specifies the location of an object in $X$ at time $t$. 

In this work, we are concerned with lifting paths on a torus to paths in $\mathbb{R}^2$. Recall that $S^1 $ represents a circle, and that $S^1 \times S^1$ represents a torus. Let $f: I \to S^1 \times S^1$ be a path on the torus, and let $p: \mathbb{R}^2 \to S^1 \times S^1$ be a covering of a torus. For example, $p: \mathbb{R}^2 \to S^1 \times S^1$ defined by
\begin{equation}
\label{eq:covering_torus}
p(x,y) = (\, (\cos2 \pi x, \sin 2 \pi x), \, (\cos 2 \pi y, \sin 2 \pi y)\,) 
\end{equation} 
is a valid covering of a torus. Note that $(\cos2 \pi x, \sin 2 \pi x)$ and $(\cos 2 \pi y, \sin 2 \pi y)$ each specify points in $S^1$. 

The following lemma states that any path on a torus can be lifted to a path in $\mathbb{R}^2$.

\begin{lemma} (Lemma 54.2 \cite{munkres2017topology}, modified)
The path $f: I \to S^1 \times S^1$ can be lifted to a path $\tilde{f}: I \to  \mathbb{R}^2$ such that the following diagram commutes. 
\[
\begin{tikzcd}
\quad & \mathbb{R}^2 \arrow[d, "p"] \\
I \arrow[ru, dashed, "\tilde{f}"] \arrow[r, "f"] & S^1 \times S^1
\end{tikzcd}
\]
Furthermore, given a $b_0 = f(0) \in S^1 \times S^1$ and $e_0 \in p^{-1}(b_0)$, the lifted path $\tilde{f}$ with $\tilde{f}(0) = e_0$ is unique.  
\end{lemma}

A constructive proof can be found in \cite{munkres2017topology}. Here, we illustrate the construction of $\tilde{f}$ in a simple example. Let $p: \mathbb{R}^2 \to S^1 \times S^1$ the covering map from Equation~\ref{eq:covering_torus}. We construct $\tilde{f}$ in pieces.

\begin{figure}[h!]
    \centering
    \includegraphics[width=0.8\linewidth]{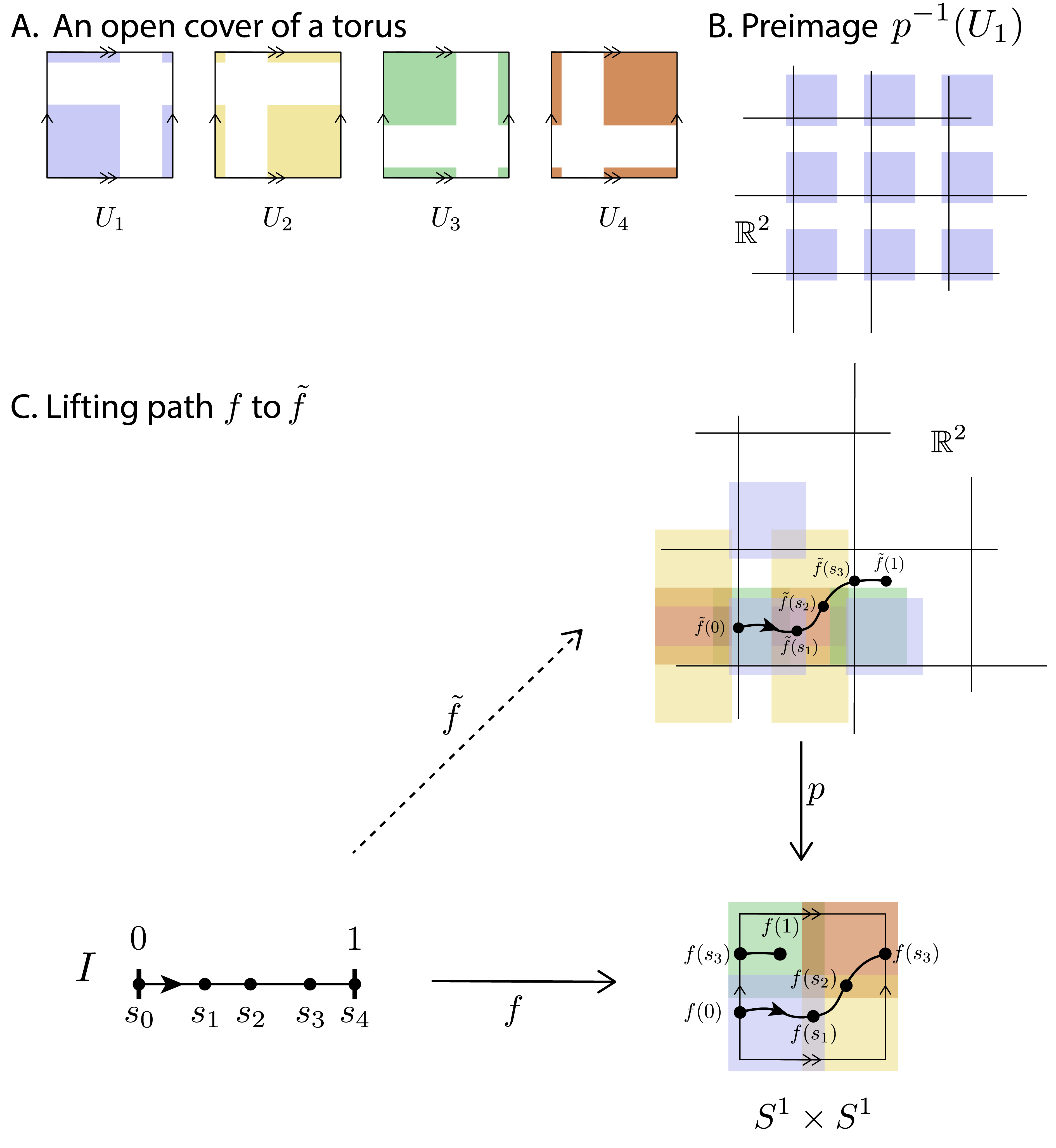}
    \caption{Lifting a path $f$ on torus to a path $\tilde{f}$ in $\mathbb{R}^2$.  \textbf{A.} The open cover $\{U_1, \dots, U_4\}$ of the torus. \textbf{B.} The preimage $p^{-1}(U_1)$ consists of homeomorphic copies of $U_1$ in $\mathbb{R}^2$. \textbf{C.} The lifted path $\tilde{f}$ is constructed piece-by-piece.}
    \label{fig:lifting-procedure}
\end{figure}

Let $U_1, \dots, U_4$ be open sets covering the torus (Fig. ~\ref{fig:lifting-procedure}A). Note that the for any $U \in \{ U_1, \dots U_4\}$, the preimage $p^{-1}(U)$ consists of infinitely-many homeomorphic copies of $U$ in $\mathbb{R}^2$ (SI Fig.~\ref{fig:lifting-procedure}B). Each copy of $U$ in $p^{-1}(U)$ is called a \emph{slice}.

We then partition $I = [0,1]$ into segments $0=s_0<s_1<\dots<s_4=1$ so that $f$ maps each segment $[s_i, s_{i+1}]$ into one of $U_1, \dots, U_4$ (Figure~\ref{fig:lifting-procedure}C). Without loss of generality, assume $f(0)$ lies in $U_1$. We choose some slice, let's say $V_1$, of $U_1$. We let $\tilde{f}(0)$ be the unique point in $V_1^1$ that maps to $f(0)$ via $p$. We then define $\tilde{f}$ on $[0, s_1]$ to be the unique path in $V_1 \subset \mathbb{R}^2$ that maps to $f|_{ [0, s_1] }$ via $p$. 

To define $\tilde{f}$ on $[s_1, s_2]$, note that $f(s_1)$ lives in both $U_1$ and $U_2$. Among the slices $p^{-1}(U_2)$ of $U_2$, there exists a unique slice, say $V_2$ of $U_2$ where $\tilde{f}(s_1) \in V_2$. We then define $\tilde{f}$ on $[s_1, s_2]$ to be the unique path in $V_2 \subset \mathbb{R}^2$ that maps to $f|_{ [s_1, s_2] }$ via $p$. 

We continue this procedure until we define $\tilde{f}$ on the entire interval $[0,1]$. By construction, $p \circ \tilde{f} = f$.

\subsection{Persistent homology} 
We provide a brief description of simplicial homology (with field coefficients) and persistent homology. 

\subsubsection{Simplicial Complexes and Simplicial Homology}

An \emph{(abstract) simplicial complex} $K = (V, F)$ is a combinatorial structure built from a set of vertices $V$. It consists of simplices, where a simplex is an unordered subset of $V$. A collection of $n+1$ elements in $V$, for example, $(v_0, \dots, v_n)$, is called an $n$-simplex. Concretely, a single vertex is a $0$-simplex, the collection $(v_0, v_1)$ is a 1-simplex, and $(v_0, v_1, v_2)$ is a 2-simplex. The collection of simplices $F$ must satisfy the following: given a simplex $\sigma \in F$, all of its non-empty subsets must also be in $F$. 

In this paper, the homology of a simplicial complex is computed with field coefficients $\mathbb{F}$. Therefore, all homology computations are done in the context of vector spaces and linear maps. 

To study the topology of $K$, we work with \emph{chains}, which are formal linear combinations of simplices. More precisely, if $F_n$ denotes the number of $n$-simplices in a complex $K$, then the vector space of $n$-chains is  
\[
C_n(K) = \Big\{ \sum_{i=1}^{F_n} c_i \sigma_i^n \;\;\Big|\;\; c_i \in \mathbb{F}, \;\sigma_i^n \in K \Big\}.
\]  

The \emph{boundary homomorphism} $\partial_n: C_n(K) \to C_{n-1}(K)$ is constructed as follows. For an $n$-simplex $\sigma = (v_0,\dots,v_n)$, the boundary is defined as  
\[
\partial_n(\sigma) = \sum_{i=0}^n (-1)^i (v_0, \dots, \hat{v}_i, \dots, v_n),
\]  
where the notation $\hat{v}_i$ means that the vertex $v_i$ has been omitted. This map is then extended linearly to all of $C_n(K)$. We then obtain the following 
\[
\cdots \to C_{n+1}(K) \xrightarrow{\partial_{n+1}} C_n(K) \xrightarrow{\partial_n} C_{n-1}(K) \to \cdots \to C_0(K) \to 0,
\]  
which is a sequence of vector spaces and linear maps. The boundary homomorphisms satisfy the key property that $\partial_n \circ \partial_{n+1} = 0$ for all $n$.  

This property ensures that $\text{im}\, \partial_{n+1} \subseteq \ker \partial_n$. Elements in $\ker \partial_n$ are called \emph{cycles}, representing potential $n$-dimensional holes, while those in $\operatorname{im} \partial_{n+1}$ are called \emph{boundaries}, representing cycles that are filled in by higher-dimensional simplices. The true $n$-dimensional holes are captured by the \emph{homology group}  
\[
H_n(K) = \ker \partial_n \big/ \operatorname{im} \partial_{n+1}.
\]  

Each homology group is a vector space, and its dimension records the number of independent $n$-dimensional holes. These invariants provide a compact algebraic summary of the underlying topological structure of the simplicial complex.

\subsubsection{Persistent homology}

We present a short overview of persistent homology. Suppose we have a population $P= \{ p_1, \dots, p_n\}$ of interest that we wish to analyze. We assume that the pairwise dissimilarities between any pair $p_i$ and $p_j$ is known. A convenient way to encode the system is through a \emph{simplicial complex} that has $P$ as its vertex set. One way of obtaining this goal is to fix a threshold $\varepsilon > 0$ and construct the simplicial complex $X_P^{\varepsilon} = (P, F_\varepsilon)$, where the vertices are given by $P$ and a subset $\sigma \subseteq P$ belongs to $F_\varepsilon$ whenever all its members are at most $\varepsilon$ apart. The first homology group $H_1(X_P^{\varepsilon})$ then records the 1-dimensional cycles in this complex, with its dimension giving the number of independent loops present.  

Choosing a single threshold $\varepsilon$ can be arbitrary, however. Persistent homology addresses this by examining how the homology evolves as $\varepsilon$ varies. Given a sequence of thresholds $\{\varepsilon_1 < \varepsilon_2 < \cdots < \varepsilon_N\}$, one obtains a filtration of simplicial complexes  
\[
X_P^{\varepsilon_1} \subseteq X_P^{\varepsilon_2} \subseteq \cdots \subseteq X_P^{\varepsilon_N}.
\]  
Applying homology to this nested sequence yields a diagram of vector spaces and linear maps,  
\begin{equation}
\label{eq:persistence}
H_1(X^\bullet_P): \; H_1(X_P^{\varepsilon_1}) \to H_1(X_P^{\varepsilon_2}) \to \cdots \to H_1(X_P^{\varepsilon_N}),
\end{equation}  
where each map arises from the inclusion of one simplicial complex into another, carrying cycles forward across scales. Persistent homology tracks when a homological feature (such as a cycle) first appears and when it disappears within this filtration.  

The lifespan of a feature is summarized by its \emph{birth} parameter $b$ and \emph{death} parameter $d$. Plotting the collection of points $(b,d)$ gives the \emph{persistence diagram}, a compact summary of the multi-scale topological structure. For a comprehensive treatment, see \cite{ghrist2008barcodes, edelsbrunner2008persistent, carlsson2009topologydata, edelsbrunner2002topological}

\begin{figure}[h!]
    \centering 
    \includegraphics[width=0.9\linewidth]{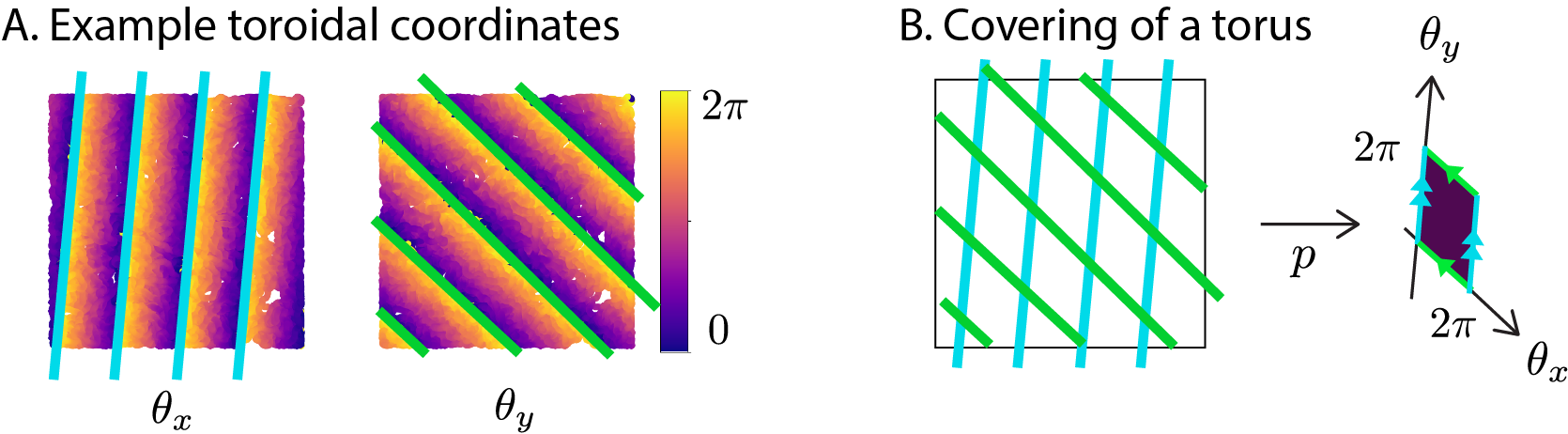}
    \caption{The toroidal coordinates define a tiling of $\mathbb{R}^2$ via parallelograms. \textbf{A}. An example visualization of the toroidal coordinates $(\theta_x, \theta_y)$. Here, given a time point $t$, let $(x, y)$ denote the location of the mouse at time $t$. Let $(\theta^t_x, \theta^t_y)$ be the toroidal coordinate assigned to population vector $P(t)$. The toroidal coordinates are visualized by a scatter plot in which a dot is placed at $(x,y)$ whose color value represents $\theta^t_x$ (left) and $\theta^t_y$ (right). \textbf{B}. The toroidal coordinates define a tiling of $\mathbb{R}^2$ via parallelograms. The map $p$ takes each parallelogram to one copy of the grid cell torus $S^1 \times S^1$. }
    \label{fig:parallelogram_tiles}
\end{figure}

\clearpage

\section{Comparison against null models of lifting}

To demonstrate that the proposed path-lifting procedure is essential to a faithful path reconstruction, we compared the paths reconstructed from our method against two null models: reconstruction without any lifting, and reconstruction with random lifting.

In the first null model (no lifting), the sequence of toroidal coordinates $\{ \Theta(t)\}$ is treated directly as the reconstructed path without any lifting. That is, $\tilde{\Theta}(t) = \Theta(t)$, and all lifted coordinates remain in a single tile (see Fig. 2 in main text). In the second null model (random lifting), at each time point $t+1$, the lifted coordinates $\Theta(t+1)$ are assigned to tile randomly chosen from the current tile and its eight neighbors.

\begin{figure}[b!]
    \centering
    \includegraphics[width=0.9\linewidth]{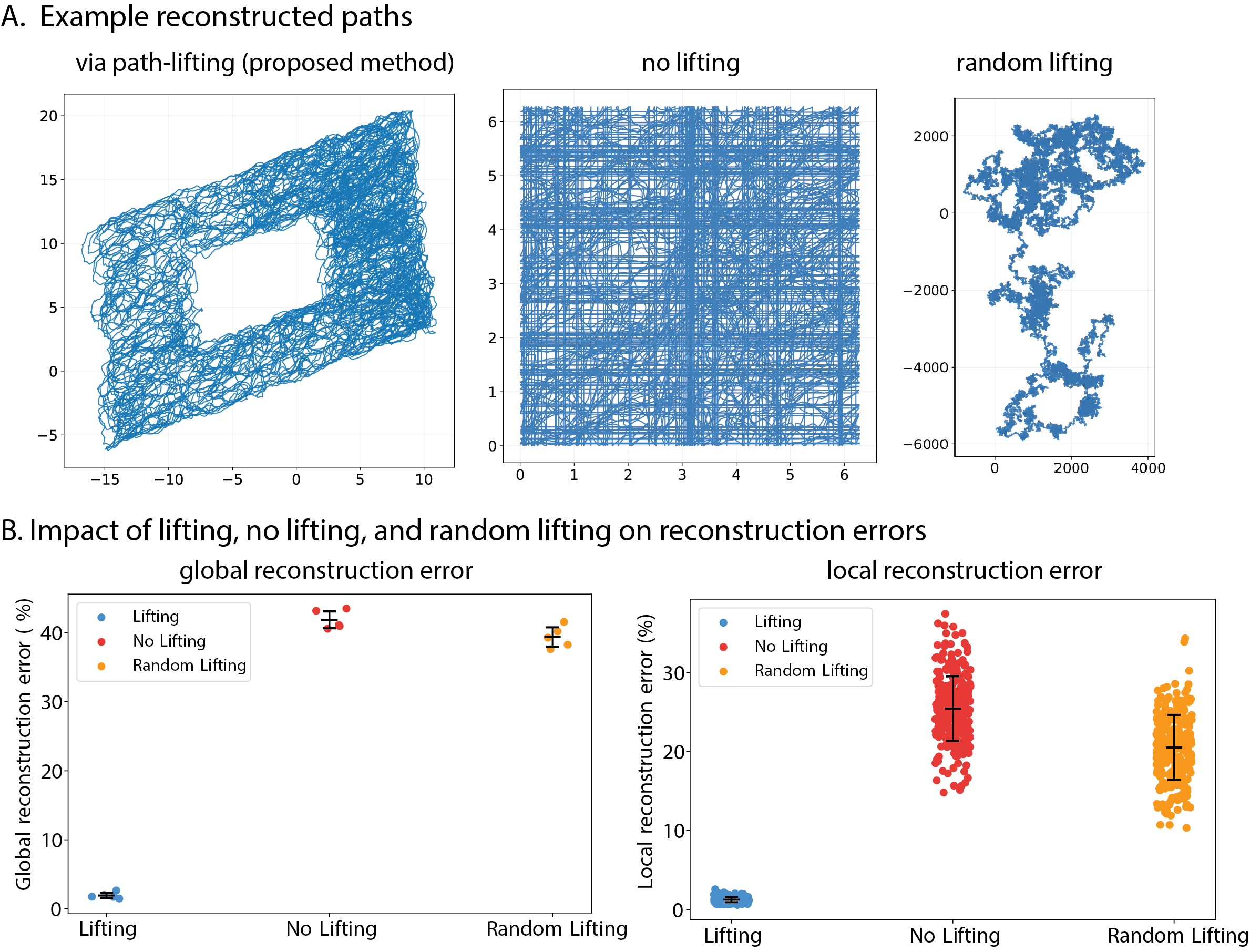}
    \caption{Comparison of proposed method against two null models (no lifting, random lifting) in the CAN-simulated data. \textbf{A.} Example  paths reconstructed via the proposed method and two null methods from a simulated trajectory in the one-hole world. \textbf{B}. Global (left) and local (right) reconstruction errors across 5 independent simulations. The proposed path-lifting algorithm achieves substantially lower error than both the no-lifting and random-lifting null models. Points show errors from individual simulations; markers and error bars indicate the mean $\pm$ 1 standard deviation (for global reconstruction error, mean and standard deviation were computed across 5 independent simulations; for local reconstruction error, they are computed across the $295$ local paths, 59 local paths per simulation).}
    \label{fig:random_lifting}
\end{figure}

We applied all three approaches — the proposed method, no lifting, and random lifting — to simulated trajectories in the one-hole environment (SI Fig.~\ref{fig:random_lifting}). As shown in SI Fig.~\ref{fig:random_lifting}A, only the proposed path-lifting algorithm recovers the hole in the environment; the two null models fail to reproduce the underlying topology of the movement path.

We repeated this experiment 5 times over independent simulations and report the global and local reconstruction errors in SI Fig.~\ref{fig:random_lifting}B. For the local reconstruction error, we take paths of length 10,000 time bins. Consistent with panel A, both null models yield reconstructions that deviate substantially from the true movement path, producing large global and local reconstruction errors.

\clearpage
\section{Robustness against noise in neural activity}

In this section, we analyze the robustness of the path reconstruction algorithm against noise in neural activity. We analyze three types of noise: spontaneous activity (SI Section~\ref{sec:spontaneous_firing}), neural activity suppression (SI Section~\ref{sec:information_deletion}), and time shifts in neural activity (SI Section~\ref{sec:time_shift}) in the CAN-simulated dataset. All reconstruction errors reported in this section are global reconstruction errors.

\subsection{Robustness against spontaneous activity}
\label{sec:spontaneous_firing}

We report the global reconstruction errors under varying levels of additional spontaneous activity of simulated grid cells. In the main text, we examine the robustness of the method under addition of one-dimensional Gaussian functions with peak height $h = 0.4$ (main text, Table 1). Here, we examine the reconstruction errors while varying the peak heights $h \in \{0.08, 0.2, 0.3,0.4\}$. SI Table~\ref{tab:spontaneous-twocol} summarizes the result. See SI Figure \ref{fig:dif-noise-data3} for example activity traces after the addition of Gaussian noise of height $h=0.4$. Note that in Table 1 of main text, we report the mean reconstruction error over 10 independent trials. Here, we report the error from a single trial.

\begin{table}[htbp]
\centering
\caption{Global reconstruction errors (\%) under added Gaussian noise with varying peak heights~$h$, proportions~$p$ of affected time points,
and standard deviations~$\sigma$ of the added Gaussian noise.
N/A indicates conditions where toroidal coordinates could not be computed.}
\label{tab:spontaneous-twocol}
 
\smallskip
\renewcommand{\arraystretch}{1.15}
 
 

  \begin{tabular}{@{}cl rrrr@{}}
    \toprule
     & & \multicolumn{4}{c}{\textbf{Std. Dev.} ($\sigma$)} \\
    \cmidrule(lr){3-6}
    $h$ & $p$ & 1 & 10 & 50 & 100 \\
    \midrule
       & 0.1\% & 1.96 & 1.68 & 1.78 & 1.78 \\
       & 0.5\% & 1.65 & 1.62 & 1.83 & 1.87 \\
      \multirow{5}{*}{0.08} & 1\% & 1.64 & 1.71 & 1.87 & 2.11 \\
       & 5\% & 1.62 & 1.84 & 43.69 & N/A \\
       & 10\% & 1.74 & 2.18 & N/A & N/A \\
    \midrule
       & 0.1\% & 1.59 & 1.61 & 1.65 & 1.72 \\
       & 0.5\% & 1.61 & 1.78 & 2.05 & 32.71 \\
      \multirow{5}{*}{0.2} & 1\% & 1.94 & 1.94 & 31.46 & 40.39 \\
       & 5\% & 1.77 & 22.45 & N/A & N/A \\
       & 10\% & 1.70 & 63.32 & 37.03 & N/A \\
    \bottomrule
  \end{tabular}\hspace{1.5em}
  \begin{tabular}{@{}cl rrrr@{}}
    \toprule
     & & \multicolumn{4}{c}{\textbf{Std. Dev.} ($\sigma$)} \\
    \cmidrule(lr){3-6}
    $h$ & $p$ & 1 & 10 & 50 & 100 \\
    \midrule
       & 0.1\% & 1.66 & 1.68 & 1.71 & 2.17 \\
       & 0.5\% & 1.62 & 1.73 & 25.79 & 45.80 \\
      \multirow{5}{*}{0.3} & 1\% & 1.66 & 1.88 & 71.48 & 41.49 \\
       & 5\% & 1.71 & 60.36 & N/A & N/A \\
       & 10\% & 16.25 & 40.72 & N/A & N/A \\
    \midrule
       & 0.1\% & 1.63 & 1.62 & 1.79 & 11.45 \\
       & 0.5\% & 1.63 & 1.82 & 43.88 & 40.82 \\
      \multirow{5}{*}{0.4} & 1\% & 1.86 & 2.06 & 39.47 & 43.11 \\
       & 5\% & 7.37 & 58.59 & N/A & N/A \\
       & 10\% & 55.23 & 38.31 & N/A & N/A \\
    \bottomrule
  \end{tabular}
\end{table}

\begin{figure}[h!]
    \centering
    \includegraphics[width=0.9\linewidth]{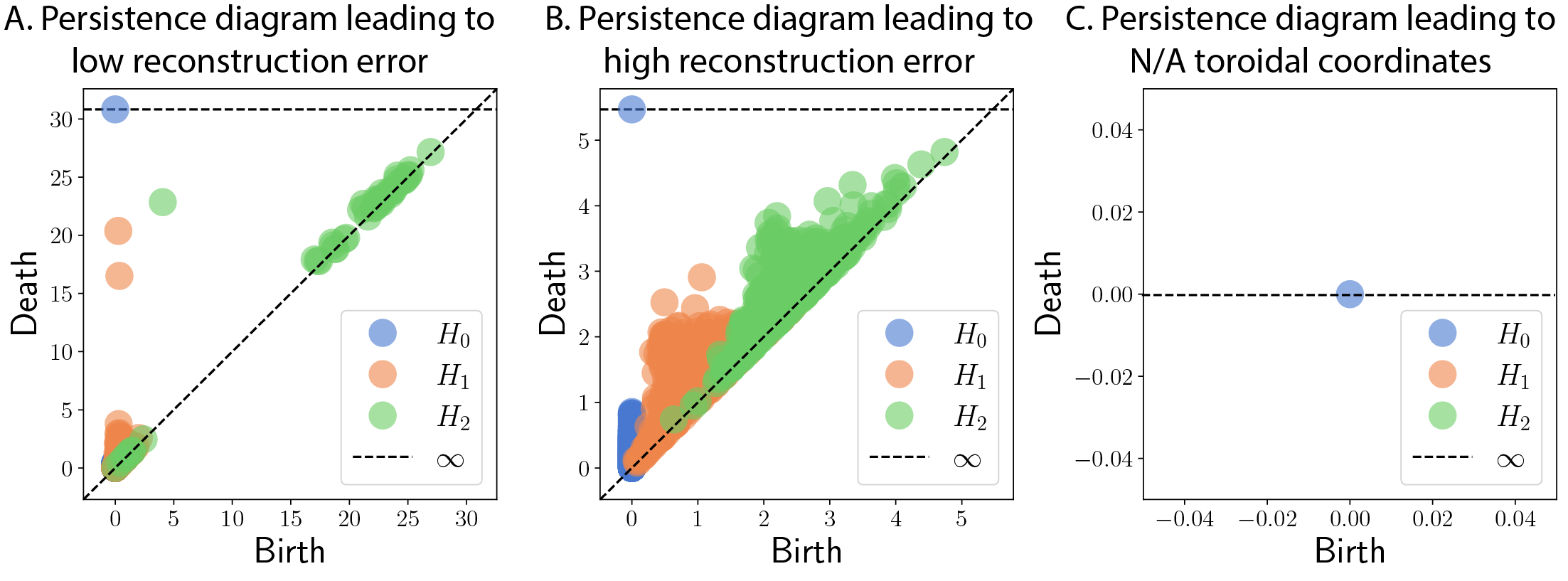}
    \caption{Example persistence diagrams that lead to low, high, and N/A reconstruction errors in the presence of spontaneous firings with peak height $h= 0.4$. \textbf{A}. Persistence diagram of the population vectors for $p = 5 \%$ and $\sigma = 1$ has a clear toroidal structure, which leads to a low reconstruction error. \textbf{B}. Persistence diagram of the population vectors for $p = 5\%$ and $\sigma = 10$ does not have a clear toroidal structure, which leads to a high reconstruction error.   \textbf{C}. Persistence diagram of the population vectors for $p = 5\%$ and $\sigma = 100$ has no non-trivial topological features. In such cases, the toroidal coordinates cannot be computed.  }
    \label{fig:noisy_experiments_PDs}
\end{figure}




\begin{figure}[h!]
    \centering
    \includegraphics[width=0.7\linewidth]{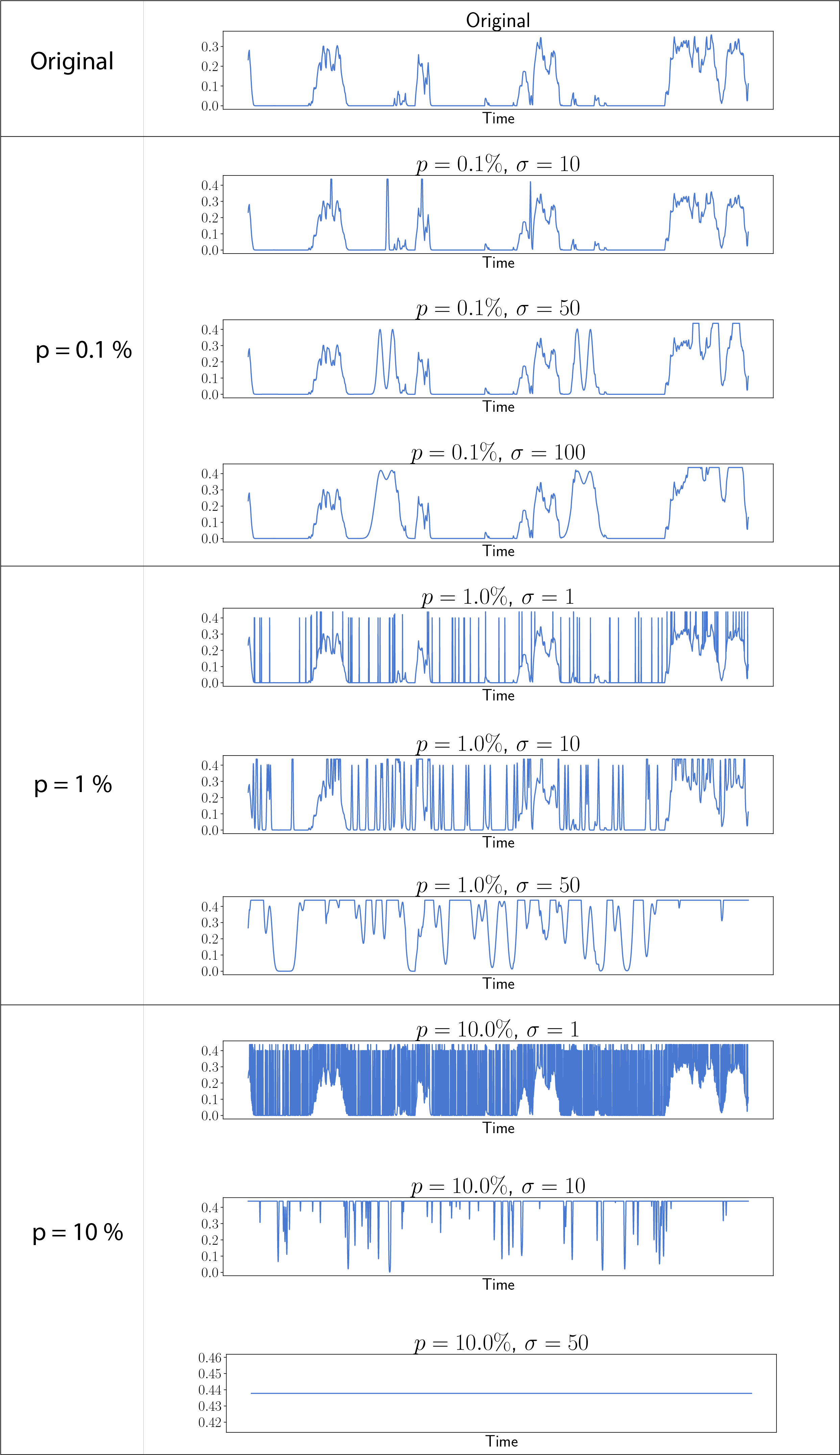}
    \caption{Simulated grid cell activities with added noise of different parameters of $p$ and $\sigma$. Peak height is fixed at  $h =0.4$.}
    \label{fig:dif-noise-data3}
\end{figure}

\clearpage

\subsection{Robustness against neural activity suppression}
\label{sec:information_deletion}

To test the robustness of the method against signal dropout or inhibitory transients in neural activity traces, we tested the path reconstruction under varying levels of suppression of simulated grid cell activity. From the CAN-simulated activity, we corrupt each neuron's activity trace as follows. A fraction $p$ (the portion parameter) of time points is selected uniformly at random without replacement. At each selected time point $t_i$, a Gaussian pulse

$$g_i(t) = h \exp\left(-\frac{(t - t_i)^2}{2\sigma^2}\right)$$
is subtracted from the activity trace, where $\sigma$ is the standard deviation controlling the temporal width of the dropout and $h$ is the peak height. The resulting trace is then clipped to $[0, r_{\max}]$, where $r_{\max}$ is the maximum magnitude of the original activity trace, so that no values become negative or exceed the original dynamic range. For varying $h, p$, and $\sigma$ values, we applied this suppression independently to each neuron's activity trace and performed path reconstruction. The results are summarized in SI Table~\ref{tab:deletion-twocol}. Note that for $h=0.4$, $p=10\%$, and $\sigma =10$, the reconstructed, transformed path occupied a space much larger than a square of size 100 $\times$ 100. The normalization factor of the reconstruction error (main text, Section 4.3.2) is $S = 100$. Because the transformed path occupies a window larger than $100 \times 100$, dividing by $S = 100$ results in a reconstruction error that is larger than $100\%$.

\begin{figure}[h!]
    \centering
    \includegraphics[width=0.7\linewidth]{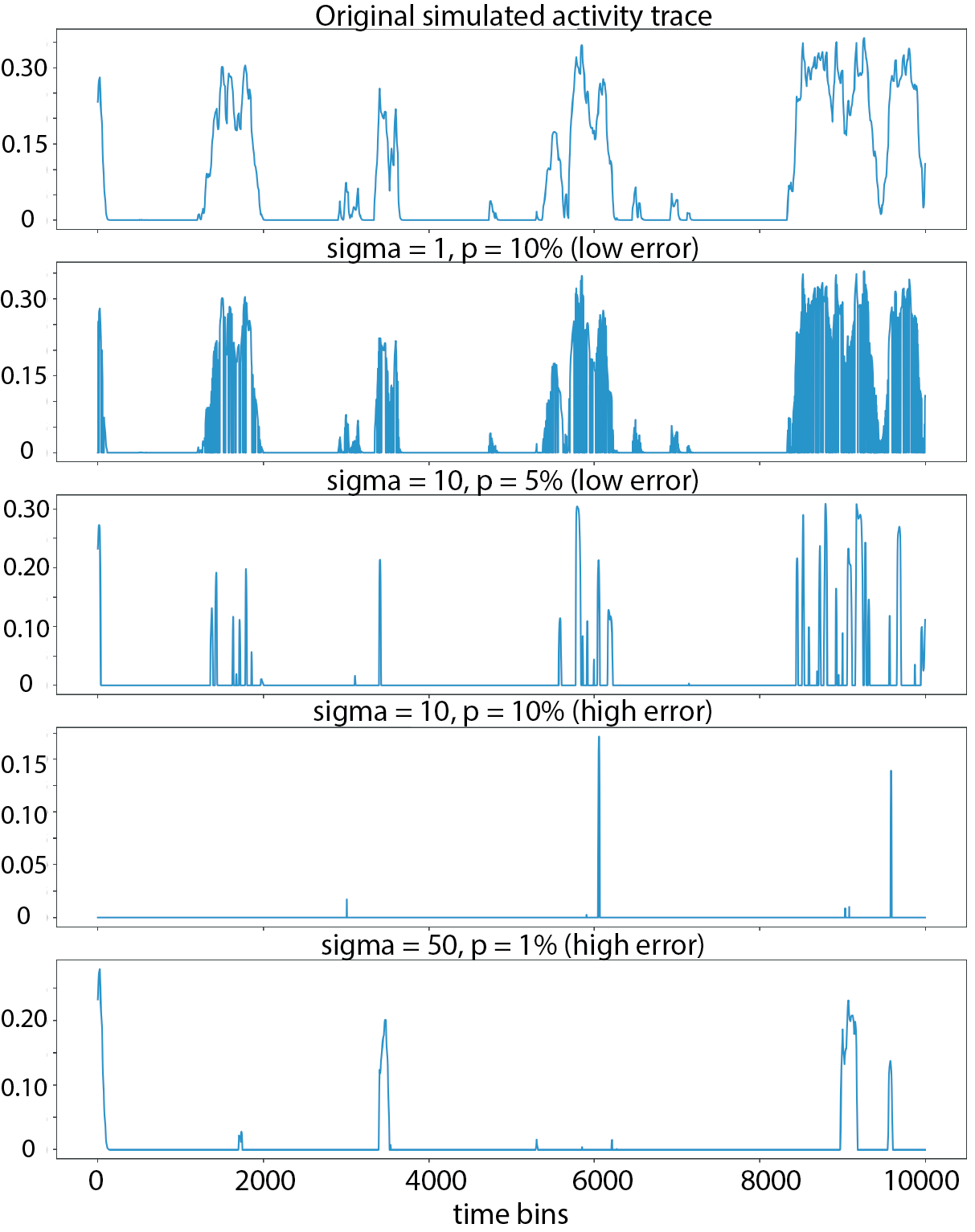}
    \caption{Original simulated neural activity trace and its modification after neural activity suppression that leads to low and high reconstruction error.}
    \label{fig:example_Gaussian_subtraction}
\end{figure}

See SI Fig.~\ref{fig:example_Gaussian_subtraction} for visualizations of simulated neural activity traces and the modified traces that lead to high and low reconstruction errors.

\begin{table}[htbp]
\centering
\caption{Reconstruction errors (\%) under neural activity suppression for varying peak heights~$h$, proportions~$p$ of affected time points, and standard deviations~$\sigma$ of the subtracted Gaussian pulses.
N/A indicates conditions where toroidal coordinates could not be computed.}
\label{tab:deletion-twocol}
 
\smallskip
\renewcommand{\arraystretch}{1.15}
  \begin{tabular}{@{}cl rrrr@{}}
    \toprule
     & & \multicolumn{4}{c}{\textbf{Std. Dev.} ($\sigma$)} \\
    \cmidrule(lr){3-6}
    $h$ & $p$ & 1 & 10 & 50 & 100 \\
    \midrule
       & 0.1\% & 1.69 & 1.70 & 1.73 & 1.71 \\
       & 0.5\% & 1.73 & 1.87 & 1.71 & 1.68 \\
      \multirow{5}{*}{0.08} & 1\% & 1.65 & 1.66 & 1.67 & 1.73 \\
       & 5\% & 1.65 & 1.63 & 44.01 & 40.24 \\
       & 10\% & 1.70 & 1.68 & N/A & N/A \\
    \midrule
       & 0.1\% & 1.68 & 1.67 & 1.66 & 1.72 \\
       & 0.5\% & 1.67 & 1.59 & 1.90 & 1.82 \\
      \multirow{5}{*}{0.2} & 1\% & 1.64 & 1.59 & 1.95 & 2.08 \\
       & 5\% & 1.66 & 1.75 & 43.18 & N/A \\
       & 10\% & 1.59 & 2.08 & N/A & N/A \\
    \bottomrule
  \end{tabular}\hspace{1.5em}
  \begin{tabular}{@{}cl rrrr@{}}
    \toprule
     & & \multicolumn{4}{c}{\textbf{Std. Dev.} ($\sigma$)} \\
    \cmidrule(lr){3-6}
    $h$ & $p$ & 1 & 10 & 50 & 100 \\
    \midrule
       & 0.1\% & 1.63 & 1.60 & 1.68 & 1.64 \\
       & 0.5\% & 1.75 & 1.64 & 1.77 & 24.09 \\
      \multirow{5}{*}{0.3} & 1\% & 1.60 & 1.66 & 1.91 & 50.40 \\
       & 5\% & 1.60 & 1.83 & 66.17 & N/A \\
       & 10\% & 1.62 & 76.02 & N/A & N/A \\
    \midrule
       & 0.1\% & 1.67 & 1.60 & 1.70 & 1.79 \\
       & 0.5\% & 1.60 & 1.61 & 1.82 & 2.01 \\
      \multirow{5}{*}{0.4} & 1\% & 1.64 & 1.66 & 24.39 & 35.65 \\
       & 5\% & 1.63 & 2.02 & N/A & N/A \\
       & 10\% & 1.67 & 188.44 & N/A & N/A \\
    \bottomrule
  \end{tabular}
\end{table}

\clearpage

\subsection{Robustness against time shifts of neural activity}
\label{sec:time_shift}

The decoding pipeline assumes that the grid cell activity matrix is temporally aligned, i.e., column $t$ of the activity matrix corresponds to time step $t$ for all neurons simultaneously. To test robustness to violations of this assumption (for instance, arising from clock drift or imprecise spike-sorting alignment across tetrodes), we applied independent random circular shifts to each simulated neuron's activity trace before path reconstruction. Specifically, given a maximum shift amount $d_{\max}$, for each neuron $i$, a temporal shift amount $\delta_i$ was drawn uniformly at random from $[-d_{\max}, d_{\max}]$ (integer-valued), and the trace was cyclically shifted by $\delta_i$ time steps via \texttt{np.roll}. Note that the shift amounts are in time bin units, out of a total of 600,000 time bins. See SI Fig.~\ref{fig:examples_shifted_neurons} for visualizations of two example neurons and their activity shifted by various time steps. 

We considered various maximum shift parameters $d_{\max} \in \{0, 10, 20, 50, 100, 200, 500, 1000\}$. For each shift amount $d_{\max}$, we shifted each neuron's activity trace as described above and performed path reconstruction on the shifted activities. We repeated this experiment on 5 independent simulations.

\begin{figure}[h!]
    \centering
    \includegraphics[width=\linewidth]{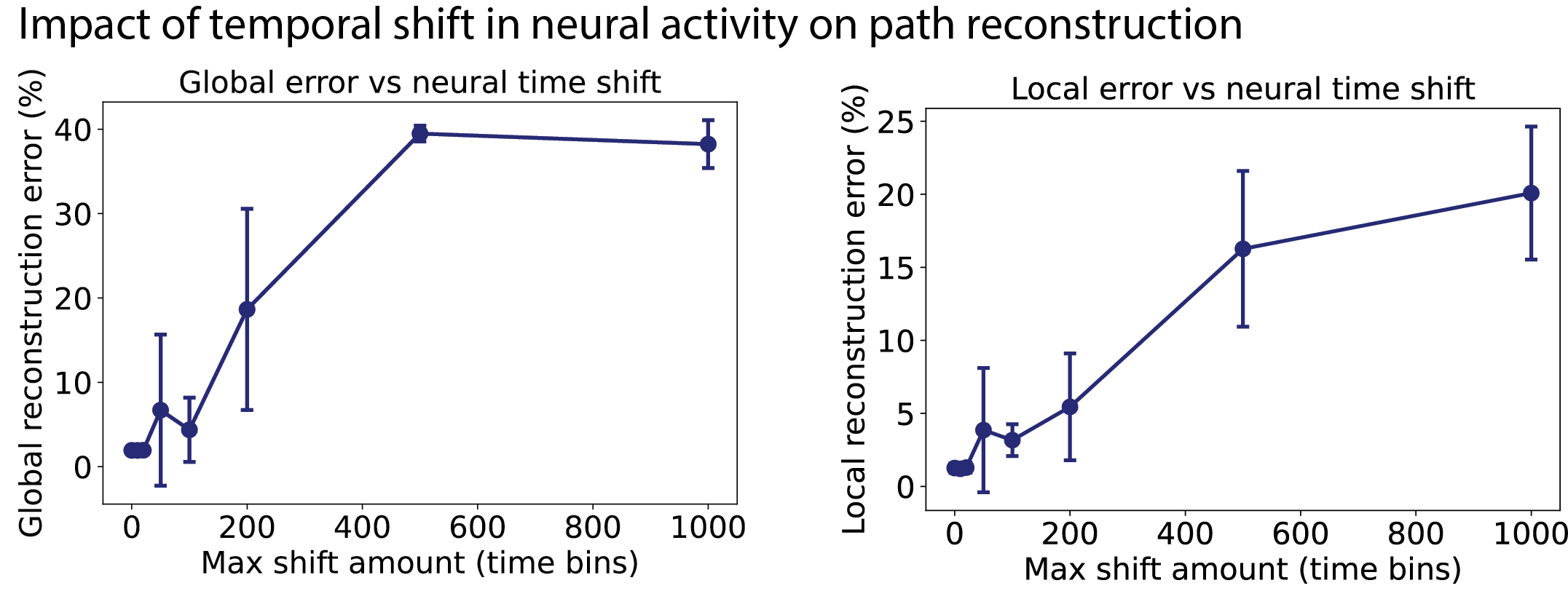}
    \caption{Effect of temporal shift in neural activity on path reconstruction in simulated data. Each neuron's activity trace was independently circularly shifted by an integer sampled uniformly from $[-d_{\max}, d_{\max}]$, and the shifted population activity was used to perform path reconstruction. Results are shown for 5 independent repeats in the 1-hole simulated environment. Error bars denote standard deviation across repeats. (Left) Global reconstruction error as a function of maximum shift magnitude $d_{\max}$. At $d_{\max} = 500$, the reconstruction error approaches that of the null models. 
    (Right) Local reconstruction error as a function of $d_{\max}$. Each local path was 10,000 time bin long.}
    \label{fig:temporal_shift}
\end{figure}

\begin{figure}
    \centering
    \includegraphics[width=0.8\linewidth]{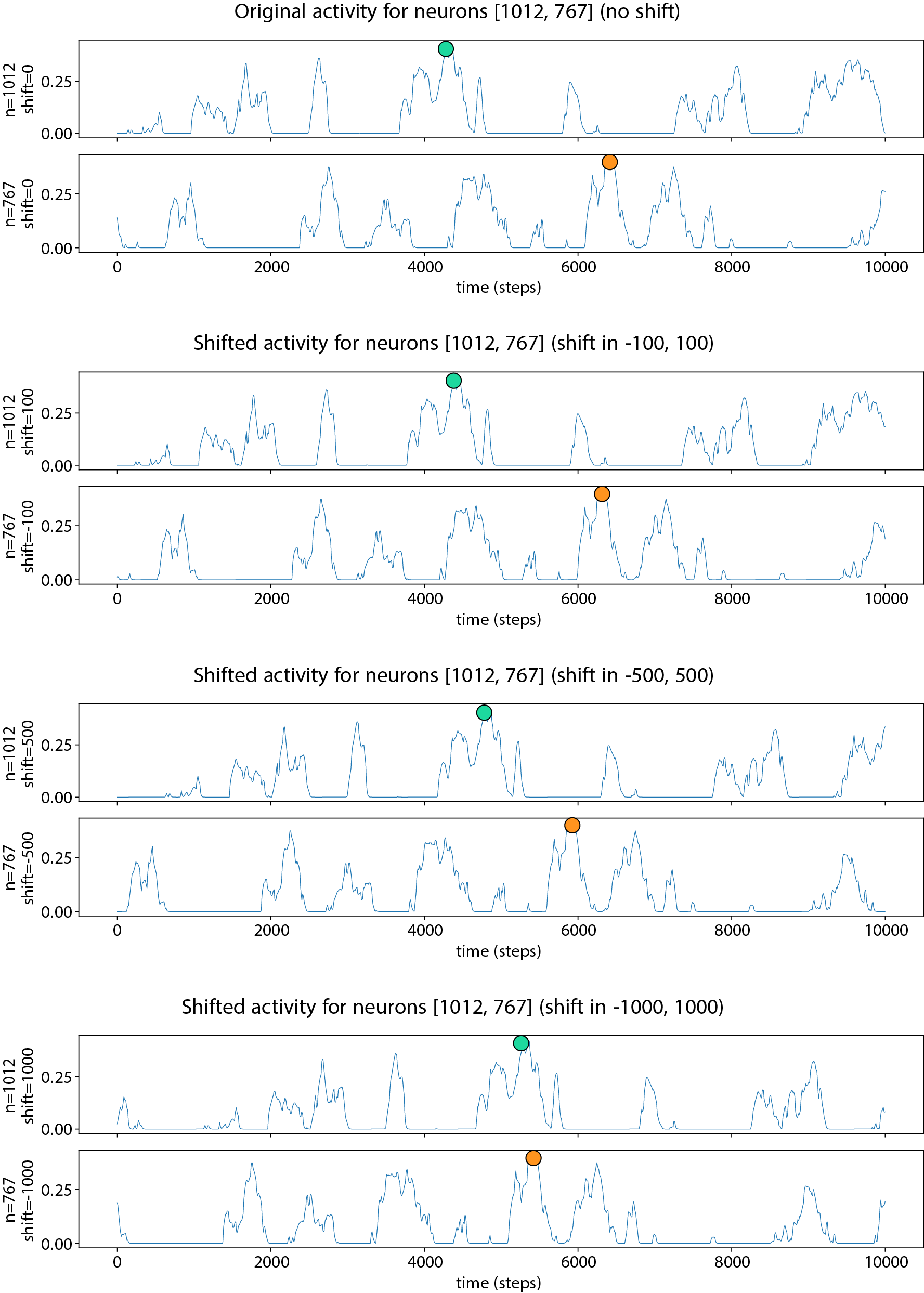}
    \caption{Example activity of simulated grid cells (neurons 1012 and 767) under deterministic temporal shifts. For each shift magnitude $d_{max} \in \{ 100, 500, 1000\}$, neuron 1012 is shifted by $+d_{max}$ and neuron 767 by $-d_{max}$, and the first 10,000 time points are shown to illustrate how increasing shift size alters the alignment of the two neural activities. For ease of comparison, the maximal neural activities of each neuron is marked with teal and orange circles. }
    \label{fig:examples_shifted_neurons}
\end{figure}

Both the global and local reconstruction error remained low for maximum shift distance of up to $d_{max} = 100$ time bins, indicating that the pipeline tolerates moderate asynchrony across neurons without appreciable degradation (SI Fig.~\ref{fig:temporal_shift}). Performance began to decline noticeably at $d_{max} = 500$. These results demonstrate that the method is robust to small-to-moderate temporal misalignment, while sufficiently large desynchronization disrupts the toroidal structure for path reconstruction.

\clearpage

\section{Impact of various factors on global and local reconstruction errors}

In this section, we analyze the various factors that can potentially impact global and local reconstruction errors. The factors analyzed are proximity parameter epsilon (SI Section~\ref{sec:epsilon}), number of time points (SI Section~\ref{sec:timepoints}), experiment duration (SI Section~\ref{sec:error_accumulation}), number of neurons (SI Section~\ref{sec:num_neurons}), metric (SI Section~\ref{sec:metric}), noise in toroidal coordinates (SI Section~\ref{sec:noise_toroidal_coord}), and smoothing of reconstructed paths (SI Section~\ref{sec:smoothing}). 

Throughout this section, all analysis was performed on five independent simulations on one-hole environment. When presenting the global reconstruction errors, the points and error bars indicate the mean and standard deviation across the five simulations. For the local reconstruction errors, all local paths had length 10,000 time bins (out of 599,999 total time bins). The mean and standard deviation are computed across all 295 local paths (59 local paths per simulation).

Whenever applicable, we also performed an analogous analysis on the two-dimensional experimental data \cite{gardner_toroidal_2022} (111 grid cells from rat R, module 1, day 2, open-field session). The local paths of the two-dimensional experimental data consisted of 20-second intervals. In all figures reporting the local reconstruction errors from the two-dimensional experimental dataset, the points and error bars indicate the mean and standard deviation across all local paths.

\subsection{Impact of proximity parameter epsilon}
\label{sec:epsilon}
\begin{figure}[h!]
    \centering
    \includegraphics[width=\linewidth]{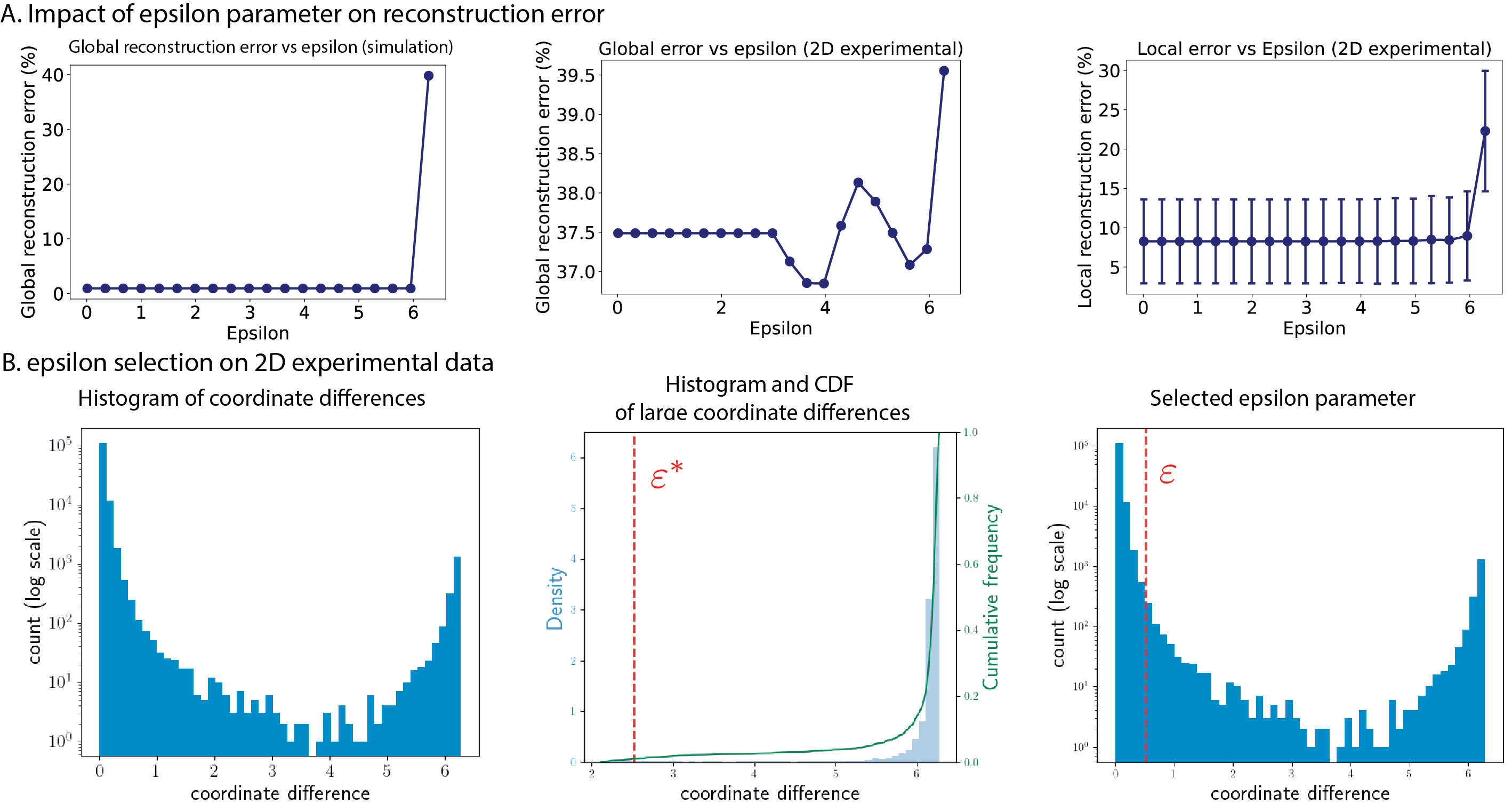}
    \caption{Impact of proximity parameter $\varepsilon$ on reconstruction error and selection of $\varepsilon$. \textbf{A.} Analysis of impact of epsilon parameter on the global reconstruction errors on the CAN-simulated dataset (left) and in two-dimensional experimental data (center). Impact of epsilon parameter on the local reconstruction error in two-dimensional experimental data (right).
    \textbf{B.} $\varepsilon$ selection for two-dimensional experimental data. (Left) A histogram of the maximal coordinate differences. (Center) A zoomed in view of the coordinate differences restricted to the range $[2, 2 \pi)$. The green curve shows the cumulative sum. The red dotted line indicates the parameter $\varepsilon^*$ at which 99\% of the maximal coordinate differences exceed $\varepsilon^*$. (Right) As observed in panel A, smaller epsilon parameters generally lead to better reconstruction. We choose $\varepsilon = \varepsilon^* - 2$. 
    }
    \label{fig:epsilon}
\end{figure}

In the path reconstruction algorithm, whether two consecutive toroidal coordinates should be tested for potential lifts is determined by a proximity parameter $\varepsilon$ (Equation~2, main text). We analyzed the impact of varying this parameter on the simulated dataset (SI Fig.~\ref{fig:epsilon}A, left) and the two-dimensional experimental dataset (SI Fig.~\ref{fig:epsilon}A, center, right). Smaller epsilon parameters lead to lower reconstruction errors, and the errors are stable for a wide range of epsilon parameters.

\subsection{Impact of number of time points}
\label{sec:timepoints}

\begin{figure}[h!]
    \centering
    \includegraphics[width=0.7\linewidth]{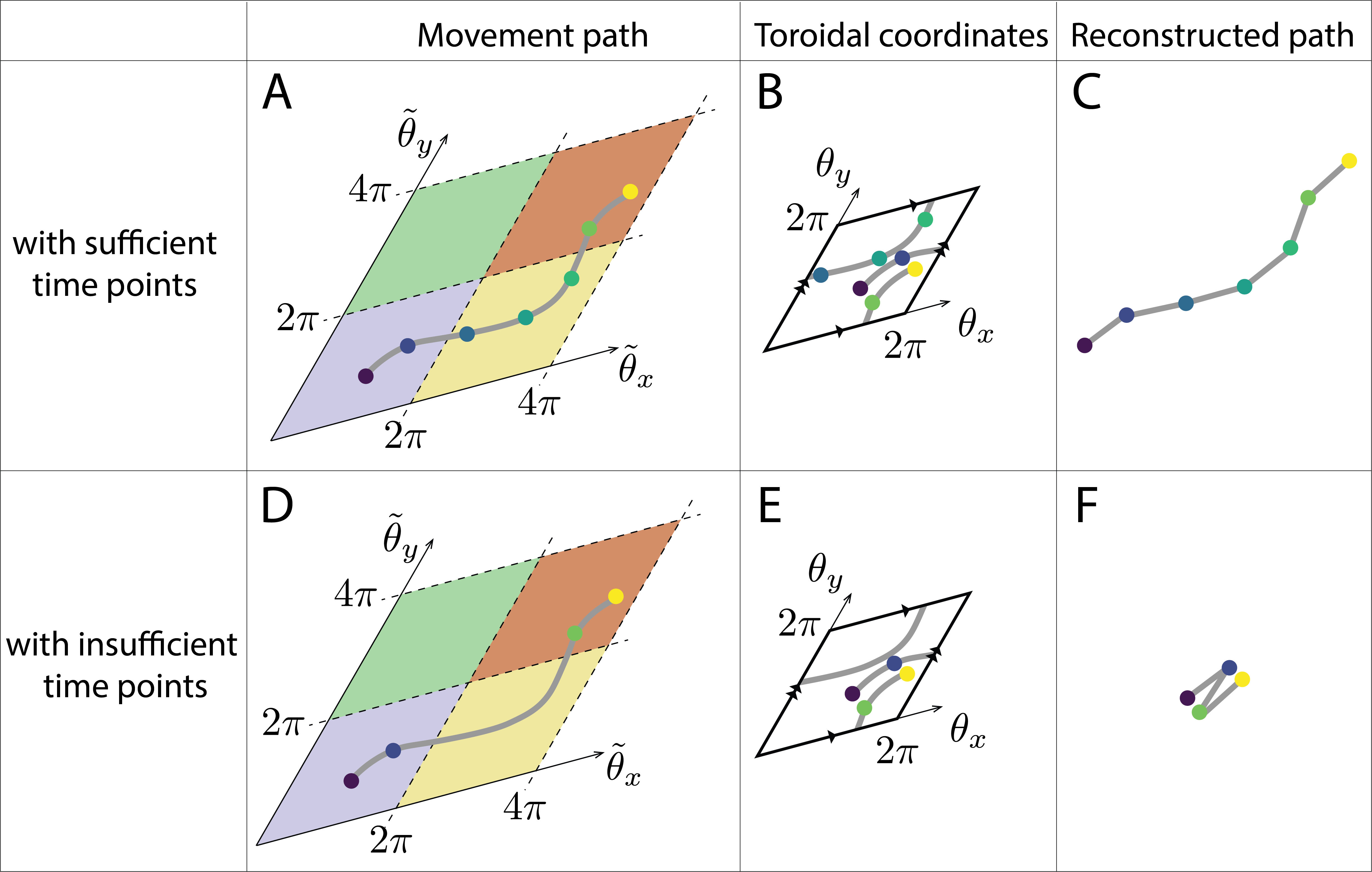}
    \caption{Having insufficient time points leads to poor path reconstruction \textbf{A.} The movement path in $\mathbb{R}^2$ is shown in grey, and the time points at which data is collected are indicated via colored circles. Here, the movement path goes from lower left corner to the upper right corner. \textbf{B.} The observed toroidal coordinates are shown in colored dots. \textbf{C.} The reconstructed path. \textbf{D.} Data is collected at four time points indicated by the colored dots. \textbf{E.} The observed toroidal coordinates are shown via colored dots. \textbf{F}. In this case, the algorithm fails to lift between the 2nd and 3rd time point, resulting in a reconstructed path that does not resemble the original movement path in panel \textbf{D}.   
    }
    \label{fig:insufficient_times_cartoon}
\end{figure}

A faithful path reconstruction requires that the grid cell population activity is observed at sufficient number of time points. Having insufficient number of time points can lead to poor path reconstructions as illustrated in SI Fig.~\ref{fig:insufficient_times_cartoon}. 

In order to analyze the impact of number of time points in the reconstruction error, we performed the following experiments on the simulated grid cell activity from the 1-hole world. 

\subsubsection{Impact of movement speed in simulated movements}
When simulating the mouse trajectory in a bounded environment of size $100 \times 100$, 
the movement speed is controlled by the parameter $s_{\max}$: at every trajectory step, 
the distance traveled is sampled uniformly from $[0, s_{\max}]$ spatial units. Larger 
values of $s_{\max}$ produce faster, more spatially diffuse trajectories, while smaller 
values produce slower, more locally concentrated paths.

\begin{figure}[h!]
    \centering
    \includegraphics[width=0.85\linewidth]{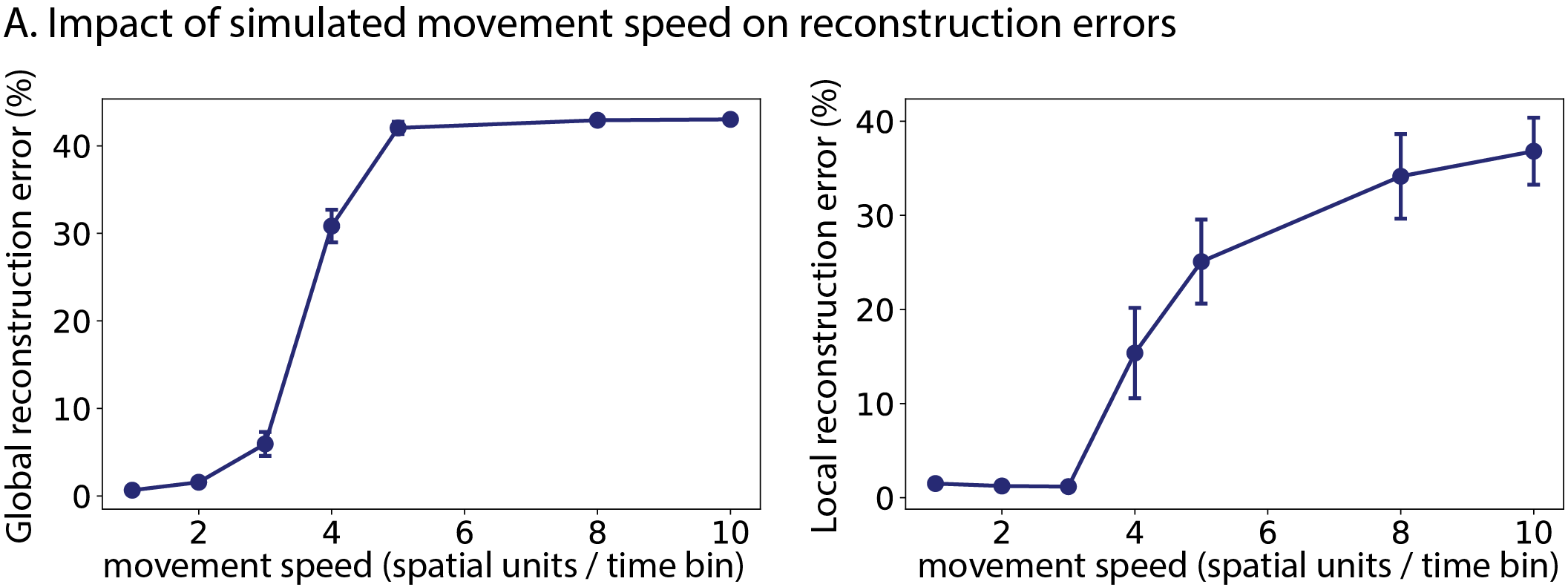}
    \caption{Global (left) and local (right) reconstruction errors as a function of movement speed $s_{\max}$ (spatial units per trajectory step) in the CAN-simualted data. Each point shows the mean $\pm$ std across 5 independent trajectories. Both errors remain near zero for $s_{\max} \leq 3$ and increase sharply for $s_{\max} \geq 4$, indicating that path reconstruction is robust to moderate movement speeds but degrades when the simulated movement speed increases.}
    \label{fig:step_size_only}
\end{figure}
To assess the sensitivity of path reconstruction to movement speed, we experimented with 
$s_{\max} \in \{1, 2, 3, 4, 6, 8, 10\}$ spatial units per trajectory step. For each $s_{\max}$ value, we simulated a movement trajectory of 25,000 steps, simulated grid cell activity (which resulted in 599,999 time bins), and performed path reconstruction using our pipeline. We repeated this process 5 times per $s_{\max}$ value. The global and local reconstruction errors are reported in SI Fig.~\ref{fig:step_size_only}. 

While this particular simulation studied movement speed in the simulated trajectory, biologically realistic speeds are unlikely to substantially impair reconstruction performance, since neural processes operate on a much finer timescale. Therefore, this analysis should be considered as a way to examine how the physical distance traveled between sampled time points affects path reconstruction. In real data, the determining factor will be the temporal sampling rate rather than animal speed.

\subsubsection{Impact of uniform temporal subsampling of grid cell activity}

Neural recordings are typically acquired at a fixed sampling rate that does not necessarily coincide with the temporal resolution of the simulations. To assess sensitivity to temporal resolution, a simulated trajectory and grid cell activity, consisting of 599,999 time points, were downsampled to retain every $k$-th time point, for $k \in \{1, 50, 100, 200, 300, 400, 500\}$. We then performed path reconstruction from the downsampled grid cell activity. Each condition was repeated $5$ times using independent simulations of a 1-hole environment. 

For the local reconstruction errors, we scaled the length of local paths proportionally as $\ell = \lfloor 10{,}000 / k \rfloor$, ensuring that each local path corresponds to the same time window regardless of $k$. 

Path reconstruction quality was robust across a wide range of downsampling factors (see SI Fig.~\ref{fig:impact_speed_timepoints}A). Both global and local reconstruction error remained stable for $k \leq 200$. A sharp increase in global reconstruction error occurred at $k = 300$ ($28\% \pm 5\%$). These results indicate that the path reconstruction pipeline tolerates substantial reductions in temporal resolution — up to 200-fold subsampling — before performance degrades, suggesting robustness to practical variations in recording frame rate.

We repeated a similar analysis on the two-dimensional experimental data \cite{gardner_toroidal_2022}. Starting from the full-resolution recording of $126,728$ time bins, we uniformly subsampled both the neural activity time bins and ground-truth trajectory by factors of $k \in \{1, 2, 5, 10, 20, 50, 100, 150\}$. To each downsampled dataset, we performed path reconstruction and reported both global and local reconstruction errors. For the local reconstruction errors, recall that the original analysis (without any downsampling) utilized local paths of length 2,000 time bins. The length of the local paths were proportionally shortened to $\lfloor 2,000 / k\rfloor $. From SI Fig.~\ref{fig:impact_speed_timepoints}A (bottom row), one can see that downsampling causes local reconstruction errors to increase beyond $k = 20$. 

\subsubsection{Impact of non-uniform temporal subsampling of grid cell activity}

We then tested robustness of the method to irregular temporal sampling. Inter-sample intervals were drawn from a Poisson distribution with mean $\lambda$, and their cumulative sum was used to identify the subsampled time points. We downsampled the grid cell activity with varying levels of $\lambda \in \{1, 50, 100, 200, 300, 400, 500\}$ and performed path reconstruction on the downsampled activity. As in the uniform downsampling analysis, the lengths of local paths were scaled proportionally as $\ell = \lfloor 10{,}000 / \lambda \rfloor$. 

\begin{figure}[h!]
    \centering
    \includegraphics[width=\linewidth]{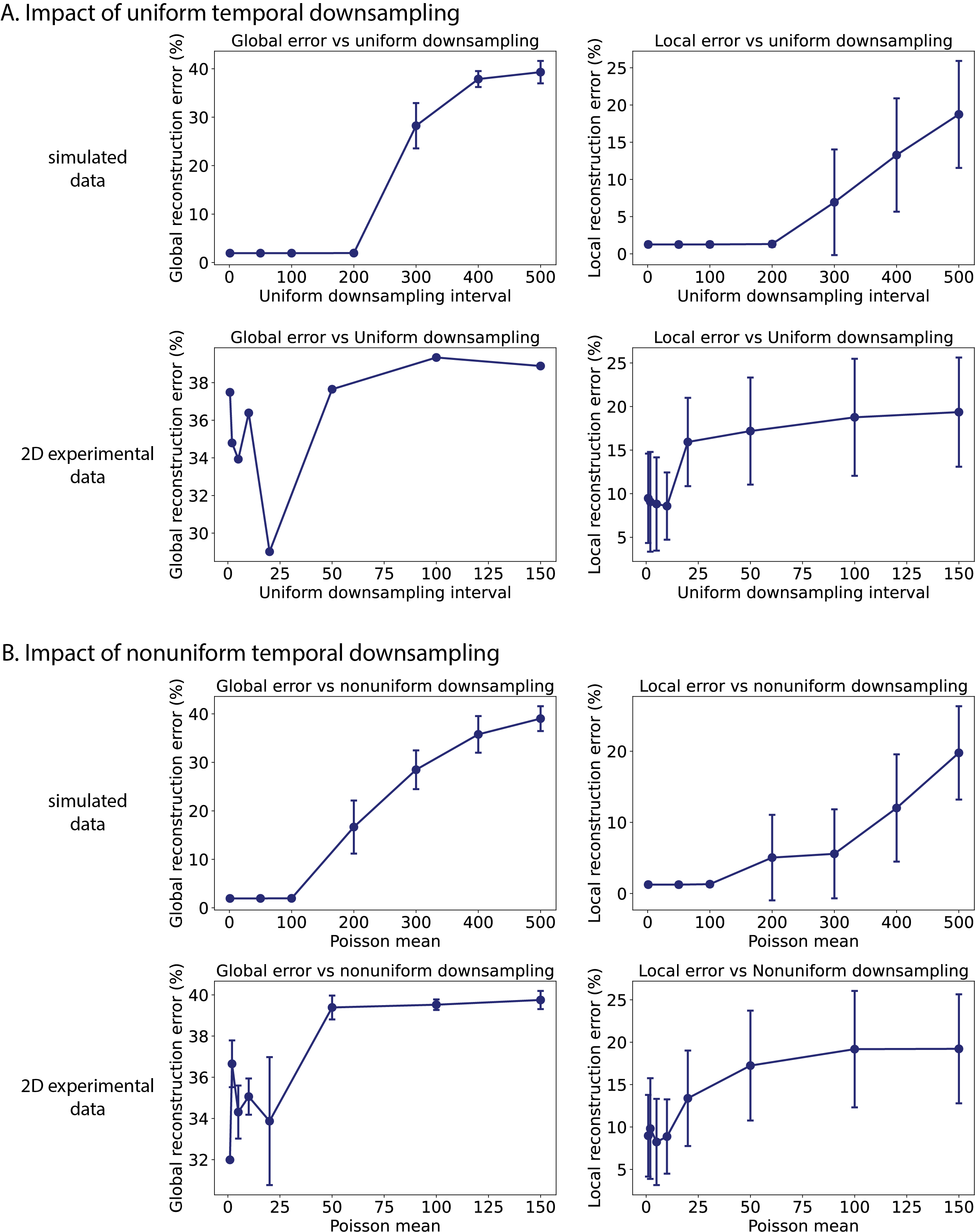}
    \caption{Impact of temporal subsampling on reconstruction errors. \textbf{A.} Global (left) and local (right) reconstruction errors as a function of uniform downsampling interval $k$, for simulated data (top row) and two-dimensional experimental data (bottom row). \textbf{B.} Errors as a function of non-uniform (Poisson) subsampling with mean interval $\lambda$. Figures show that simulated data can tolerate up to 200-fold temporal subsampling in the simulated dataset and 20-fold temporal subsampling in the experimental dataset. }
    \label{fig:impact_speed_timepoints}
\end{figure}

Similarly to the uniform-downsampling experiments, the global and local reconstruction error remained stable for a wide range of $\lambda$, with performance degrading starting at $\lambda = 200$ (see SI Fig.~\ref{fig:impact_speed_timepoints}B, top). Compared to uniform downsampling (SI Fig.~\ref{fig:impact_speed_timepoints}A, top), the degradation onset is somewhat earlier. 

We repeated a similar analysis on the two-dimensional experimental data \cite{gardner_toroidal_2022}. For each $\lambda \in \{1, 2, 5, 10, 20, 50, 100, 150\}$, we generated 5 independent subsampling of the recorded neural activity by drawing inter-sample intervals from a Poisson distribution with mean $\lambda$. We then performed path reconstruction on each subsampled dataset.

For the analysis of local reconstruction error, we scaled the local path length proportionally as $\ell_{\text{seg}} = \lfloor 2000 / \lambda \rfloor$ as before. Similarly to the uniform downsampling analysis, both errors remained stable up to $\lambda = 20$ (see SI Fig.~\ref{fig:impact_speed_timepoints}B, bottom). Beyond $\lambda = 20$, both global and local reconstruction errors increased gradually. This analysis indicates a modest degradation in local path reconstruction as the neural recording becomes more sparse.

\clearpage

\subsection{Error accumulation for long paths and impact of experiment duration}
\label{sec:error_accumulation}

In the analysis of two-dimensional experimental data \cite{gardner_toroidal_2022}, the reconstruction of the rat's full trajectory failed to resemble its actual movement, even though shorter segments of the same trajectory were reconstructed faithfully. This discrepancy has two possible sources. First, experimental recordings are noisy, and this noise can cause the path-lifting algorithm to make errors. Second, and more importantly, even a small number of local errors can accumulate, producing a global reconstruction whose shape differs substantially from the original path. 


A reconstruction can be a decent reconstruction on local paths and still be globally distorted, because errors during path reconstruction can propagate. SI Fig.~\ref{fig:error_accumulation_cartoon} illustrates how the two types of lifting error (main text, Fig. 8) distort the global shape of a reconstructed path, even when each local segments is lifted correctly except at the segment boundaries. When the first type of error occurs, toroidal coordinates that should be lifted to the same tile are instead lifted to different tiles. This causes the reconstructed path to be more stretched out than the true movement path. When the second type of error occurs, toroidal coordinates that should be lifted to different tiles are instead placed in the same tile, leading to an overly compressed reconstruction. In both cases, errors occur at only a small number of points, yet the resulting reconstruction differs in global geometry from the true movement path.

\begin{figure}[h!]
    \centering
    \includegraphics[width=\linewidth]{figures/error_accumulation_cartoon.png}
    \caption{A small number of lifting errors can distort the global shape of a reconstructed path. Each panel shows an example movement path and the two types of lifting error that can occur during reconstruction. Throughout, the dotted parallelograms indicate the tiles produced by the covering map $p: \mathbb{R}^2 \to S^1 \times S^1$. Each parallelogram is mapped onto a torus under $p$. \textbf{A}. An example movement path. \textbf{B}. Errors of the first type (over-stretching). Assume that each colored segment of the path is reconstructed correctly, except at the two locations marked by red crosses. At those points, the toroidal coordinates should have been lifted into the same tile as the neighboring segment, but the algorithm lifted them into different tiles. (Left) The correct reconstruction. (Right) The reconstruction produced under this error, which is more stretched out than the correct reconstruction.
  \textbf{C}. Errors of the second type (over-compression). Conversely, assume that at the two marked locations the toroidal coordinates were lifted into the same tile, when they should have been lifted into distinct tiles. (Left) The correct reconstruction. (Right) The reconstruction produced under this error, which is more compact than the correct reconstruction.}
    \label{fig:error_accumulation_cartoon}
\end{figure}

\begin{figure}
    \centering
    \includegraphics[width=0.7\linewidth]{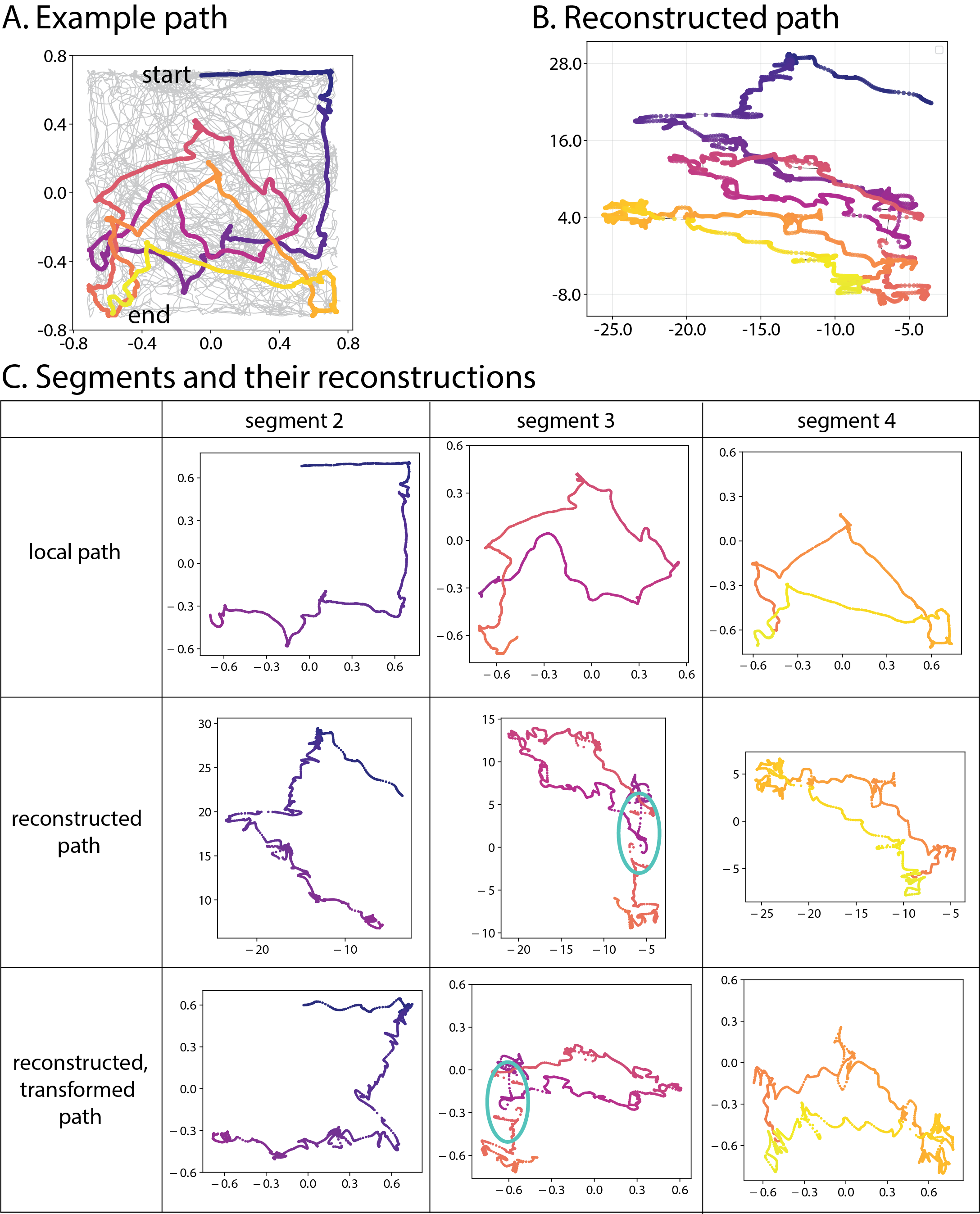}
    \caption{
    Error accumulation in the two-dimensional experimental data \cite{gardner_toroidal_2022}. We highlight a portion of the rat's trajectory consisting of three consecutive 20-second segments (labeled 2, 3, 4), and compare its reconstruction segment-by-segment and as a whole. The rat moved from navy to purple to yellow across the three segments. \textbf{A}. The full movement path in $\mathbb{R}^2$ (gray), with segments 2–4 highlighted. \textbf{B}. The reconstruction of the highlighted portion. Its global geometry is clearly distorted relative to the original: for example, segment 4 (yellow) visits regions occupied by segments 2 (navy) and 3 (purple) in the original path, but does not intersect their reconstructions here. \textbf{C}. Segment-by-segment comparison. For each of segments 2, 3, and 4: (top row) the original local path, (middle row) its reconstruction, and (bottom row) the reconstruction after aligning to the original coordinate frame. Each segment's local reconstruction faithfully preserves the per-segment shape. The reconstruction of segment 3, however, contains a long ``discontinuous" tail (teal highlights in middle and bottom rows), indicating that a portion of the segment may have experienced the first type of error. This lengthened displacement in segment 3 propagates forward, causing segment 4 to be reconstructed in a region that no longer overlaps with segments 2 and 3, thereby distorting the global shape shown in panel B.}  
    \label{fig:error_accumulation}
\end{figure}

We observe this form of error accumulation in the two-dimensional experimental data. In SI Fig.~\ref{fig:error_accumulation}A, we highlight a portion of the rat's movement path consisting of three consecutive 20-second segments (labeled 2, 3, 4). SI Fig.~\ref{fig:error_accumulation}B shows the reconstruction of the highlighted portion; its geometry is clearly distorted relative to the original. For example, in the original path, segment 3 (yellow) visits regions occupied by segment 1 (navy) and segment 2 (purple), whereas in the reconstruction these segments no longer overlap.

The source of the distortion becomes apparent in SI Fig.~\ref{fig:error_accumulation}C, which shows each segment together with its reconstruction. Each segment is reconstructed faithfully locally: the per-segment shape is preserved. However, the reconstruction of segment 3 appears to have been lifted into a different set of tiles than it should have been, producing a long tail (teal highlight in middle and bottom rows) that pulls segment 4 away from segments 2 and 3 in the global reconstruction. A handful of such lifting errors is sufficient to destroy the global geometry.

These observations highlight that even when local reconstruction errors are small, their cumulative effect can lead to large global discrepancies. Therefore, it is important to evaluate both the local reconstruction error, which measures fidelity of reconstruction within short segments, and the global reconstruction error, which captures long-range consistency across the full trajectory. A low local error and a high global error indicates that the pipeline is reconstructing each short segment faithfully, but  that the errors are accumulating in a way that causes the overall shape of the reconstructed path to differ from the original movement path.

\subsubsection{Impact of duration of experiment}

The length of the experiment can impact the reconstruction quality in several ways. If the recording is too short, the toroidal structure in grid cell population activity may not be sufficiently clear, leading to poor toroidal coordinate computation or failure of computation. However, if the experiment is long enough to yield reliable toroidal coordinates but short enough that the path requires few non-trivial lifts, the reconstruction quality will be high. 

In principle, longer experiments should produce clearer toroidal structure and therefore better path reconstructions. However, in the presence of noise, longer experiments are also prone to error accumulation during path lifting. These two effects -- improved toroidal coordinates and accumulated lifting errors -- compete. To determine which effect dominates, we performed the following experiments on both simulated and experimental data.
 
\begin{figure}[h!]
    \centering
    \includegraphics[width=\linewidth]{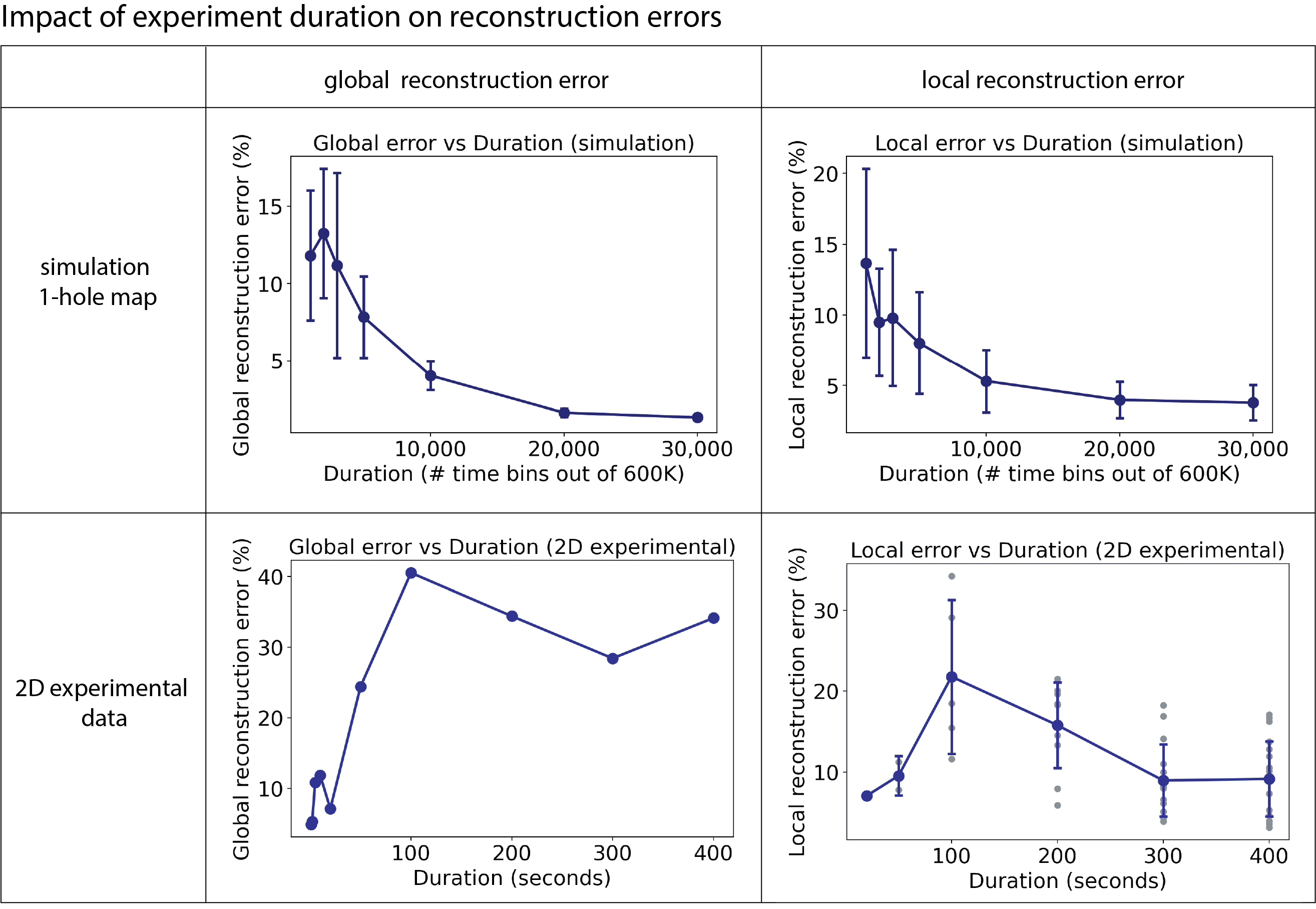}
    \caption{Impact of experiment duration on reconstruction errors. Global (left) and local (right) reconstruction errors for simulated data (top) and two-dimensional experimental data (bottom).} 
    \label{fig:experiment_window_length}
\end{figure}

Using simulated data on 1-hole environment (full simulation: 599,999 time bins), we truncated the grid cell activity to the first 1,000, 2,000, 3,000, 5,000, 10,000, 20,000, and 30,000 time bins and performed path reconstruction on each truncation. Global and local reconstruction errors were computed over $n = 5$ independent simulations. Local reconstruction errors were evaluated using segments of length 10,000. Both global and local reconstruction errors decreased with longer durations (SI Fig.~\ref{fig:experiment_window_length}, top row), suggesting that in the simulated setting, the dominant effect is improved toroidal coordinate quality.

We repeated the analysis on the two-dimensional experimental data
(\cite{gardner_toroidal_2022}, rat~R, module~1, day~2, open-field session; full recording: 21.1 minutes, 126{,}728 time bins).  Grid cell activity was truncated to the first 0.5, 1, 2, 5, 10, 20, 50, 100, 200, 300, and 400 seconds. Because local reconstruction errors were computed using 20-second segments, truncations shorter than 20 seconds were excluded from the local error analysis. The toroidal coordinate computation failed for the 0.5-second truncation.
 
In contrast to the simulated results, the global and local
reconstruction errors exhibited opposite trends (SI Fig.~\ref{fig:experiment_window_length}, bottom row). The local reconstruction error decreased with duration, consistent with the expectation that longer recordings yield better toroidal coordinates. However, the global reconstruction error increased with duration, likely due to accumulation of lifting errors over longer trajectories.

\subsection{Impact of number of neurons}
\label{sec:num_neurons}
 
The number of simultaneously recorded neurons can affect reconstruction quality. To quantify this effect, we subsampled the simulated grid cell population (2,464 neurons total) by randomly selecting $n \in \{50, 100, 200, 500, 1000, 2000, 2464\}$ neurons and performing the full reconstruction pipeline on the reduced population. Each condition was repeated 5 times with independent random draws of neurons. Note that this differs from simulating grid cell activity using a smaller network: the grid cell activity for 2,464 neurons is simulated using the CAN-model, and the subsampling of neurons is performed on the simulated 2,464 neurons.

We carried out the same analysis on the two-dimensional experimental data of Gardner et al.~\cite{gardner_toroidal_2022}, subsampling $n \in \{10, 20, 40, 60, 80, 100, 111\}$ neurons.
 
\begin{figure}[h!]
    \centering
    \includegraphics[width=\linewidth]{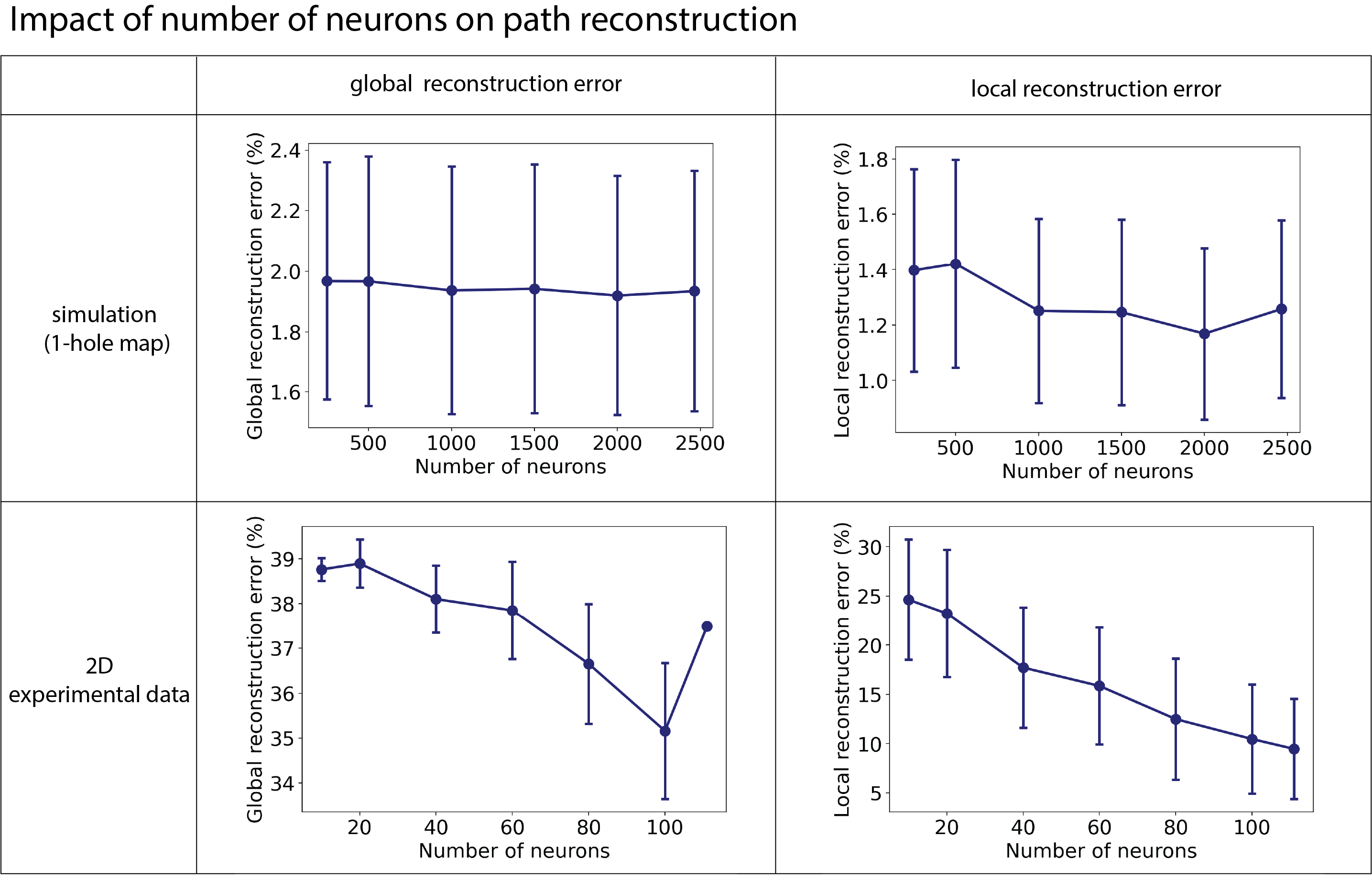}
    \caption{Impact of neuron count on reconstruction errors. Global (left) and local (right) reconstruction errors as a function of the number of neurons used for path reconstruction, for simulated data (top) and two-dimensional experimental data (bottom).}
    \label{fig:neuron_count}
\end{figure}

SI Fig.~\ref{fig:neuron_count} reports the results.  In the simulated data, the toroidal coordinates could not be computed for $n=50$ and $n=100$, so we report the result only for $n=200$ and higher. When there were 200 or more neurons, both reconstruction errors were quite low and remained stable. In the experimental data, both global and local reconstruction errors decreased with increased number of neurons.

\subsection{Impact of metric}
\label{sec:metric}

The persistent cohomology computation is based on a notion of dissimilarity between population vectors, which can be computed via various choices of metric. Furthermore, the toroidal coordinates can be computed using two distinct methods (DREiMac \cite{DREIMAC_Perea2023} and cohomological decoding from Gardner et al. \cite{gardner_toroidal_2022}). To assess the sensitivity to these choices, we compared the performance of path reconstruction across the two toroidal coordinate computation method and three dissimilarity metrics --- cosine, Euclidean, and correlation --- on both simulated and experimental data.
 
\begin{figure}[t!]
    \centering
    \includegraphics[width=1\linewidth]{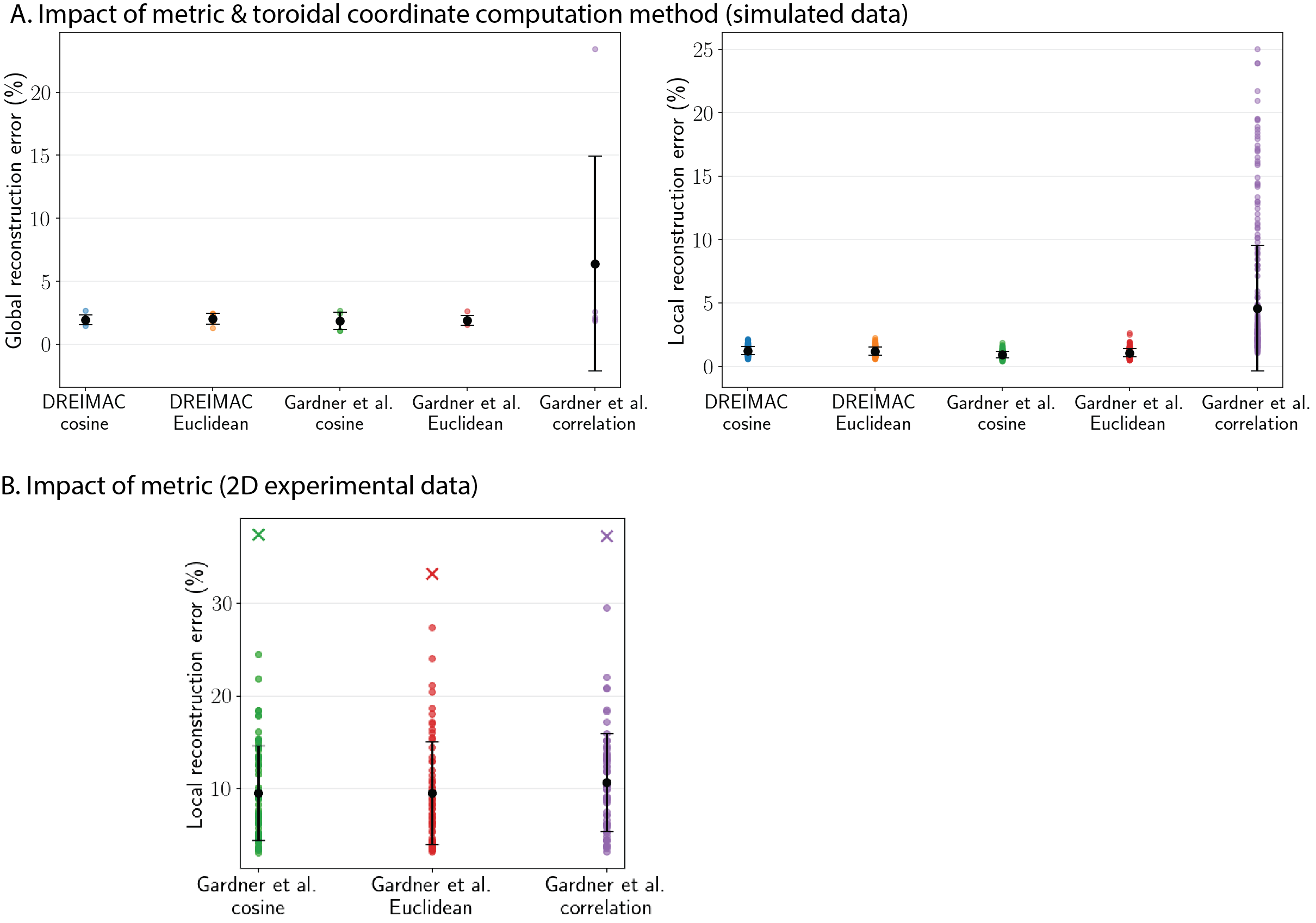}
    \caption{Impact of toroidal coordinate computation methods (DREiMac \cite{DREIMAC_Perea2023} vs Gardner et al. \cite{gardner_toroidal_2022}) and dissimilarity metric on reconstruction errors. For simulated data, the default method is to use DREiMac with Euclidean dissimilarity. For experimental data, we use the toroidal coordinate computation from Gardner et al. \cite{gardner_toroidal_2022} with cosine dissimilarity. \textbf{A.} On the CAN-simulated data on one-hole environment, we compared the performance of path reconstruction using two methods of computing toroidal coordinates (DREiMac and Gardner et al. \cite{gardner_toroidal_2022}) and three dissimilarity metrics (cosine, Euclidean, and correlation) over five independent simulations. Both global (left) and local (right) reconstruction errors were similar, except for using correlation dissimilarity. \textbf{B.}~Global ($\times$) and local ($\circ$) reconstruction errors for two-dimensional experimental data \cite{gardner_toroidal_2022}. }
    \label{fig:metric_comparison}
\end{figure}

For the simulated data, we observed no significant difference between the two toroidal coordinate computation methods and cosine and Euclidean dissimilarity (SI Fig.~\ref{fig:metric_comparison}A). Correlation dissimilarity resulted in noticeably higher errors. 

For the two-dimensional experimental data (Gardner et al.~\cite{gardner_toroidal_2022}, rat~R, module~1, day~2, open-field), where we utilized the cohomological decoding of Gardner et al. \cite{gardner_toroidal_2022}, all three dissimilarity metrics yielded comparable global and local reconstruction errors (SI Fig.~\ref{fig:metric_comparison}B).

Together, these results suggest that the pipeline is relatively robust to the choice of dissimilarity metric and the toroidal coordinates computation method.

\subsection{Impact of noise in toroidal coordinates}
\label{sec:noise_toroidal_coord}

The path-lifting algorithm takes as input a sequence of toroidal
coordinates $ \{ \Theta(t)\} = \{(\theta_x^t, \theta_y^t)\}$ that have been computed via persistent cohomology.  To characterize how sensitive the reconstruction is to noise in toroidal coordinates, we perturbed the toroidal coordinates with additive Gaussian noise and measured the resulting reconstruction quality.
 
Specifically, we first simulated the grid population activity in the one-hole environment and computed the toroidal coordinates. Given toroidal coordinates $(\theta_x^t, \theta_y^t)$ at time $t$, we replaced the coordinates by $(\theta_x^t + n_x,\; \theta_y^t + n_y)$, where $n_x$ and $n_y$ were independently drawn from $\mathcal{N}(0, \sigma^2)$. The perturbed values were then wrapped to $[0, 2\pi)$.  We considered
$\sigma \in \{0, 0.1, 0.2, 0.3, 0.5, 0.75, 1.0, 1.5, 2.0, 3.0\}$ and repeated each condition 5 times.
 
\begin{figure}[h!]
    \centering
    \includegraphics[width=\linewidth]{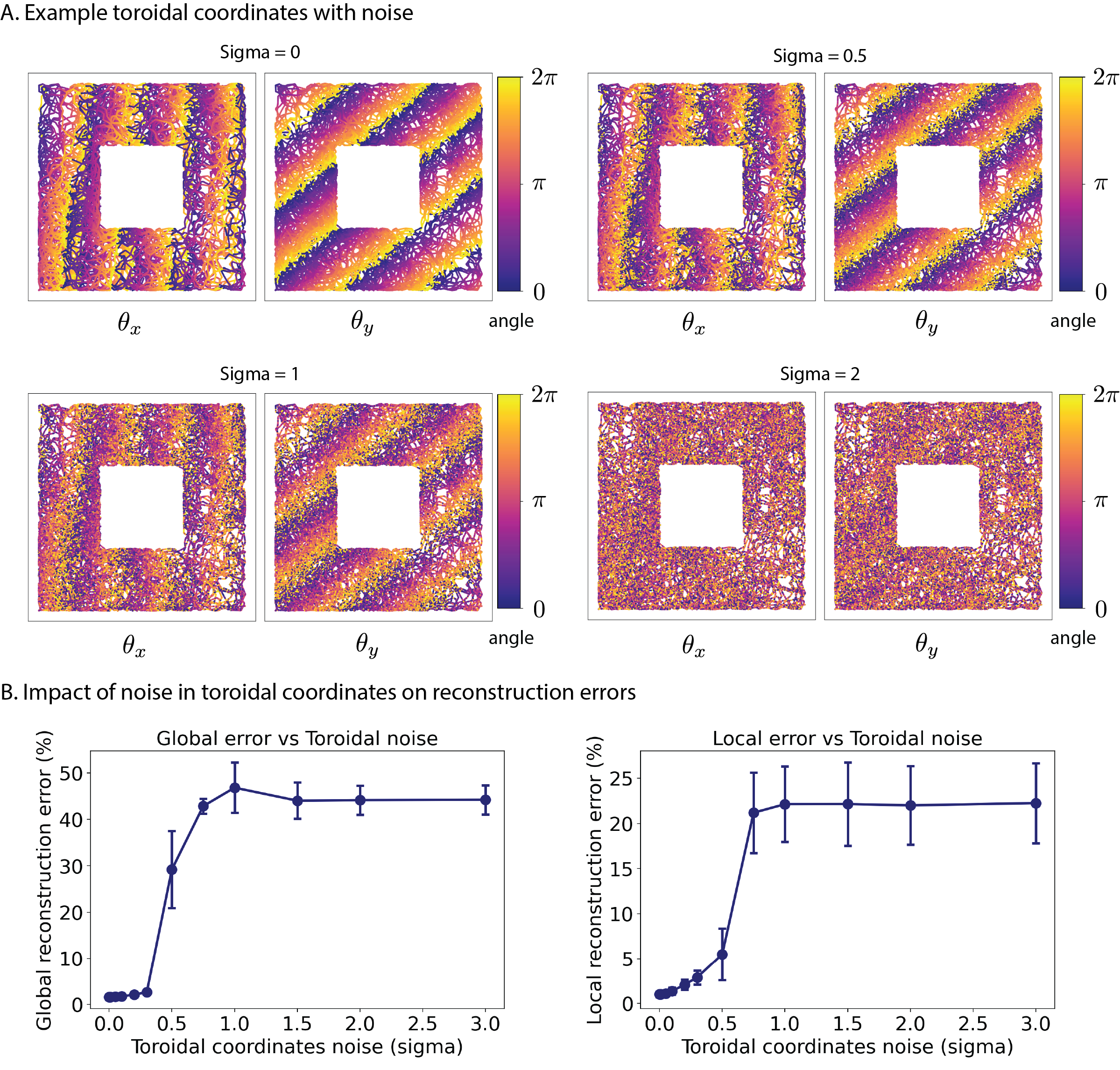}
    \caption{Impact of noise in toroidal coordinates on reconstruction error in simulated data (1-hole environment, $n=5$ repeats). \textbf{A.}~Example toroidal coordinate visualizations at noise levels $\sigma \in \{0, 0.5, 1, 2\}$. \textbf{B.}~Global (left) and local (right) reconstruction errors as a function of $\sigma$.}
    \label{fig:toroidal_coords_noise}
\end{figure}

SI Fig.~\ref{fig:toroidal_coords_noise}A visualizes the perturbed toroidal coordinates for $\sigma \in \{0, 0.5, 1, 2\}$.  As $\sigma$ increases, the coordinate structure becomes progressively less well-defined.  SI Fig.~\ref{fig:toroidal_coords_noise}B reports the global and local reconstruction errors as a function of~$\sigma$. Both error measures remained low for $\sigma \leq 0.3$ and increased sharply around $\sigma = 0.5$.  At $\sigma \geq 1.0$, errors plateaued at high values, indicating that the toroidal coordinate signal had been largely destroyed.

\subsection{Impact of smoothing reconstructed paths from experimental data}
\label{sec:smoothing}

\begin{figure}
    \centering
    \includegraphics[width=\linewidth]{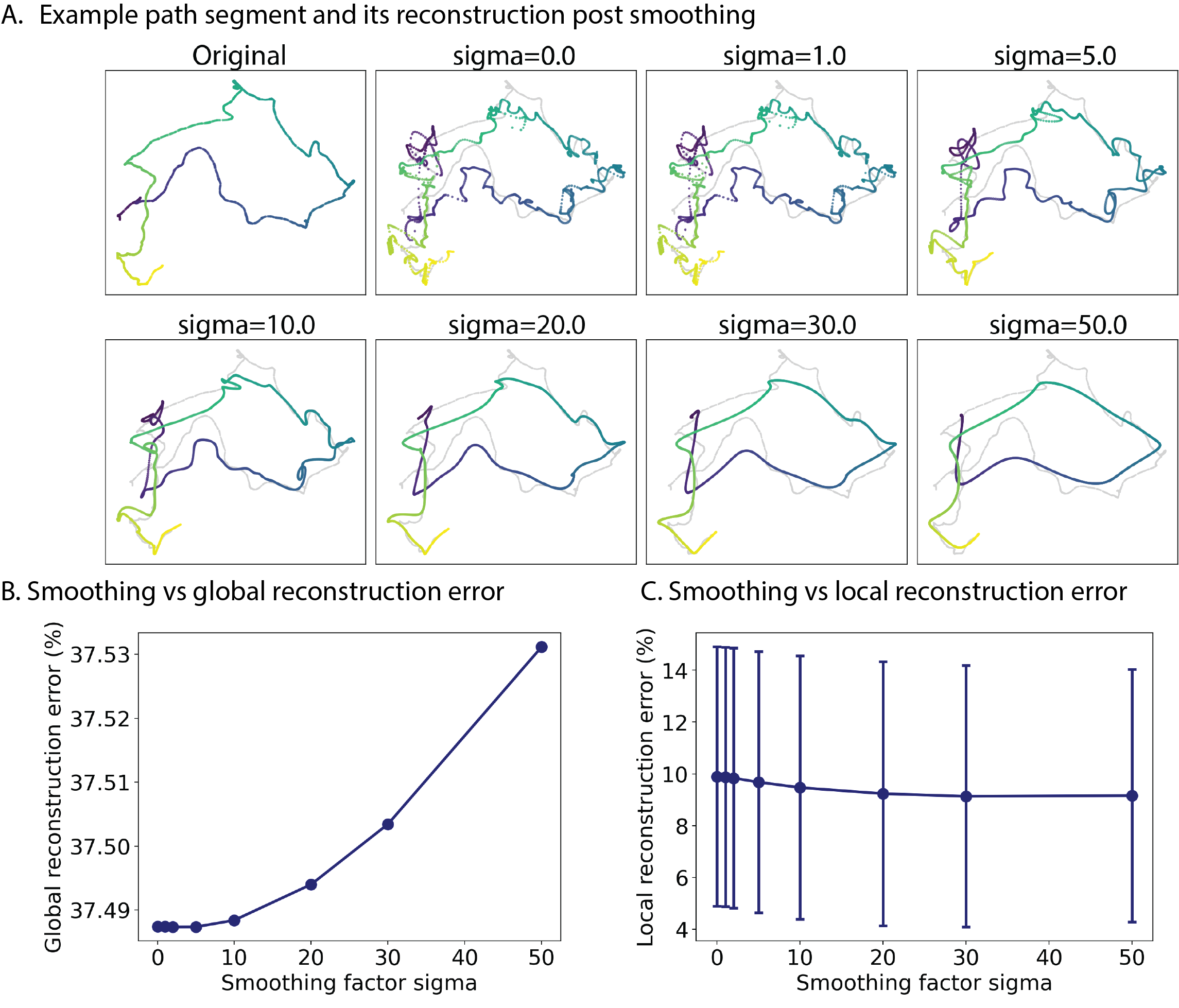}
    \caption{Impact of smoothing on an example reconstructed path from two-dimensional experimental data (rat R, module 1, day 2, OF; \cite{gardner_toroidal_2022}). \textbf{A.} An example local path segment (left, "Original") and its reconstruction after applying Gaussian smoothing with varying standard deviations ($\sigma$ = 0, 1, 5, 10, 20, 30, 50). \textbf{B.} Moderate levels of smoothing (up to $\sigma = 50)$ has minimal impact on global reconstruction error, which stays around 37\%. \textbf{C.} Increased smoothing improves local reconstruction error slightly. }
    \label{fig:path_smoothing}
\end{figure}

Paths reconstructed from two-dimensional experimental data (rat R, module 1, day 2, OF; \cite{gardner_toroidal_2022}) contain a lot of jitter, as illustrated in Fig. 7 of main text, as well as SI Fig.~\ref{fig:path_smoothing}A. One may thus choose to report the reconstructed path after smoothing the path to a certain degree. For example, one might choose to apply Gaussian smoothing with varying standard deviations ($\sigma$ = 0, 1, 5, 10, 20, 30, 50). Applying such smoothing preserves the overall trajectory shape while reducing jitter. Such smoothing has minimal impact on both global and local reconstruction errors (see SI Fig.~\ref{fig:path_smoothing}B, C).  

\clearpage

\section{Supplementary Figures}

\begin{figure}[h!]
    \centering
    \includegraphics[width=0.95\linewidth]{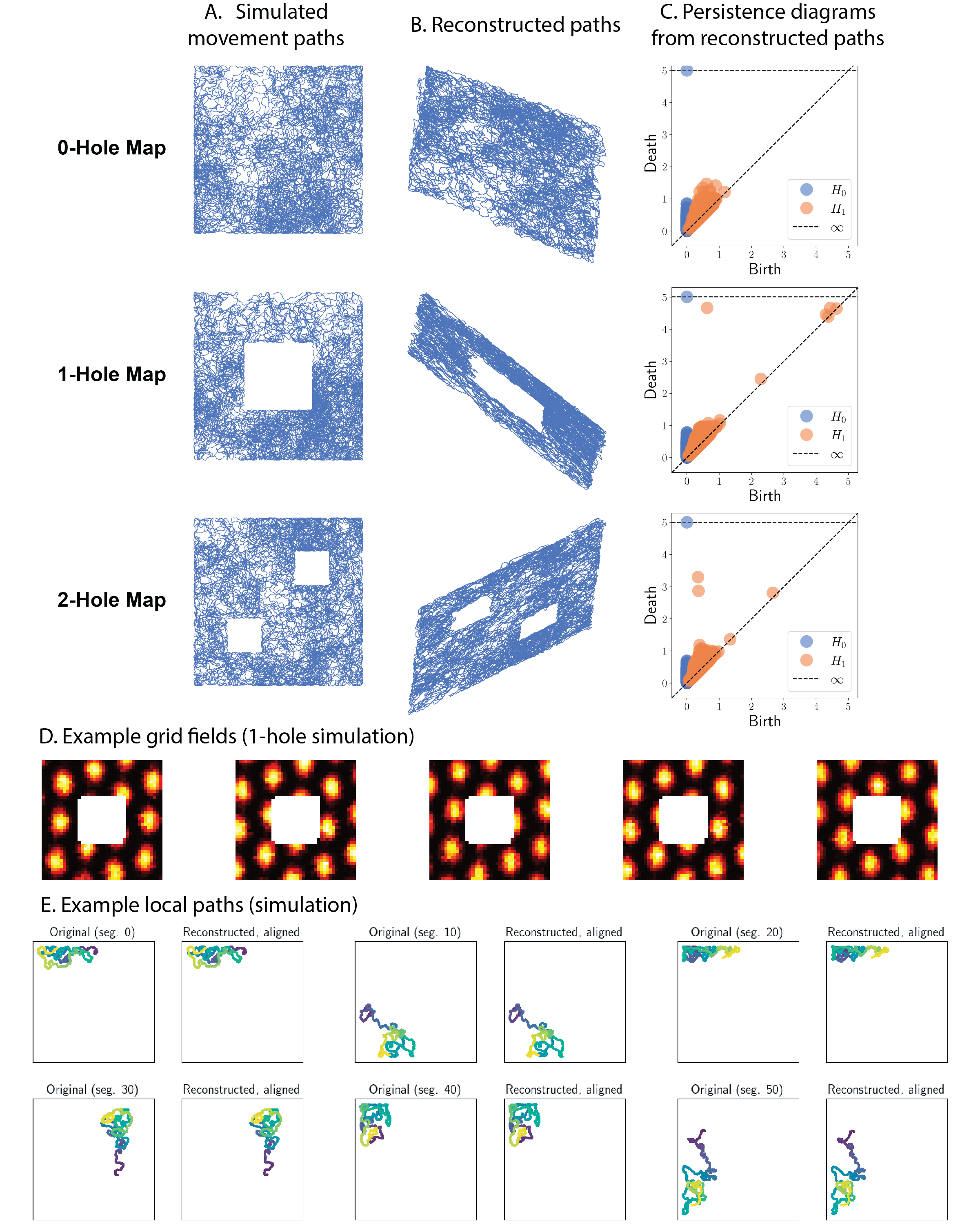}
    \caption{ Path lifting on simulated grid cell activity preserves the global topology of the environment. \textbf{A}. Simulated movement trajectories in environments with 0, 1, and 2 holes.  \textbf{B}. Reconstructed paths from simulated grid cell activity. \textbf{C}. To show that the number of holes in the environment can be recovered from the reconstructed path, we computed pairwise Euclidean dissimilarity between every pair of points in the reconstructed path and computed persistent homology of the Vietoris-Rips filtration. The persistence diagrams computed on the reconstructed path recover the correct number of holes in the environment. \textbf{D.} Example grid fields from simulated grid cells.
    \textbf{E.} Example local paths corresponding to $10,000$ time bins and their reconstructions, post alignment. The original trajectory corresponds to $599,999$ time bins. }
    \label{fig:simulated_SI}
\end{figure}

\clearpage 




\clearpage 

\begin{figure}[h!]
    \centering
    \includegraphics[width=0.9\linewidth]{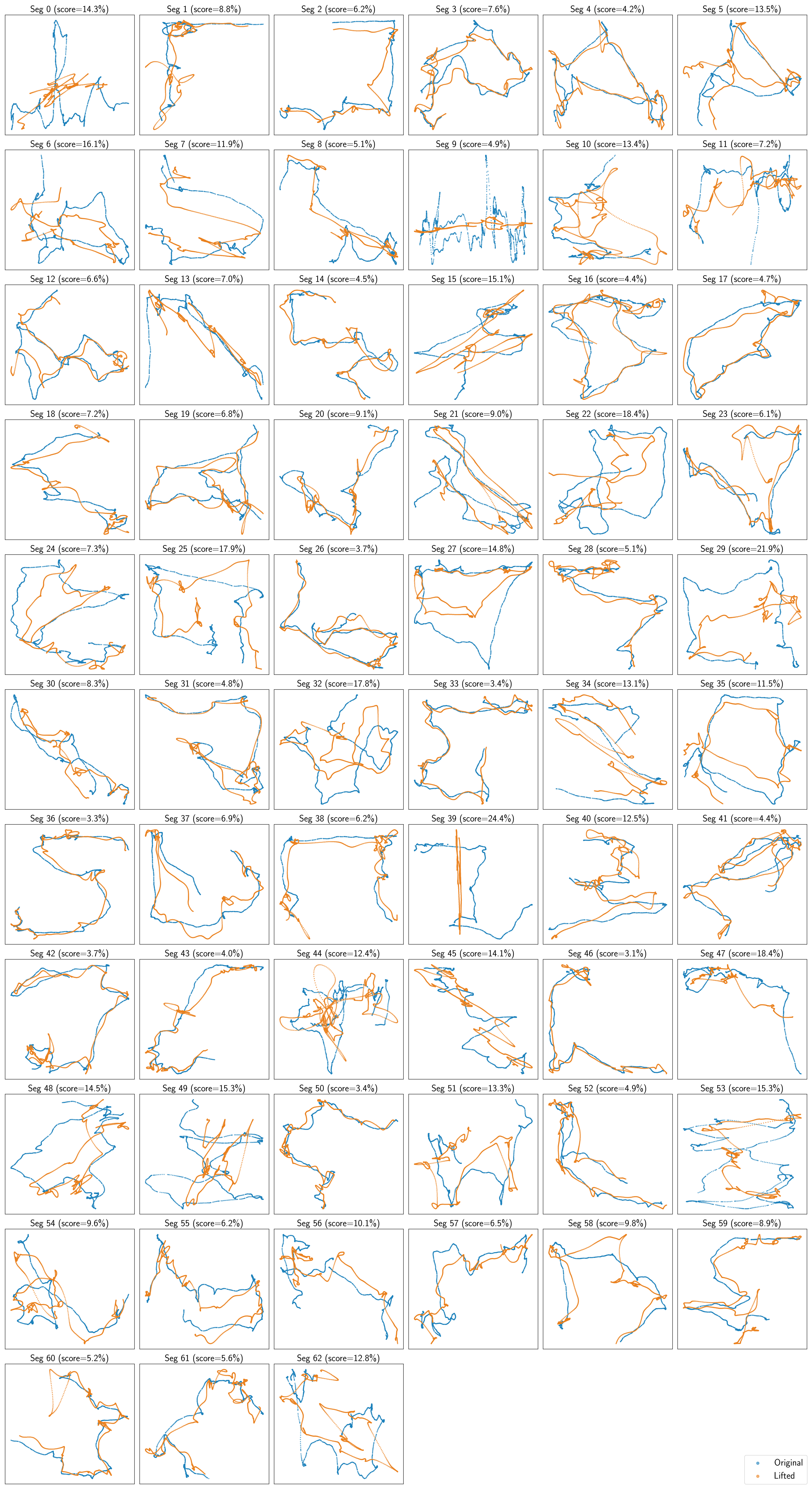}
    \caption{Complete local path reconstruction for two-dimensional experimental data from \cite{gardner_toroidal_2022}. In each panel, blue shows the original movement path, and orange visualizes the reconstructed movement path.} 
    \label{fig:Gardner-data-all}
\end{figure}



\clearpage
\newpage
\printbibliography